\documentclass[11pt]{article}

\usepackage{amsmath, amsfonts, amsthm, amssymb, graphicx}
\usepackage{xspace}
\usepackage{epsfig}
\usepackage{psfrag}
\usepackage{hyperref}
\numberwithin{equation}{section}

\renewcommand{\labelitemi}{\bfseries --}

\newcommand{\op}{\ensuremath{\mathcal{O}}\xspace}

\newcommand{\vev}[1]{\ensuremath{\langle #1 \rangle}\xspace}

\def\ie{{\it i.e.\ }}

\def\A{\mathcal{A}}

\let\a=\alpha \let\b=\beta \let\g=\gamma \let\d=\delta \let\e=\epsilon
\let\z=\zeta    \let\k=\kappa
\let\m=\mu \let\n=\nu  \let\r=\rho
\let\s=\sigma \let\t=\tau    
  \let\D=\Delta 
 \let\F=\Phi 
    \let\G=\Gamma
\let\del=\partial

\let\cdel=\nabla
\renewcommand{\l}{\lambda}
\newcommand{\hf}{\frac{1}{2}}
\newcommand{\half}{\frac12}

\newcommand{\non}{\nonumber}
\newcommand{\qt}{\frac{1}{4}}
\newcommand{\ph}[1]{\phantom{#1}}
\newcommand{\fdel}[2][]{\ensuremath{\frac{\delta #1}{\delta #2}}}

\def\dalemb#1#2{{\vbox{\hrule height .#2pt
        \hbox{\vrule width.#2pt height#1pt \kern#1pt
                \vrule width.#2pt}
        \hrule height.#2pt}}}
\def\square{\mathord{\dalemb{6.8}{7}\hbox{\hskip1pt}}}

\newcommand{\p}{\partial}

\def\0{{\sst{(0)}}}
\def\1{{\sst{(2)}}}
\def\2{{\sst{(2)}}}
\def\3{{\sst{(3)}}}
\def\4{{\sst{(4)}}}
\def\5{{\sst{(5)}}}
\def\6{{\sst{(6)}}}
\def\7{{\sst{(7)}}}
\def\8{{\sst{(8)}}}

\def\ep{\epsilon}
\def\td{\tilde}

\def\nn{\nonumber}

\let\pa=\partial 
\newcommand{\be}{\begin{equation}}
\newcommand{\ee}{\end{equation}}
\def\ba{\begin{array}}
\def\ea{\end{array}}

\def\del{\partial}
\def\sst#1{{\scriptscriptstyle #1}}
 
\def\ie{{\it i.e.\ }}

\def\bo1{ \left | B^0 (p^+) \right \rangle}

\def\<{ \langle }
\def\>{ \rangle }

\newcommand{\bea}{\begin{eqnarray}}
\newcommand{\eea}{\end{eqnarray}}

\newcommand{\tr}{{\rm tr} }

\begin{document}

\begin{flushright}
\end{flushright}

\vspace{25pt}

\begin{center}

{\Large \bf{Holography for Schr\"{o}dinger backgrounds}}

\vspace{20pt}

%\auth
{\large Monica Guica${}^{\spadesuit}$, Kostas Skenderis${}^{\clubsuit, \diamondsuit}$, Marika Taylor${}^{\clubsuit}$  and Balt van Rees${}^{\clubsuit}$}

\vspace{15pt}

\begin{itemize} \renewcommand{\labelitemi}{${}^\spadesuit$}
\item{\it  Laboratoire de Physique Th\'{e}orique et Hautes Energies (LPTHE) \\
                          CNRS, UPMC, Univ. Paris 6, Boite 126, 4 Place Jussieu, \\
75252 Paris Cedex 05, France}
  \renewcommand{\labelitemi}{${}^\clubsuit$}
\item  {\it Institute for Theoretical Physics \\
Science Park 904, Postbus 94485, 1090 GL Amsterdam, The Netherlands}
\renewcommand{\labelitemi}{${}^\diamondsuit$}
\item {\it Korteweg-de Vries Institute for Mathematics \\
Science Park 904, Postbus 94248, 1090 GE Amsterdam, The Netherlands}
  \end{itemize}

\bigskip
 {\it E-mail:}
 {\tt {mmguica@lpthe.jussieu.fr, k.skenderis@uva.nl, m.taylor@uva.nl, b.c.vanrees@uva.nl}}

%\vspace{15pt}

%{\ams}

\vspace{20pt}

\begin{abstract}

We discuss holography for Schr\"{o}dinger solutions of both topologically massive gravity in three dimensions
and massive vector theories in $(d{+}1)$ dimensions. In both cases
the dual field theory can be viewed as a $d$-dimensional conformal field theory (two dimensional in the case of TMG)
deformed by certain operators that respect the Schr\"{o}dinger symmetry.
These operators are irrelevant from the viewpoint of the relativistic conformal group but they are exactly marginal with respect to the
non-relativistic conformal group. %Had one of the lightcone directions been
%compactified the corresponding quantized momentum $k_v$ would be the discrete particle number in the resulting $(d{-}1)$-dimensional theory, but in this paper we do not compactify any of the lightcone directions.
The spectrum of linear fluctuations around the background solutions corresponds to operators
that are labeled by their scaling dimension and the lightcone momentum $k_v$. We set up the
holographic dictionary and compute 2-point functions
of these operators both holographically and in field theory using conformal perturbation theory
and find agreement. The counterterms needed for holographic renormalization
are non-local in the $v$ lightcone direction.

\end{abstract}

\end{center}

\newpage
{\small \tableofcontents}
\newpage

\section{Introduction and summary}

Gauge/gravity dualities have become an important new tool in extracting strong coupling physics.
The best understood examples of such dualities involve relativistic quantum field theories.
Strongly coupled non-relativistic QFTs are common place in condensed matter physics and as
such there would be many interesting applications had one had under control holographic
dualities involving non-relativistic QFTs.
Motivated by such applications \cite{Son:2008ye,Balasubramanian:2008dm} initiated a discussion of holography\footnote{Note that
a Kaluza-Klein framework for the geometric realization of Schr\"{o}dinger symmetries was introduced much earlier, in \cite{Duval:1990hj}; the
relation of this work to the holographic framework is discussed in \cite{Duval:2008jg}.} for
$(d+1)$ dimensional spacetimes with metric,
\be
\label{eq:nrmetric}
ds^2 = - \frac{b^2 du^2}{r^4} + \frac{2 du dv + dx^i dx^i  + dr^2}{r^2}\,,
\ee
with $i \in \{1, \ldots, d-2\}$. The isometries of this metric form the so-called \emph{Schr\"odinger group},
whose generators are given by:
\begin{align}
&\mathcal H : u \rightarrow u + a, \nn \\
&\mathcal M : v \rightarrow v + a, \nn \\
&\mathcal D : r \rightarrow (1 - a) r, & & u \rightarrow (1-a)^2 u, & & v \rightarrow v, && x^i \rightarrow (1-a) x^i \label{group} \\
&\mathcal C : r \rightarrow (1 - a u) r, & &u \rightarrow (1 - a u) u, & &v \rightarrow v + \frac{a}{2} (x^i x^i + r^2), && x^i \rightarrow (1 - a u) x^i\nn
\end{align}
plus rotations, translations and Galilean boosts in the $x^i$ directions. Here $(\mathcal D,\mathcal C)$ are
the generators of non-relativistic scale transformations
with dynamical exponent $z=2$ and special conformal symmetries, respectively.

It was initially suggested that the metric \eqref{eq:nrmetric} could play the role of a background for the holographic study of critical non-relativistic
systems in $(d-1)$ spacetime dimensions, for example fermions at unitarity, which have the same symmetry group. For such theories the specific
realization of the Schr\"odinger group on \eqref{eq:nrmetric} dictates that the $(d-1)$ dimensional theory should live on the spacetime spanned by the
coordinates $(u,x^i)$ with $u$ playing the role of time. In this setup one considers operators ${\cal O}_{\Delta_s,m} (u,x^i)$ of definite scaling
dimension $\Delta_s$ and of charge $m$ under the symmetry ${\mathcal M}$. This charge $m$, which corresponds to momentum in the $v$ direction, would
then have to be identified with a discrete quantum number like particle number. In order to discretize the possible values of $m$ one
therefore needs to compactify the $v$ direction. This procedure is however very nontrivial as in general quantum corrections become
important and one cannot trust the metric \eqref{eq:nrmetric} with a compact null direction \cite{Maldacena:2008wh}.
Recent work aiming at obtaining solutions without such a compact direction can be found in \cite{Balasubramanian:2010uw}.

%A peculiarity of this proposed holographic duality is that it realizes the Schr\"{o}dinger group as the symmetry group of a spacetime with $(d+1)$
%asymptotically `large' dimensions and the usual holographic arguments then predict that the dual field theory lives in $d$ rather than
%$(d-1)$ spacetime dimensions.
%Another oddity is the fact that although

The metric \eqref{eq:nrmetric} has constant negative scalar curvature but it is \emph{not} an asymptotically
locally AdS (or AlAdS) spacetime. (We review this further below.)
This property sets it apart from other gauge/gravity dualities,
which are based upon the notion of AlAdS spacetimes. It should be stressed that this implies that a priori one cannot extend any of the
standard holographic results to such spacetimes. For example, there are no guarantees that the dual description has the form
of an ordinary local, renormalizable quantum field theory.

The principal aim of this paper is to understand how holography works for these spacetimes.
To avoid the complications of a compact null direction, we consider \eqref{eq:nrmetric} with $v$
non-compact. The effects of such a compactification may be considered afterwards but this issue will
be for the most part suppressed in this paper. We will present evidence that
the spacetime \eqref{eq:nrmetric} is dual to a $d$-dimensional quantum field theory that is
non-local in the $v$ direction.
More precisely, our viewpoint is that the dual quantum field theory can be obtained from a
$d$-dimensional conformal field theory by deforming with operators that respect the Schr\"odinger
symmetry. These operators are irrelevant from the perspective of the relativistic
conformal group but they are exactly marginal from the perspective of the non-relativistic
conformal group.
As a first consistency check, notice that an irrelevant deformation generally changes the UV properties of the theory and this explains why the dual gravity solution is no longer asymptotically anti-de Sitter.
%Moreover, the deformation (\ref{eq:nrmetric}) becomes small
%in the interior of the spacetime, corresponding to the IR of the field theory, as expected for an irrelevant deformation.
In our context the UV properties are now governed by the Schr\"odinger group and indeed the solution
 realizes this symmetry geometrically.

Since the deformed QFT is Schr\"odinger symmetric, we should organize it in a way that makes this
symmetry manifest. To discuss this let us first recall that any relativistic $d$-dimensional
CFT is also invariant
under the $(d-2)$ dimensional Schr\"odinger group. This follows from the fact the conformal group in $d$ dimensions
has as a subgroup the $(d-2)$-dimensional Schr\"odinger group (as first discussed in
\cite{CMP34(85)} for $d=4$)\footnote{The embedding
is the following. Choosing lightcone coordinates
$u,v$, the relativistic momentum generators $P_u$ and $P_v$ are identified with ${\cal H}$ and ${\cal M}$,
respectively, the non-relativistic scaling generator ${\cal D}$ is a linear combination of the
relativistic scaling generator and a boost in the $uv$ direction and ${\cal C}$ is related to a relativistic
special conformal generator. Translations, rotations and Galilean boosts and related to translations and
rotations in the relativistic theory. More details can be found, for example, in \cite{Son:2008ye} or \cite{Maldacena:2008wh}.}. It follows that it should be possible to rewrite the
correlation function in a form that manifests the Schr\"odinger symmetry. Indeed, as discussed
recently in \cite{Barnes:2010ev}, the mixed representation
of the 2-point function where one Fourier transforms over the lightcone coordinate $v$ brings the
relativistic 2-point function into the form dictated by the
Schr\"odinger symmetry \cite{Henkel:1993sg}.

Schr\"{o}dinger solutions have arisen as solutions of various
gravitational theories; in this paper we will focus on topologically
massive gravity (TMG) in three dimensions and the massive vector model
used by Son \cite{Son:2008ye}. Schr\"{o}dinger solutions of TMG arise
for a specific value of the coupling, namely $\mu =3$, and in earlier
literature have been referred to as null warped $AdS_3$ backgrounds. The
operator $X_{vv}$ which deforms the dual theory has relativistic
scaling dimensions $(h,\bar{h})=(3,1)$, breaks the Lorentz invariance
and is dual to a null component of the extrinsic curvature, as discussed
in \cite{Skenderis:2009nt,Skenderis:2009kd}. Gravity coupled to a
cosmological constant and a massive vector with $m^2 = 2d$
also admits Schr\"{o}dinger solutions in general dimension $d$;
moreover the solution for $d=4$ can be understood as a consistent truncation
of a decoupling limit of TsT transformed branes
\cite{Maldacena:2008wh}. In this case the deforming
operator $X_v$ has relativistic scaling dimension $d+1$ and
is dual to a null component of the massive vector.

As mentioned above, in theories with Schr\"odinger invariance
operators are labeled by the non-relativistic scaling dimension $\D_s$
and the eigenvalue of ${\cal M}$, which in our case is the
right-moving momentum $k_v$. We will therefore Fourier transform in
the right moving sector and consider the operators with different $k_v$
as independent operators.  If $v$ is compact then $k_v$ would be a
discrete label, but as mentioned above we do not compactify $v$ and
therefore $k_v$ is a continuous variable. The
deformation of the original CFT for the
case of the massive vector is of the
form,
\be \label{deformation}
S_{\rm CFT} \to S_{\rm CFT} + \int d^{d{-}2}\!x du dv \
b^\mu X_\mu(v,u,x^i)
= S_{\rm CFT} + \int d^{d{-}2}\!x du\  b^\mu \tilde{X}_\mu(k_v{=}0, u, x^i),
\ee
where $S_{\rm CFT}$ is the original CFT action,
$b^\mu$ is a constant null vector whose only non-vanishing component
is $b^v= b$ and  $\tilde{X}_\mu$ is the Fourier transform of
$X_\mu$ in the $v$ direction. In the rest of this paper we will often drop the
tilde and (with abuse of notation) denote an operator and its
Fourier transform with respect to $v$ by the same name.
We see from (\ref{deformation}) that the theory is deformed by an
operator of zero lightcone momentum. The operator $X_\mu$ has dimension
$d+1$ from the perspective of the relativistic CFT  and breaks the
Lorentz symmetry. Since its (relativistic) dimension is different
from $d$, one can use relativistic scaling\footnote{Alternatively, 
one can set $b=1$ by rescaling of the lightcone \label{ft:sc}
coordinates, $u \to u/b, \ v \to v b$. 
This transformation
is a composition of a relativistic scaling with parameter $b$ and a
Schr\"odinger scaling with parameter $1/b$.} to 
set $b=1$.
This operator however is exactly marginal from the perspective of the
Schr\"odinger symmetry, i.e. its non-relativistic scaling dimension is
$\D_s=d$ and this implies that the action (\ref{deformation}) has 
Schr\"odinger invariance. These facts have an exact counterpart in the 
bulk: one can set $b=1$ in (\ref{eq:nrmetric})
by either by a bulk diffeomorphism that scales
all coordinates by $b$ (corresponding to the relativistic rescaling)
or by rescaling the lightcone coordinates as in footnote \ref{ft:sc} 
and the metric has 
Schr\"odinger isometries. 
The discussion for the case of TMG is similar: one replaces $b^\mu$
by the symmetric null tensor $b^{\mu \nu}$
with the only non-zero component being $b^{vv}=-b^2$ and
$X_\mu$ by $X_{\mu \nu}$, where $X_{\mu \nu}$ has relativistic scaling $(3,1)$
and non-relativistic scaling $\D_s=2$.

To show that the deformation is exactly marginal we have to show the
non-relativistic scaling dimension of deforming operator, which was
equal to $\D_s=d$ in the original CFT, remains
equal to $\D_s=d$ in the deformed theory. This amounts to showing that
the 2-point function of this operator in the deformed
theory is the same as the 2-point function in the original CFT and
this can be proven using conformal perturbation theory.
In more detail, consider for concreteness the case of the
massive vectors, then one needs to establish that for any $n$
\be \label{marg}
\lim_{k_v \to 0} \left \langle X_v(k_v) \left(\prod_{i=1}^n
\int d^{d{-}2}\!x_i du_i\  b^\mu X_\mu(k_v{=}0)\right) X_v(-k_v)
\right \rangle_{CFT}
=0
\ee
where the expectation value is taken in the original CFT and to avoid
clutter we only display the $k_v$ dependence. One can show
%scaling argument (which we explain in detail in section \ref{sec:fieldthy})
 that
\be \label{corr}
 \left \langle X_v(k_v) \left(\prod_{i=1}^n b^\mu X_\mu(k_v{=}0)\right)
X_v(-k_v) \right \rangle_{\rm CFT} =
 \left \langle X_v(k_v) X_v(-k_v) \right \rangle_{\rm CFT}
(b^v k_v)^n  f(\log k_v,...)
\ee
where  $f(\log k_v,...)$ is a dimensionless function that
carries the dependence on the positions of the operators and is
at most logarithmically dependent on $k_v$. 
Taking the limit $k_v \to 0$
we indeed find that the rhs of (\ref{corr}) vanishes and thus
(\ref{marg}) is satisfied. Consequently the
operator $X_v(k_v{=}0)$ is exactly marginal.

The same computation shows that operators with $k_v \neq 0$
will in general acquire an anomalous dimension,
\be
\Delta_s = \Delta_s(b=0) + \sum_{n> 0} c_{n} (b k_v)^n. \label{pert-dimb}
\ee
where $\Delta_s(b=0)$ is the non-relativistic scaling dimension
in the original CFT and the $c_n$ are computable numerical coefficients.
This discussion holds not only for $X_v(k_v)$ but also for general
composite operators ${\cal O}(k_v)$,
i.e. in general, when $k_v \neq 0$, they
acquire anomalous dimensions in the deformed theory.
We will discuss explicit examples in section \ref{sec:fieldthy}.

Note that $b^\mu$ appears always in the combination $b^v k_v$
and this quantity is invariant under the rescalings 
(discussed above) that can set $b$ to any non-zero value. 
Later on, when we study in section 
\ref{sec:fieldthy} specific examples using conformal perturbation 
theory to leading order the small parameter will be $b^v k_v$.

The fact that $\Delta_s$
depends on $k_v$ has several important consequences. Correlation functions of composite operators
in general require renormalization and the corresponding
counterterms contain poles when $\D_s$ takes integer values because
new infinities arise when $\D_s$ is an integer. The dependence of
$\Delta_s$ on $k_v$ then implies that the
counterterms are non-polynomial in $k_v$ (when $b k_v \ll 1$, they are polynomial but
contain an infinite number of $k_v$ factors) and thus non-local in the $v$ direction.

 The stress energy tensor in the field theory is somewhat more subtle
because there are actually two related stress energy operators. One is
the operator that couples to the metric; this is a natural operator
when considering the theory as a deformation of the CFT. This operator
however is not conserved except at $b=0$.  The second operator is the
one that couples to the vielbein; this tensor is not symmetric but it
is conserved and it is the natural operator to consider in the
deformed theory. We will illustrate this point with a toy
Schr\"{o}dinger invariant field theory in section \ref{sec:fieldthy}.

We now move to the gravitational side.
%Next we explore the holographic dictionary for Schr\"{o}dinger backgrounds.
In all other examples of holography one sets up a
holographic dictionary as follows: one first derives the most general
asymptotic solutions of the field equations consistent with the
boundary conditions. Substituting these solutions into the on-shell
action $S$ allows one to regulate the volume divergences and derive a
covariant local boundary counterterm action $S_{\rm ct}$ which
renormalizes the action, via $S_{\rm ren} = S + S_{\rm ct}$
\cite{Skenderis:2002wp}.  Renormalized one point functions in the
presence of sources are then derived by applying the GKPW prescription
\cite{Gubser:1998bc,Witten:1998qj} to the renormalized action, namely:
\be
\< {\cal O} \> = \frac{1}{\sqrt{g}} \frac{\delta S_{\rm ren}}{\delta \Phi_{(0)}},
\ee
where the boundary value $\Phi_{(0)}$ acts as a source for the
operator ${\cal O}$. Higher correlation functions are obtained by
further functional differentiation; to obtain $n$-point functions one
will need exact regular solutions of the bulk field equations to order
$(n-1)$ in fluctuations.

The first question that we address is whether such a holographic
dictionary can be set up for probe scalar operators, dual to minimally
coupled scalar fields. We find that indeed there is such a dictionary,
but with a key conceptual difference to all earlier examples of
holography: {\it the boundary counterterms are non-local
in the lightcone direction $v$}. This is consistent with the
the field theory discussion which implies that, once we turn on a source
for an operator with non-zero $k_v$, there would generically be counterterms
that depend on $(b k_v)^n$ for all $n$, and thus provides structural evidence for the
holographic duality. As previously noted in
\cite{Son:2008ye,Balasubramanian:2008dm,Goldberger:2008vg} the
Schr\"{o}dinger dimension for a scalar operator has the closed form:
\be
\Delta_s = \frac{d}{2} + \sqrt{ \left (\frac{d}{2} \right )^2
+ m^2 + b^2 k_v^2}, \label{resum-dimb}
\ee
where $m^2$ is the mass squared of the bulk scalar field (in units of
the curvature radius). This indicates that the series expansion like the
one in (\ref{pert-dimb}) should resum %into the surd of 
the square root form (\ref{resum-dimb}).
%it would be interesting to derive this resummation using a null dipole
%realization of the dual field theory at finite $b$.

We then focus on the gravitational sector of TMG and the massive vector
theories. As a first step we consider linearized fluctuations about
the Schr\"{o}dinger backgrounds.
%Recall that the Schr\"{o}dinger
%solutions have anisotropic bulk stress energy tensors, which are
%supported by higher derivative terms (in TMG) and the massive
%vector. Since the metric is coupled to the bulk fields dual to the
%irrelevant deforming operators, their linearized fluctuations are also
%coupled and the fluctuation equations need to be diagonalized.
We consider here the linearized equations %in three bulk dimensions
in radial axial gauge ($h_{r i} = h_{rr}= 0$) such that
\be
\label{eq:nrfluc}
ds^2 = - \frac{b^2 du^2}{r^4} + \frac{2 du dv + dr^2}{r^2} +
\frac{h_{ij}}{r^2} dx^{i} dx^{j} \,
\ee
along  with vector fluctuations in
the vector model.
%being ${\cal A}_{\m}$.
We give the general solutions for these
linearized fluctuations in sections \ref{sec:linearizedtmg} and
\ref{sec:linearizedvector}. In particular, we note that the number of
independent solutions corresponds to the correct number of sources for
dual operators and their expectation values (subject to constraints
related to dual operator Ward identities).

These solutions have a number of interesting features which we now
briefly summarize. Both models admit two distinct sets of solutions to
the linearized equations, which we distinguish with superscripts `T'
and `X'. The `T' solutions are associated with the dual stress
energy tensor, whilst the `X' solutions are associated with the dual
deforming operator. Note however that all bulk field fluctuations are
non-zero in both sets of solutions.

Looking first at the `X' solutions, we see that indeed these exhibit
the behavior expected for a bulk field dual to an operator with a
$k_v$ dependent scaling dimension. In the case of TMG, the leading
component of the fluctuation of the extrinsic curvature,
$\bar{\kappa}_{uu}$, acts as a source for the dual operator
$X_{vv}$. The asymptotic expansion of this linearized fluctuation is
%derived from $H^T_{mn}$ and is:
\bea
\bar{\kappa}_{uu} &=& \bar{\kappa}_{(0) uu} r^{-\Delta_s -2} (1 + \cdots) + \bar{\kappa}_{(2 \Delta_s - 2) uu} r^{\Delta_s - 4} ( 1 + \cdots), \\
\Delta_s &=& 1 + \sqrt{1 + b^2 k_v^2}, \nn
\eea
where the ellipses denote subleading terms as $r \rightarrow 0$. Then
$\bar{\kappa}_{(0) uu}$ acts as a source for the dual operator
$X_{vv}$, and $\Delta_s$ is its scaling dimension; moreover the two
point function of this operator is indeed found to be of the expected
Schr\"{o}dinger form. The scaling dimension depends on the lightcone
momentum, as shown in the field theory, and resums into the square root. At
$b=0$ the Schr\"{o}dinger dimension is in agreement with that of a
relativistic dimension $(3,1)$ operator, as required by the AdS/CFT
dictionary.

The case of the massive vector is analogous: the `X' solutions are
associated with the dual operators $X_u$ and $X_v$, the latter of
which is the deforming operator. The asymptotic expansions indicate
that the Schr\"{o}dinger dimensions of these operators are
respectively:
\be
\Delta_s (X_v) = 1 + \sqrt{1 + b^2 k_v^2}; \qquad
\Delta_s (X_u) = 1 + \sqrt{9 + b^2 k_v^2},
\ee
which are consistent with the CFT operator dimensions at $b=0$ and are
again expressed as closed forms in $(b k_v)$.

An important feature of both these solutions %and of the `T' modes
is that they diverge faster as $r \rightarrow 0$ than the background
solution. This is because the operators with non-zero $k_v$ renormalize
and while these operators were marginal (w.r.t. Schr\"odinger scaling) at $b=0$
they become irrelevant when $b \neq 0$. Irrelevant operators change
the UV behavior of the theory
and should hence modify the leading asymptotics of the holographic
dual. This is precisely what happens here: the faster rate near
$r \to 0$ is precisely that dictated by the anomalous dimension.

Now let us turn to the `T' modes.
In TMG the metric perturbation takes the form:
\bea
h^T_{uu} &=& \frac{1}{r^2} h_{(-2)uu} + \tilde h_{(0)uu} \log(r^2)
+ h_{(0)uu} + r^2 h_{(2)uu} \nn \\
h^T_{uv} &=& \frac{1}{r^2} h_{(-2)uv}  + \tilde h_{(0)uv} \log(r^2)
+ h_{(0)uv}+ r^2 h_{(2)uv}\\
h^T_{vv} &=& h_{(0)vv} + r^2 h_{(2)vv}, \nn
\eea
with analogous results for the massive vector model. The precise form
of the coefficients $h_{(a) ij}$ as functions of $(u,v)$ is given in
section \ref{sec:linearizedtmg}; there are six independent
coefficients subject to three constraints.
The metric perturbation diverges faster as $r \rightarrow 0$
than the background solution; it does not respect the falloff
conditions discussed in \cite{Horava:2009vy}. This is associated with
the fact that certain components of the dual stress tensor are
irrelevant with respect to the Schr\"{o}dinger dilatation symmetry.
It is possible to impose by hand the constraint that the metric
perturbations should not blow up faster than the background metric as
$r \rightarrow 0$, but this generically imposes constraints on the
operator sources in the dual field theory. In previous works such as
\cite{Son:2008ye,Herzog:2008wg,Ross:2009ar} only such solutions with
constrained asymptotics were discussed, and indeed the fact that the
sources were apparently constrained was already noticed in
\cite{Herzog:2008wg}.

These `T' modes should correspond to the stress energy
tensor in the dual field theory. However, as noted above, there are
subtleties in setting up the holographic dictionary in this case. One
issue is the fact that one component of the stress energy tensor is an
irrelevant operator, as mentioned above.  The other is the fact that
the conserved stress energy tensor for the field theory should couple to
the vielbein, rather than the metric, and therefore, as noted in
\cite{Ross:2009ar}, the appropriate variational problem in the bulk
corresponds to specifying boundary conditions for the vielbein.  This
involves reformulating the bulk theories in a vielbein formulation and
will be discussed elsewhere.

The fact that the Schr\"{o}dinger spacetime is obtained from anti-de
Sitter by irrelevant deformations was already briefly noted in
\cite{Son:2008ye}.  %The main goal here is to explore the implications of this fact for holography using field theoretic as well as bulk computations.
In \cite{Maldacena:2008wh} the relation of
Schr\"{o}dinger to anti-de Sitter via TsT transformations was used to
argue that the dual field theory should be a null dipole theory; see
also \cite{Herzog:2008wg,Adams:2008wt} for additional discussion of
the TsT transformations required.  This observation is
consistent with the view espoused here: the irrelevant deformations
should resum into a null dipole theory which respects Schr\"{o}dinger
invariance. Note that all terms in the dipole theory are
exactly marginal w.r.t. the Schr\"odinger symmetry.
We postpone to subsequent work full exploration of the
null dipole structure, since null dipole theories have not been
developed in earlier literature and their properties differ
qualitatively from the spacelike dipole theories developed in
\cite{Bergman:2000cw,Dasgupta:2000ry,Bergman:2001rw}.

%Throughout this paper we will accumulate evidence that the holographic dual
%to these backgrounds is a $d$-dimensional theory which is the deformation a CFT by null, irrelevant operators.
Whilst our viewpoint is consistent with the earlier proposals of
\cite{Son:2008ye} and \cite{Maldacena:2008wh} in the massive vector
case, a somewhat different proposal has been made for the holographic
dual to null warped backgrounds of TMG. In \cite{Anninos:2008fx} it
was suggested that the holographic dual to the null warped background
should be a two-dimensional CFT with certain central charges
$(c_L,c_R)$. Applying the Cardy formula at finite temperature using
these central charges gives an entropy in agreement with that of a
black hole which has null warped asymptotics. This agreement is
certainly thought-provoking, but there are conceptual challenges with
the suggestion that the dual field theory is a conformal field
theory. %As discussed above, Schr\"{o}dinger backgrounds are not
%asymptotically AdS and therefore the asymptotic symmetry group does
%not match that of a conformal field theory. Cardy's formula should
%also only be applied to unitary theories and TMG is not unitary.
%Furthermore, when $b^2$ is small and treated as a linear perturbation
When $b^2$ is small we can treat the solution as a linear perturbation
around AdS and the spacetime (\ref{eq:nrmetric}) can be interpreted using
the standard AdS/CFT dictionary at the linearized level. Using the TMG
holographic dictionary derived in \cite{Skenderis:2009nt},
the $b^2$ term acts as a source for the dimension $(3,1)$
irrelevant operator in the dual CFT. Therefore, the dual
interpretation of null warped solutions of TMG at finite $b^2$ should
indeed be in terms of a CFT deformed by irrelevant operators, which
are however exactly marginal from the perspective of the Schr\"odinger
symmetry.

The outline of the paper is as follows.  In the next section we review
two gravitational models which admit Schr\"{o}dinger solutions, namely
topologically massive gravity in three dimensions and the massive
vector model. In section \ref{sec:asymptotics} we discuss the modified
asymptotics of the Schr\"{o}dinger metrics and how these show that the
dual theory is obtained from specific irrelevant deformations of a
relativistic conformal field theory. We show in section \ref{sec:fieldthy} that
these deformations are exactly marginal with respect to the Schr\"odinger
group and use conformal perturbation theory to discuss how the dimensions
of operators change in the deformed theory. In section
\ref{sec:hol-prel} we begin the holographic analysis by setting up the
correct variational principle for TMG and deriving the dilatation
operator at linearized level.  In section
\ref{sec:bulkcomputations} we discuss the bulk computation of the
two-point function of a scalar operator. We consider the
linearized analysis for TMG in section \ref{sec:linearizedtmg}, and
present the general solutions of the linearized equations. We set up a
holographic dictionary for the deforming operator, and show that its
two point function is of the expected form. We treat the
linearized problem in the massive vector model in $d=2$ and give the
general solution of the linearized equations of motion in section
\ref{sec:linearizedvector}. We discuss our conclusions and open
questions in section \ref{sec:conclusions}.
%In the appendix we extend
%the AdS/CFT dictionary derived for TMG in \cite{Skenderis:2009nt} to
%the case of $\mu =3$.

\section{Gravity theories}
\label{sec:theories}
Schr\"{o}dinger backgrounds arise as solutions to a variety of gravitational theories. In this paper we will consider
topologically massive gravity and massive vector models and here we briefly review both theories. Throughout this paper we use the
following conventions for the curvatures:
\be
R_{\m \n \r}^{\phantom{\m \n \r}\s} = \del_\n \G_{\m \r}^\s + \G_{\m \r}^{\lambda}\Gamma_{\n\lambda}^\s - (\m \leftrightarrow \n),
\qquad \qquad R_{\m \r} = R_{\m \s \r}^{\phantom{\m \s \r}\s}
\ee
so that for the metric $G_{\m \n}$ given by \eqref{eq:nrmetric} one obtains:
\be
R_{\m \n} = - d G_{\m \n} + (d+2) \frac{b^2}{r^4} \delta_\m^u \delta_\n^u
\ee
which can be used to verify the formulae below.

\subsection{Topologically massive gravity}
Topologically massive gravity (TMG) is a three-dimensional theory of gravity where the usual Einstein-Hilbert action is supplemented with a
gravitational Chern-Simons term. The total action reads:
\be
S = \frac{1}{16 \pi G_N} \int d^3 x \, \sqrt{-G}\Big( R-2\Lambda+ \frac{1}{2\m} \e^{\lambda\m\n} \big( \G_{\lambda \s}^\r \del_\m \G_{\r \n}^\s
+ \frac{2}{3}\G_{\lambda \s}^\r \G_{\m \t}^{\s}\G_{\n \r}^\t \big) \Big)
\ee
where $\Gamma_{\m \n}^\lambda$ are the connection coefficients associated to the metric $G_{\m \n}$ and where we use the covariant $\e$-symbol
such that $\sqrt{-G}\epsilon^{uvr} = 1$, with $r$ the radial direction in \eqref{eq:nrmetric}. Variation of the action results in the equations of motion:
\be
\label{eq:eomtmg}
{R_{\m \n} - \hf G_{\m \n}R + \Lambda G_{\m \n} + \frac{1}{2\m} \Big(\epsilon_\m^{\phantom{\m}\r\s}\cdel_\r R_{\s \n}
+ \epsilon_\n^{\phantom{\n}\r\s}\cdel_\r R_{\s \m}\Big) = 0.}
\ee
We henceforth set $\Lambda = -1$. Taking the trace of this equation results in $R = -6$, so all solutions to TMG have a constant negative Ricci scalar.
Furthermore, any Einstein metric in three dimensions has $R_{\m \n} = - 2 G_{\m \n}$ and is easily seen to be a solution of \eqref{eq:eomtmg} as well.
In particular, AdS$_3$ is a solution of TMG for all values of $\mu$. For generic values of $\mu$ there also exist so-called warped $AdS_3$ spaces,
see \cite{Anninos:2008fx} for their properties. In the specific case $\mu = 3$ we find null warped $AdS_s$ as a solution,
\be \label{null_warped}
ds^2 = - \frac{b^2 du^2}{r^4} + \frac{2 du dv + dr^2}{r^2},
\ee
which is precisely \eqref{eq:nrmetric} with $d = 2$. Therefore null warped $AdS_3$ is equivalent to the three-dimensional
Schr\"{o}dinger spacetime.
In solving the TMG field equations $b^2$ is an arbitrary real parameter, so, in particular, $b^2$ can have either sign.
In the massive vector model we discuss below this is not the case and $b^2$ is necessarily positive. Furthermore, the analysis in section
\ref{sec:bulkcomputations} seems to indicate that $b^2 > 0$ is necessary for stability. For these reasons we will continue to use the
notation $b^2$ even when we consider the metric as a solution to TMG.

In \cite{Skenderis:2009nt} details of the holographic dictionary for TMG were presented. The most important feature for our purposes
is that, since the
equations of motion of TMG are third order in derivatives, we need to specify not only the boundary metric but also (a component of) the
extrinsic curvature in order to find a unique bulk solution. When we apply gauge/gravity duality to TMG with a negative
cosmological constant, the extra boundary data corresponds to the source of an extra operator. Therefore, besides the boundary
energy-momentum tensor $T_{ij}$, which couples to the boundary metric $g_{(0)ij}$, we also have a new operator $X_{vv}$ which couples
to the leading coefficient of the radial expansion of the $(uu)$ component of the extrinsic curvature. It was shown in
\cite{Skenderis:2009nt} that the operator $X_{vv}$ has
weights $(h_L, h_R)  = \hf (\m + 3, \m -1)$. (Strictly speaking, the analysis of \cite{Skenderis:2009nt} was
for $0 < \mu < 2$, but the extension to general $\mu$ is straightforward.)
Moreover, a precise relation between the extra boundary data and the presence of a new operator in the
dual field theory was established and the two-point functions of these operators around the state dual to an AdS background
were computed.
%In the appendix we extend this analysis to the case of interest here, namely $\m = 3$, and

Working to leading order in $b^2$, we can interpret the Schr\"{o}dinger solution (\ref{null_warped}) using the linearized AdS/CFT dictionary. Noting that the metric
\be
ds^2 = \frac{h_{ij} dx^{i} dx^j + 2 du dv + dr^2}{r^2}
\ee
describes linearized perturbations about an AdS background, then the results of \cite{Skenderis:2009nt} indicate that
the general solution to the linearized equations of motion at $\mu =3$ in which $h_{ij}$ depends only on the radial coordinate is:
\bea
h_{uv} &=& h_{(0) uv}; \\
h_{uu} &=&   \frac{1}{r^2} h_{(-2) uu } + h_{(0) uu} +  r^2 h_{(2) uu}; \nn \\
h_{vv} &=& h_{(0) vv} +  r^2 h_{(2) vv} + r^4 h_{(4) vv}, \nn
\eea
which is expressed in terms of seven independent integration constants.
If only the constant $h_{(-2) uu} \equiv - b^2$ is non-zero, the linearized solution is precisely the Schr\"{o}dinger solution, which of
course also solves the full non-linear equations of motion. However, applying the holographic dictionary derived at the linearized level, $-b^2$
acts as a (constant) source for the dimension $(3,1)$ operator $X_{vv}$. Therefore, the dual field theory, at least to leading order in $b^2$, must be
a null irrelevant deformation of the original conformal field theory.

Applying the holographic one point functions applicable at the linearized level which are given in \cite{Skenderis:2009nt} yields:
\be
\< T_{uu} \> = \< T_{vv} \> = \< X_{vv} \> = 0,
\ee
for this background. This indicates that the background corresponds to the vacuum of the deformed theory.
It is also interesting to note that a linearized solution with constant
$h_{(-2) uu} =  h_{(2) uu}$ also solves the non-linear equations of motion. The resulting background:
\be
ds^2 = \frac{d r^2}{r^2} + \frac{1}{r^2} \left ( - b^2 (\frac{1}{r^2} + r^2) du^2 + 2 du dv \right )
\ee
is Schr\"{o}dinger in the global coordinates introduced by \cite{Blau:2009gd}. Applying the linearized holographic
one point functions given in \cite{Skenderis:2009nt} to this background, we note that there is still a constant source for the dimension $(3,1)$
operator $X_{vv}$ but:
\be
\< T_{uu} \> = - \frac{1}{6 G_N} b^2,
\ee
with $\< T_{vv} \> = \< X_{vv} \> = 0$. This suggests that the background
should correspond to the deformed field theory in a different state, with $\< T_{uu} \>$ presumably related
to the Casimir energy. We should emphasize however that these formulae for the holographic one point functions apply only at
leading order in $b^2$. Understanding the dictionary at finite $b^2$ is far more subtle; it requires
us to go beyond asymptotically locally AdS spacetimes and is the focus of section \ref{sec:linearizedtmg}.

Note that the holographic dictionary for TMG reflects the various
problems of the theory.  The theory contains negative norm
states and it is thus unlikely that TMG in
itself could be a consistent theory of quantum gravity. Nevertheless
it remains a rich and interesting toy model offering gravitational
dynamics in three bulk dimensions, logarithmic correlation functions
(for $\mu = 1$) and, as we exploit here, it allows Schr\"{o}dinger
backgrounds as solutions.

\subsection{Massive vector model}
The massive vector model consists of Einstein gravity coupled to a massive vector field. The action is
\be
S = \frac{1}{16 \pi G_N} \int d^{d+1} x \, \sqrt{-G} (R - 2 \Lambda - \qt F_{\m \n}F^{\m \n} - \hf m^2 A_\m A^\m)
\ee
The equations of motion take the form
\be
\begin{split}
&R_{\m \n} - \hf R G_{\m \n} + \Lambda G_{\m \n} = \hf F_{\m \r} F^{\ph{\n}\r}_{\n} + \hf m^2 A_{\m} A_{\n}
- G_{\m \n} \Big( \frac{1}{8} F_{\r \s} F^{\r \s} + \qt m^2 A_\r A^\r\Big)\\
&\cdel^\n F_{\n \m} - m^2 A_\m = 0 \label{eomgf}
\end{split}
\ee
% where we note that the latter equation can be more conveniently written as:
% \be
% \frac{1}{\sqrt{-G}} \del_\m \Big(\sqrt{-G} F^{\m \n} \Big) - m^2 A^\n = 0\,.
% \ee
The vector equations of motion also imply the identity $\nabla^{\mu} A_{\mu} = 0$.

For $\Lambda = - d(d-1)/2$ and $m^2 = 2 d$ we find the metric \eqref{eq:nrmetric} and
\be
\label{eq:nrgaugefield}
A = \frac{b}{r^2} du.
\ee
as a solution. Note that the dimension of the dual vector operator $X_i$ is $(d+1)$ and the linearized AdS/CFT dictionary implies that
$b$ acts as a source for the operator $X_v$.

The solution with $d = 4$ requires a massive bulk vector field with $m^2 = 8$. Such a vector field arises as one
of the Kaluza-Klein modes in type IIB compactifications of the form AdS$_5  \times Y$ with $Y$ a Sasaki-Einstein manifold.
A consistent truncation including this mode was found in \cite{Maldacena:2008wh}, and
this can be used to uplift the solution to a ten-dimensional solution of type IIB supergravity. The five-dimensional consistent truncation
contains the metric, the massive vector field with $m^2 = 8$ and three scalars (one of which is massless) with a nontrivial potential. Further
discussions of embeddings of Schr\"{o}dinger solutions into string theory may be found in \cite{Hartnoll:2008rs,Gauntlett:2009zw,Donos:2009en,Colgain:2009wm,Bobev:2009mw,
Donos:2009xc,Donos:2009zf}.

The solution presented above in the case of $d=4$
is related by a so-called TsT transformation to AdS$_5  \times Y$ \cite{Herzog:2008wg,Maldacena:2008wh}. Following the chain
of transformations, this implies that the field theory dual to a Schr\"{o}dinger spacetime should be a null dipole theory. Dipole theories were originally
introduced in \cite{Bergman:2000cw} and arise from taking decoupling limit of branes in a background with a $B$ field with one leg longitudinal
to the brane and one leg transverse to the brane. The structure of dipole theories
is analogous to the better known non-commutative theories that arise on decoupling branes in a background with constant $B$ longitudinal
to the brane worldvolume \cite{Seiberg:1999vs}. In general, to obtain a dipole theory,
one starts with a local and Lorentz invariant field theory in $d$ dimensions. Then to every field $\Phi$
is assigned a vector $L_{\mu}$ where $\mu = 1, \cdots d$; this is the dipole vector of the field.
The fields can be scalars, fermions or have higher spin. Next, one defines a dipole product which is non-commutative:
\be \label{dipole}
\Phi_1 \ast \Phi_2 \equiv \Phi_1 (x - \half L_2 ) \Phi_2 (x + \half L_1 ).
\ee
This defines an associative product provided that the vector assignment
is additive, that is, $\Phi_1 \ast \Phi_2$  is assigned the dipole vector $L_1 + L_2$. For CPT symmetry, one
requires that if $\Phi$ has dipole vector $L$ then the charge conjugate field, $\Phi^{\dagger}$, is assigned the
dipole vector $-L$. Also gauge fields have zero dipole length. In most of the earlier literature, the case where the dipole vector
$L$ is spacelike was considered. In the current context, however, the dipole vector is null, as we review below. In the case where
$Y = S^5$, the TsT transformations change the dual field theory from $\mathcal N = 4$ super Yang-Mills to a null dipole version of this theory.
The analytic structure of quantities computed holographically should therefore match those expected for a null dipole theory. However, there is
very little literature on null dipole theories. The analytic structure of spacelike dipole theories is believed to be closely related to that of non-commutative
theories, which in turn is extremely subtle due to UV/IR mixing \cite{Minwalla:1999px}. The analytic structure of null dipole theories should however be more
similar to theories with light-like non-commutativity, which were argued in \cite{Aharony:2000gz} to be unitary, and this is consistent with what we find here.

\section{Asymptotics in the gauge/gravity duality}
\label{sec:asymptotics}
As already mentioned in the introduction, the Schr\"{o}dinger spacetime is not an
asymptotically locally AdS spacetime. As is reviewed
in  \cite{Skenderis:2002wp}, any asymptotically locally AdS spacetime admits
in the neighborhood of the conformal boundary $r \to 0$ a metric of the form:
\be
\label{eq:alads}
ds^2 = \frac{dr^2}{r^2} + \frac{1}{r^2} g_{ij} dx^i dx^j \qquad \qquad g_{ij}(r,x^k) = g_{(0)ij}(x^k) + \ldots
\ee
where the dots represent terms that vanish as $r \to 0$ and $g_{(0)ij}$ is an arbitrary non-degenerate metric.

The relevance of this structure to holography is the following.
Suppose we would like to use holography to compute certain field
theoretic quantities. Just as in field theory computations, the holographic
computation of these correlation functions suffers from divergences
which need to be regularized and renormalized
\cite{Skenderis:2002wp}. In this procedure of \emph{holographic
  renormalization} the asymptotics of the spacetime play a crucial
role: they are, via the equations of motion, directly related to the
form of the divergences which need to be subtracted
\cite{deHaro:2000xn,Bianchi:2001kw}.  For example, for asymptotically
AdS spacetimes the supergravity equations of motion, combined with the
asymptotics \eqref{eq:alads}, guarantee that all divergences are local
in the boundary data \cite{Papadimitriou:2004ap,Papadimitriou:2004rz}.
These divergences
can therefore be subtracted by local counterterms as well. This is in
agreement with the fact that the dual field theory is local and
renormalizable, and the fact that the divergences on the gravity side
have this form is strong structural evidence for gauge/gravity
dualities.

For the case at hand, we find ourselves outside of the usual framework, since for
nonzero $b$ the spacetime is no longer AlAdS. The standard results of holographic renormalization therefore do not directly carry
over to this setting and in order to obtain correlation functions one has to build a holographic dictionary from first principles.
Just as in field theory, it is imperative to understand systematically the structure of these divergences in order to discuss renormalization
and obtain correct finite correlation functions. Notice that there is a priori \emph{no} guarantee that the divergences will be
local in the boundary data and in fact we will encounter certain nonlocality in the divergences below.

%In recent work \cite{Henneaux:2010fy} it was also pointed out that null warped solutions of TMG are not asymptotically AdS. However, the perspective
%of that work was to consider only TMG solutions compatible with the two-dimensional conformal group as an asymptotic symmetry group. The view taken here
%is rather different. While solutions which are not asymptotically AdS are manifestly not dual to a conformal field theory, the question addressed
%here is whether holography can be developed for such solutions and, if so, what is the nature of the dual quantum field theory.

\subsection{Interpretation as a deformation}
\label{sec:deformation}

In this paper we will explore the analytic structure on both sides of the duality.  As a first step in this exploration it is interesting
to take the limit $b \to 0$, for which the metric \eqref{eq:nrmetric} reduces to that of empty AdS, and expand correlation functions
perturbatively in the small parameter $b$. When $b$ is zero we simply recover the AdS metric in Poincar\'e coordinates, and results to leading order
around $b = 0$ may be obtained via the usual AdS/CFT dictionary. In particular,
for small perturbations around the conformal vacuum we find the following radial expansion for the massive vector field $A_\mu$:
\be
\label{eq:pertgaugefield}
\begin{split}
A_i &= \frac{1}{r^2} \Big( A_{(0)i} + \ldots + r^{d+2} A_{(d+2)i} + \ldots \Big).
% \\
% A_z &=
\end{split}
\ee
Holographically, $A_{(0)i}$ is interpreted
as the source for a dual operator $X^i$ and its expectation value is related to the normalizable mode $A_{(d+2)i}$.
Given the mass of the bulk vector field, the corresponding scaling dimension $\Delta$ of the dual operator $X^i$ is given by $(d+1)$.

By comparing the explicit gauge field solution \eqref{eq:nrgaugefield} with \eqref{eq:pertgaugefield} we find that, to first order,
switching on $b$ can be interpreted holographically as an irrelevant deformation of the original CFT of the form:
\be
\int d^{D+2} x \, b^v X_v\,
\ee
where $b^v = b$. Similarly, as we discussed earlier, for TMG the corresponding deformation is:
\be
\int d^2 x \, b^{vv} X_{vv}\,.
\ee
with $b^{vv} = - b^2$.

These results are based on an analysis for small $b$. Given that the
deforming operators are irrelevant (so the deformed theory appears
non-renormalizable and thus uncontrollable in the UV) it would seem
hard to extend these results to finite $b$.  We have seen however that
the linearized solution automatically solves the non-linear equations
of the motion. This is related to the fact that the
linearized solution has a new scaling symmetry, the dilatation of the
Schr\"ondinger group, which controls the UV behavior of the theory
and the value of $b$ can be set to any value (provided
it is non-zero) by an appropriate scaling of the lightcone 
coordinates (see footnote \ref{ft:sc}) while maintaining the 
new scaling symmetry.
Thus the solution derived for small $b$ is automatically
also a solution for large $b$. The counterpart of this statement
on the QFT side is that the deforming operator is exactly marginal and
after the deformation the theory finds itself at a non-relativistic
fixed point.

One of the main questions is then
to understand how the deformation changes the spectrum of operators.
We will analyze this question on the field theory side using
conformal perturbation theory. On the
gravitational side, the same question amounts to solving the linearized
equations of the motion.

\section{Field theory analysis}
\label{sec:fieldthy}

From the previous discussion, Schr\"{o}dinger backgrounds are dual to
conformal field theories deformed by operators that respect the
Schr\"{o}dinger symmetry. In this section we will use conformal
perturbation theory to show that the deforming operator is exactly
marginal and then study how the spectrum of scaling dimensions
changes in the deformed theory. We will only use general results that
follow from conformal invariance, so our results are valid for
any relativistic CFT, weakly or strongly coupled, that has in its
spectrum operators of the right type. Similar arguments can be made for
scale invariant theories with other dynamical exponents, $z \neq 2$, and
these are discussed briefly in section \ref{zneq2}.

\subsection{Marginal Schr\"{o}dinger invariant deformations}

Any relativistic $d$-dimensional conformal field theory is also
invariant under the $(d-2)$-dimensional Schr\"{o}dinger group.
In particular the non-relativistic scaling dimension $\D_s$
is related to the relativistic scaling dimension $\D$ via
\cite{CMP34(85),Son:2008ye,Maldacena:2008wh}:
\be
{\D_s} = \D  + M_{vu},
\ee
where $M_{vu}$ is the eigenvalue of the boost operator in the lightcone
directions normalized such that $v$ and $u$ have
eigenvalues +1 and -1, respectively. This implies that
one can break the relativistic conformal group while preserving
the $(d-2)$-dimensional Schr\"{o}dinger subgroup. Such deformations
will necessarily also break the Lorentz invariance since the deforming
operator must have a non-zero eigenvalue $M_{vu}$.

Consider now a conformal field theory deformed by an operator with
Schr\"{o}dinger scaling dimension $\Delta_s$.
Such a deformation is irrelevant with respect to the Schr\"{o}dinger
symmetry when $\Delta_s > d$, marginal when $\Delta_s = d$ and
relevant when $\Delta_s < d$. Consider first the case where we deform
with a scalar operator. Then the deformation
respects not only Schr\"{o}dinger symmetry, but also the relativistic
conformal symmetry, as it does not break rotational invariance in the
$(u,v)$ directions; such marginal deformations of CFTs have been
extensively explored. Suppose now that the operator is not a scalar,
but rather a vector $X_i$ or a
tensor $X_{ij}$, of Schr\"{o}dinger dimension $d$. Such deformations
can respect Schr\"{o}dinger symmetry, but break the relativistic
symmetry, provided that the sources are constant null vectors or
tensors respectively, with components only along the $v$
directions. An example of such a deformation is:
\be
{S}_{\rm CFT}  \rightarrow {S}_{\rm CFT} + \int d^{d-2} x du dv b^i {X}_{i} (v,u,x^i) = {S}_{\rm CFT} + \int d^{d-2} x du b^i \td{X}_{i}(k_v=0,u,x^i),
\ee
with $b^i$ a constant null vector with non-vanishing component $b^v \equiv b$ and $\td{X}$ the Fourier transform of $X$ in the $v$ direction. This example is
realized in the duality with the massive vector theory. In the case of TMG,
the deformation of interest is by a tensor operator, namely:
\be
{S}_{\rm CFT}  \rightarrow {S}_{\rm CFT} + \int du dv b^{ij} {X}_{ij} (v,u,x^i) = {S}_{\rm CFT} + \int du b^{ij} \td{X}_{ij}(k_v=0,u,x^i),
\ee
where the only non-vanishing component of $b^{ij}$ is $b^{vv} = - b^2$.
In both cases the deformations are marginal from the perspective of the Schr\"{o}dinger symmetry, i.e. whilst both operators are irrelevant from the perspective
of the relativistic conformal symmetry, they have non-relativistic scaling dimensions $\Delta_s = d$. The deformations break the Lorentz symmetry, but respect
all rotations, translations and boosts of the Schr\"{o}dinger group. In the specific case of two dimensions, the deformations in addition respect the infinite-dimensional algebra associated with analytic coordinate transformations $u \rightarrow u'=u'(u)$.

To illustrate the general idea let us give a simple explicit example: consider the two-dimensional action
\be
S = \int du dv \left ( \pa_u \Phi \pa_v \Phi + b \pa_u \Phi (\pa_v \Phi)^3 \right ).
\ee
The first term preserves full two-dimensional relativistic conformal invariance whilst the second corresponds to a deformation of the relativistic conformal field
theory by a dimension four operator, with scaling weights $(3,1)$. The deformation manifestly breaks both relativistic conformal invariance and two-dimensional Lorentz
invariance but preserves the Schr\"{o}dinger invariance (at least at the classical level). More generally, one can observe that any deformation of the type:
\be
\delta S = \int du dv (\pa_u \Phi \pa_v \Phi) f(\pa_v \Phi, \pa_v^2 \Phi, \cdots)
\ee
respects the Schr\"{o}dinger symmetry at the classical level for any functional $f(\pa_v \Phi, \pa_v^2 \Phi,\cdots)$.

Such deformations are manifestly marginal with respect to the Schr\"{o}dinger symmetry, but one also
needs to show that the deformations are exactly marginal. This requires
proving that the non-relativistic dimensions of $X(k_v=0,u,x^i)$ remain equal to $\Delta_s = d$ in the deformed theory, which amounts to showing that the 2-point function
of  $X(k_v=0,u,x^i)$ in the deformed theory is the same as in the original theory. We can demonstrate this property using conformal perturbation theory as follows. Let us
consider first the case of deformation by a vector operator with a constant null source. In conformal
perturbation theory, the correction to the two point function in the deformed theory is expressed in terms of higher point functions in the conformal theory as:
\bea
&&\delta \< \td{X}_{v} (k_v,k_u,k_i) \td{X}_{v} (-k_v,-k_u,-k_i) \>  \nonumber \\
&& \qquad \qquad = \sum_{n \ge 1} \frac{1}{n!} \left\< \td{X}_{v} (k_v,k_u,k_i) (b \td{X}_{v}(0))^n \td{X}_{v}
(-k_v,-k_u,-k_i) \right\>_{\rm CFT}.
\eea
Note that the operator insertions are at zero momentum. In the original CFT, the operator $\td{X}_{v}(0)$ has relativistic conformal dimension $(d+1)$ and
Schr\"{o}dinger dimension $d$, and transforms as a vector. This implies that the deformed theory is invariant under relativistic conformal symmetry provided
that $b$ is transformed contravariantly, with relativistic scaling dimension minus one. Each additional insertion of $(b \td{X}_{v}(0))$ therefore adds
another factor of $b k_v$, so that:
\bea
&&\left\< \td{X}_{v} (k_v,k_u,k_i) \prod_{n} (b \td{X}_{v}(0)) \td{X}_{v}
(-k_v,-k_u,-k_i) \right\>_{\rm CFT} = \\
&& \qquad (b k_v)^n \left\< \td{X}_{v} (k_v,k_u,k_i) \td{X}_{v} (-k_v,-k_u,-k_i) \right\>_{\rm CFT}
f \left (\ln (k^2/\mu^2) \right ). \nn
\eea
As we will show below, relativistic conformal invariance implies that the scalar
function $f$ can depend only logarithmically on the momentum. Since the CFT two point function is necessarily finite or vanishing at zero momentum,
the operator at zero $k_v$ momentum receives no corrections to its two point function since:
\be
\left\< \td{X}_{v} (0,k_u,k_i) \prod^{n} (b \td{X}_{v}(0)) \td{X}_{v}
(0,-k_u,-k_i) \right\>_{\rm CFT} \rightarrow 0
\ee
for all $n \ge 1$. A similar argument implies that the operator at zero $k_v$ momentum remains orthogonal
to all other operators.
Thus the deforming operator is exactly marginal. However, the operators at non-zero $k_v$ momenta do receive corrections to their two
point functions and thus to their scaling dimensions under Schr\"{o}dinger symmetry; these corrections will be explicitly computed below.

\subsection{Proof of marginality}

The key identity required to prove that the operator $X(k_v{=}0)$ is exactly marginal is:
\bea
&&  \left \langle X_v (k_v) \left(\prod_{i=1}^n b^{\mu} \cdot X_{\mu} (k_v{=}0)\right)
X_v(-k_v) \right \rangle_{\rm CFT} =
\\
&& \qquad  \left \langle X_v(k_v) X_v(-k_v) \right \rangle_{\rm CFT}
(b^v k_v)^n  f(\log k_v,...) \nn
\eea
where $f(\log k_v,...)$ is a dimensionless function that
carries the dependence on the positions of the operators and is
at most logarithmically dependent on $k_v$. The corresponding identity for
TMG involves the operator $X_{vv}$ in the two dimensional conformal field theory.

In this section we will prove this identity in the case of two-dimensional CFTs.
In two dimensions, both the massive vector and TMG cases involve deriving the structure
of $(n+2)$-point functions of $(q,1)$ operators, $X_{v \cdots v}$ ($q$ $v$-indices), with themselves, in which $n$ of the operators
are at zero momentum. We will do the computation in position space and in Euclidean signature and then
Fourier transform afterwards. We thus need to compute
\be \label{marg_pr}
\int \left(\prod_{i=2}^{n+1} d^2w_i\right)
\left \langle X_{v \cdots v}(w_1,\bar{w}_1) \left(\prod_{i=2}^{n-1} X_{v \cdots v}(w_i,\bar{w_i})\right) X_{v \cdots v}(w_{n+2},\bar{w}_{n+2}) \right \rangle_{\rm CFT}. %\equiv \prod_{i=2}^{n+1} d^2w_i G^{(n+2)}
\ee

Recall that conformal invariance constrains the $(n+2)$-point function, $G^{(n+2)}$, of operators of weight $(h,\bar{h})$ to be of the form
\be \label{npt}
G^{(n+2)}(w_1,\bar{w_1}; \cdots; w_{n+2},\bar{w}_{n+2}) =
\prod_{i<j} w_{ij}^{-2 h/(n+1)} \bar{w}_{ij}^{-2 \bar{h}/(n+1)} Y(w_{ij}^{kl};\bar{w}_{ij}^{kl})
\ee
where $w_{ij}=w_i-w_j$ and $Y$ is an arbitrary function of the $2((n+2)-3)$ cross ratios,
$w_{ij}^{kl}=\frac{w_{ij} w_{kl}}{w_{il} w_{kj}}, \bar{w}_{ij}^{kl}=\frac{\bar{w}_{ij} \bar{w}_{kl}}{\bar{w}_{il} \bar{w}_{kj}}$. The case of interest for us is $(h,\bar{h})=(q,1)$ but we will not yet impose this.
We need to integrate over the position of $n$ operators, as in (\ref{marg_pr}). To proceed we first
use translational invariance to set $w_{n+2}=\bar{w}_{n+2}=0$ and rescale,
\be
w_i = \omega_i w_1, \qquad \bar{w}_i = \bar{\omega}_i w_1, \qquad i=2, \ldots, n+1
\ee
This yields
\bea
&&\int \left(\prod_{i=2}^{n+1} d^2w_i\right) G^{(n+2)}(w_1,\bar{w_1}; \cdots; w_{n+2},\bar{w}_{n+2}) \nonumber \\
&& \qquad \qquad =\frac{1}{w_1^{2h + n(h-1)}} \frac{1}{\bar{w}_1^{2 \bar{h} + n(\bar{h}-1)}}
\int \left(\prod_{i=2}^{n+1} d^2\omega_i\right) Y(w_{ij}^{kl};\bar{w}_{ij}^{kl}) \nonumber \\
&& \qquad \qquad \times \prod_{i=2}^{n+1} \left( \omega_i^{-2h/(n+1)} (1 - \omega_i)^{-2h/(n+1)} \right)
\prod_{i < j} (\omega_i - \omega_j)^{-2h/(n+1)}
\nonumber \\
&&\qquad \qquad  \times \prod_{i=2}^{n+1} \left( \bar{\omega}_i^{-2\bar{h}/(n+1)}
(1 -\bar{\omega}_i)^{-2\bar{h}/(n+1)} \right)
\prod_{i < j} (\bar{\omega}_i - \bar{\omega}_j)^{-2\bar{h}/(n+1)}.
\eea
Power counting shows that the integrals diverge as $\omega_i \sim \omega_j \sim 0$
with degree of divergence $(h-1) n$,  and similarly there is a divergence when
$\bar{\omega}_i \sim \bar{\omega}_j \sim 0$ with degree of divergence $(\bar{h}-1) n$.
When $h=1$ and/or $\bar{h}=1$ the integral has logarithmic divergences.

Specializing to our case we find
\bea
&&\int \left(\prod_{i=2}^{n+1} d^2w_i\right)
\left \langle X_{v \cdots v}(w_1,\bar{w}_1) \left(\prod_{i=2}^{n-1} X_{v \cdots v}(w_i,\bar{w_i})\right) X_{v \cdots v}(0,0)
\right \rangle_{\rm CFT} \nonumber \\
&&\qquad =  \left \langle X_{v \cdots v}(w_1,\bar{w}_1) X_{v \cdots v}(0,0) \right \rangle_{\rm CFT} \frac{1}{w_1^{(q-1) n}}
\tilde{f}(\log |w_1|^2 \mu^2)  \label{npt_pos}
\eea
where $\tilde{f}(\log |w_1|^2 \mu^2)$ is the (dimensionless) function that results from the evaluation of the integrals (after
appropriate regularization/renormalization) and we
explicitly indicate that it will depend on $\log |w_1|^2 \mu^2$ (at most polynomially) because the integrals are logarithmically
divergent. We will compute such integrals in the next few subsections.

Fourier transforming and reinstating  the coupling $b^{v \cdots v}$ we find
\bea
&&  \left \langle X_{v \cdots v} (k, \bar{k}) \left(\prod_{i=1}^n b^{v \cdots v} X_{v \cdots v} (k{=}0, \bar{k}=0)\right)
X_{v \cdots v} (-k, -\bar{k}) \right \rangle_{\rm CFT} =
\nn \\
&& \qquad  \left \langle X_{v \cdots v} (k, \bar{k}) X_{v \cdots v} (-k, -\bar{k}) \right \rangle_{\rm CFT}
(b^{v \cdots v}  (k)^q)^n  f(\log |k|/m^2).
\eea
Wick rotating to Lorentzian signature $k$ becomes $k_v$ and this is the identity we wanted to prove.

\subsection{Deformations of two-dimensional CFTs: TMG example}

In this subsection we consider more explicitly deformations of two dimensional conformal field theories that respect Schr\"{o}dinger symmetry. In particular, we would like to understand how the spectrum of operators changes as
we move from the relativistic fixed point to the Schr\"{o}dinger fixed point, i.e. we would like to compute
the non-relativistic dimension $\D_s$ of the operator. As in the previous subsection
we will use conformal perturbation theory; note that this does not require a weakly coupled realization of the CFT.
In contrast to the previous subsection, we will only work to the first non-trivial order. We will, however, explicitly evaluate all integrals.  Keeping in mind the example of TMG,
we begin with the case where the CFT is deformed by a $(3,1)$ operator $X_{3,1}$ (called $X_{vv}$ in the
previous subsection) so that
\be \label{TMG_def}
S_{\rm CFT} \rightarrow S_{\rm CFT} - b^2 \int d^2 w X_{3,1},
\ee
and from here onwards we work in Euclidean signature with $\sqrt{2} u \rightarrow \bar{w}$ and $\sqrt{2} v \rightarrow w$. The numerical factors are included for computational convenience in what follows.

Suppose that the original CFT has primary operators ${\cal O}_{h,\bar{h}}$ of dimension $(h,\bar{h})$
(one of which is $X_{(3,1)}$).
We would like to compute their dimension in the theory (\ref{TMG_def}).
Let their two-point functions in the original CFT be normalized as
\be \label{rel_2pt}
\< {\cal O}_{h,\bar{h}} (w, \bar{w}) {\cal O}_{h,\bar{h}} (0,0) \> = \frac{c_{\cal O}} {w^{2h} \bar{w}^{2 \bar{h}}},
\ee
where the Euclidean metric is
\be
ds^2 = dw d \bar{w}.
\ee
Now let us calculate the corrections to the two point functions in the deformed theory, working perturbatively in the deformation
parameter $b^2$. To leading order in $b^2$ the corrections to the two
point functions are computable from three point functions in the conformal theory
\be
\delta \< {\cal O}_{h_1,\bar{h}_1} (w, \bar{w}) {\cal O}_{h_3,\bar{h}_3} (0,0) \> = - b^2 \int d^2 y
\< {\cal O}_{h_1,\bar{h}_1} (w, \bar{w}) {X}_{3,1} (y, \bar{y}) {\cal O}_{h_3,\bar{h}_3} (0,0) \>. \label{integra}
\ee
Note in particular that the correction to the two point function does not in general preserve the orthogonality of the basis; there is
operator mixing and the operator basis needs to be diagonalized. For simplicity we will consider here the case of operators
which do not mix with other operators.

The three-point functions with the operator $X_{3,1}$ are given by the familiar expression
\bea
\< {\cal O}_{h_1,\bar{h}_1} (w_1, \bar{w}_1) X_{3,1} (w_2, \bar{w}_2)
{\cal O}_{h_3,\bar{h}_3} (w_3, \bar{w}_3) \> &=& C_{1X3} \frac{1} { w_{12}^{h_1 + 3 - h_3} w_{23}^{3 + h_3 - h_1} w_{13}^{h_3 + h_1 - 3}} \nn
\\
&& \cdot \frac{1}{ \bar{w}_{12}^{\bar{h}_1 + 1 - \bar{h}_3} \bar{w}_{23}^{1 + \bar{h}_3 - \bar{h}_1} \bar{w}_{13}^{\bar{h}_3 + \bar{h}_1 - 1}}, \label{3pf1}
\eea
where $w_{ij} = w_i - w_j$ and $C_{1X3}$ are constants.
However, there is a subtlety, as this expression holds only at separated points, whilst (\ref{integra}) involves an integration which receives
contributions from contact terms. This means that (\ref{3pf1}) needs to be replaced by a renormalized expression which is well-defined at coincident
points. Since the integration is over position space it is convenient to use differential regularization techniques \cite{Freedman:1991tk}
and write down an expression
for the three point function which coincides with (\ref{3pf1}) at separate points but is well-defined at coincident points. In $d$ dimensions
a distribution $|x|^{-2 \lambda}$ behaves for $x \rightarrow 0$ as:
\be
\frac{1}{|x|^{2 \lambda}} \sim \frac{1}{d + 2 n - 2 \lambda} \frac{1}{2^{2n } n!} \frac{\Gamma( \frac{d}{2})}{\Gamma (\frac{d}{2} + n)}
S_{d-1} \pa^{2n} \delta^{(d)}(x), \label{dis}
\ee
where $S_{d-1} = 2 \pi^{d/2}/\Gamma(d/2)$ is the volume of the unit $(d-1)$ sphere. This implies that the distribution has poles at
$\lambda = d/2+n$ where $n=0,1,\cdots$. To obtain a well-defined distribution one subtracts this pole; it suffices to work out the case
of $\lambda = d/2$ since the others can be obtained by differentiation. In particular,
in two dimensions one replaces $1/|x|^2$ by
\be
D(x) = {\cal R} \left ( \frac{1}{|x|^2} \right ) = \frac{1}{8} \Box \ln^2 (m^2 |x|^2) \label{2d}
\ee
where $m^2$ is the renormalization scale. The renormalized expression differs from $1/|x|^2$ by the infinite term $\delta (x) \log (m^2 |x|^2)$
localized at $x =0$. This form for the renormalized quantity is specific to two dimensions, and can be obtained
taking the $d \to 2$ limit of the corresponding expression for $d > 2$ given
in (\ref{ren-higher}). The Fourier transform of the renormalized expression was computed in the appendix of \cite{Karch:2005ms}
and is given by
\be
\tilde{D}(k) = \int d^2 x e^{- ik x} D(x) =  - \pi \ln \left ( \frac{k^2}{\mu^2} \right ), \label{ft-k}
\ee
where $\mu^2 = 4 e^{-2 \gamma} m^2$.

The three point function of interest has $h_1 = h_3 = h$ and $\bar{h}_1 = \bar{h}_3 = \bar{h}$, and can be rewritten in the form:
\bea
\< {\cal O}_{h,\bar{h}} (w_1, \bar{w}_1) {X}_{3,1} (w_2, \bar{w}_2)
{\cal O}_{h,\bar{h}} (w_3, \bar{w}_3) \> &=&
\frac{C}{ 4 w_{13}^{2 h - 3} \bar{w}_{13}^{2 \bar{h}  - 1 }} \label{3pf2}
\\
&& \times \partial_{{w}_1}^2 \left( \frac{1}{ |{w}_{12}|^{2}} \right)
\partial_{{w}_3}^2 \left( \frac{1}{|w_{23}|^{2}} \right). \nn
\eea
The structure constant $C$ is given by $C \equiv C_{1 X 1}$.
The singularities as $w_2 \rightarrow w_1$ and $w_2 \rightarrow w_3$ are removed via the renormalized expression:
\bea
\< {\cal O}_{h,\bar{h}} (w_1, \bar{w}_1) {X}_{3,1} (w_2, \bar{w}_2)
{\cal O}_{h,\bar{h}} (w_3, \bar{w}_3) \> &=&
\frac{C}{ 4 w_{13}^{2 h - 3} \bar{w}_{13}^{2 \bar{h}  - 1}} \label{3pf3}
\\
&& \times \partial_{{w}_1}^2 {\cal R} \left( \frac{1}{ |{w}_{12}|^{2}} \right)
\partial_{{w}_3}^2 {\cal R} \left( \frac{1}{|w_{23}|^{2}} \right), \nonumber
\eea
which manifestly agrees with the previous expression away from contact points. For simplicity we will compute
the correction to the two point function at separated points, i.e. we will disregard contact terms
as $w_1 \rightarrow w_3$, but it is straightforward to generalize the analysis to include these. Note also that contact terms are
in any case absent when the scaling weights are such that $2h$ and $2 \bar{h}$ are not integral.

The leading correction to the two point function is:
\be
\delta \< {\cal O}_{h, \bar{h}} (w, \bar{w}) {\cal O}_{h,\bar{h}} (0,0) \> =
- \frac{b^2 C} { 4 w^{2 h - 3} \bar{w}^{2 \bar{h}  - 1}} \int d^2 y  \partial_{w}^2 {\cal R} \left ( \frac{1}{ |w - y|^{2}} \right )
\partial_{x}^2 {\cal R} \left ( \frac{1}{|x - y|^{2}} \right )_{x=0}.
\ee
To compute the integral first note that it is a convolution and therefore:
\be
\int d^2 y {\cal R} \left ( \frac{1}{ |w - y|^{2}} \right ) {\cal R} \left ( \frac{1}{|x - y|^{2}} \right ) =
\frac{1}{(2 \pi)^2} \int d^2 k e^{ik (w - x)} \tilde{D}^2 (k),
\ee
where the Fourier transform $\tilde{D}(k)$ was given in (\ref{ft-k}). Thus:
\bea
\int d^2 y {\cal R} \left ( \frac{1}{ |w - y|^{2}} \right ) {\cal R} \left ( \frac{1}{|x - y|^{2}} \right ) &=&
\frac{1}{4} \int  d^2 k e^{ik (w - x)} \ln^2 (k^2/\mu^2), \\
&=& 2 \pi \frac{1}{|w-x|^2} \ln (m^2 |w-x|^2), \nn
\eea
where we again use the Fourier transform (\ref{ft-k}). The complete two point function to first order in $b^2$ is then:
\be
 \< {\cal O}_{h, \bar{h}} (w, \bar{w}) {\cal O}_{h,\bar{h}} (0,0) \> =  \frac{1}{w^{2h} \bar{w}^{2 \bar{h}}} \left (
c_{\cal O} - \frac{12 \pi b^2 C}{w^2} \ln (\hat{m}^2 |w|^2) \right ), \label{complete}
\ee
where $\hat{m}^2 = m^2 e^{-25/12}$. This correction respects Schr\"{o}dinger symmetry and is of the form (\ref{npt_pos}) discussed in the previous subsection.

As discussed earlier, it is natural to work with operators of fixed right moving momentum, so
we now Fourier transform this expression. For simplicity, we will work out the case $h=\bar{h}$.
The general case is a straightforward extension. Recall that
the general expression for the Fourier transform of a polynomial in $d$ dimensions is
\be
\int d^d x e^{-i \vec{k} \cdot \vec{x}} (|x|^2)^{- \lambda} = \pi^{d/2} 2^{d- 2 \lambda} \frac{\Gamma(d/2 - \lambda)}{\Gamma (\lambda)} (|k|^2)^{\lambda - d/2}, \label{ft-imp}
\ee
which is valid when $\lambda \neq (d/2 + n)$, where $n$ is zero or a positive integer. (These are the cases in which the distribution
in $x$ is ill-defined, as discussed in (\ref{dis}), and one needs to subtract poles.) Differentiating this expression with respect to $\lambda$ results
in the identity:
\be
\int d^d x e^{-i \vec{k} \cdot \vec{x}} (|x|^2)^{- \lambda} \ln (M^2 |x|^2)  = - \pi^{d/2} 2^{d- 2 \lambda}
\frac{\Gamma(d/2 - \lambda)}{\Gamma (\lambda)} (|k|^2)^{\lambda - d/2} \ln (|k|^2/\mu^2),
\ee
with
\be
\mu^2 = 4 M^2 \exp [\psi(d/2 -\lambda) + \psi(\lambda)],
\ee
where $\psi(x)$ is the digamma function.
In the two-dimensional case, $d^2 w = 1/2 dw d \bar{w}$, and
applying (anti) holomorphic derivatives leads to further identities such as:
\be
\int dw d\bar{w}  e^{-i k w - i \bar{k} \bar{w}}
w^{-2} |w|^{- 2\lambda} = - \pi 2^{3 - 2 \lambda} \frac{\Gamma(1- \lambda)}{\Gamma ( \lambda + 2)} k^2 |k|^{2 (\lambda -1)},
\ee
where $|k|^2 = 4 k \bar{k}$. After differentiating with respect to $\lambda$ one obtains:
\be
\int d w d \bar{w}  e^{-i k w - i \bar{k} \bar{w}} w^{-2} |w|^{- 2\lambda} \ln (M^2 |w|^2) =
\pi 2^{3 - 2 \lambda} \frac{\Gamma(1- \lambda)}{\Gamma ( \lambda + 2)} k^2 |k|^{2 (\lambda -1)} \ln (|k|^2/\mu_1^2),
\ee
where $\mu_1^2 = 4 M^2 \exp [ \psi(1- \lambda) + \psi(2 + \lambda)]$.

Using these identities the Fourier transform of the corrected two point function of a scalar is:
\be
 \< {\cal O}_{h, h} (k, \bar{k}) {\cal O}_{h, h} (- k, -\bar{k}) \> = A k^{2h-1} \bar{k}^{2 h - 1} \left (1 + B k^2 \ln (|k|^2/\mu_1^2) \right ),
\ee
where
\be
A = c_{\cal O} \pi 2^{2 - 4 h} \frac{\Gamma(1- 2 h)}{\Gamma (2 h)}; \qquad
B = - \frac{6 \pi C}{c_{\cal O} h (2h +1)} b^2.
\ee
Note that the normalization of scalar operators in the CFT required to match the standard holographic normalization is \cite{Skenderis:2002wp}:
\be
c_{\cal O} = (2h - 1) \frac{\Gamma(2h)}{\pi \Gamma(2h -1)}; \label{hom}
\ee
in other words, a canonically normalized bulk scalar field will lead to two point functions in the CFT with this normalization.
In momentum space the general expression for the two point function in a Schr\"{o}dinger invariant theory takes the form:
\be
 \< {\cal O}_{\Delta_{s}} (k, \bar{k}) {\cal O}_{\Delta_{s}} (- k, -\bar{k}) \> = \td{g} (k) \bar{k}^{\Delta_{s} - 1}, \label{mome-sp}
\ee
The corrected two point function is consistent with this form where at leading order in $b^2$
\bea
\Delta_{s} &=& 2 \bar{h} + \gamma(k); \qquad
\gamma(k) = - \frac{6 \pi C}{c_{\cal O} h (2h +1)} b^2 k^2; \label{adim} \\
\td{g}(k) &=& - 2^{3 - 4 h} \frac{\Gamma(2 - 2 h)}{\Gamma (2 h - 1)} k^{\Delta_s - 1}, \nn
\eea
and we have used the normalization (\ref{hom}) in the latter equality. The normalization factor $\td{g} (k)$ is
written in this way in anticipation of the holographic result in section \ref{sec:bulkcomputations}, which indeed takes this form to all orders in $b$.
As anticipated, the anomalous dimension depends explicitly on the holomorphic momentum, and vanishes when this momentum is zero.
Note that on general grounds it is also clear that correlation functions in these theories should depend on the combination $b^2 k^2$,
as this is the quantity which is invariant under rescalings of the bulk lightcone coordinates. In other words,
by rescalings of the lightcone coordinates one can rescale $b^2$ to any non-zero value but $b^2 k^2$ is independent of such rescalings.

\bigskip

Analogous corrections would be expected for the two point functions of all operators, but in general, as mentioned above, operators will mix and
one will need to rediagonalize the basis of operators. The leading order corrections are all determined by the three point function coefficients
$C_{1 X 3}$. This means in particular that, because the form of the OPE for the stress energy tensor $(T, \bar{T})$ in a conformal field theory implies that
the stress energy tensor does not have a three point function with the operator $X_{3,1}$,
the correlation functions of $(T, \bar{T})$ are not corrected at order $b^2$. This is as one might have expected, since
the stress energy tensor should not acquire an anomalous dimension. However, as we will discuss shortly, the conserved stress energy tensor of
a non-relativistic theory is necessarily not symmetric, and thus the operators $(T, \bar{T})$ are not natural operators in the deformed theory.

As discussed earlier, the deforming operator $X$ itself generally acquires an anomalous dimension when its lightcone momentum $k_v$ is non-zero.
The leading correction to the two point function is given by \eqref{complete}, and
generically corrections to the two point functions occur at all orders $b^{2n}$, and are related to $(n+2)$-point functions
in the CFT.
%and the contributions at each order do not a priori vanish. Here for comparison with section \ref{sec:linearizedtmg} we move back
%to Lorentzian signature with $k \leftarrow k_v/\sqrt{2}$ and $\bar{k} \rightarrow k_u/\sqrt{2}$.
The holographic computation in
section \ref{sec:linearizedtmg} indicates that this series in $b^2$ can be resummed into the simple form:
\be
\< {X} (k)  {X}(-k) \>_{b} =  \td{g}(k_v) k_u^{\Delta_s - 1}, \label{Scr-form2}
\ee
where the normalization $\td{g}(k_v)$ is lightcone
momentum dependent and so is the non-relativistic scaling dimension, $\Delta_s = 1 + \sqrt{1 + b^2 k_v^2}$. Notice
in particular the scaling dimension is larger than two, for $b^2 > 0$ and non-zero $k_v$, and thus the operator is irrelevant.
This expression is completely analogous to the expression for the probe scalar operator (\ref{mome-sp}).
Expanding this expression perturbatively in $b^2$ results in:
\be
\< {X} (k)  {X}(-k) \>_{b} =  \td{g}(k_v) k_u \left (1 + (\frac{1}{2} b^2 k_v^2 - \frac{b^4 k_v^4}{8}) \ln(k_u) + \frac{b^4 k_v^4}{8} (\ln(k_u))^2
+ \cdots \right )
\ee
to order $b^4$. It would be interesting to see whether the terms at order $b^4$ follow from generic features of the four-point function of ${X}$,
or whether these coefficients are specific to the theory dual to TMG.

\subsection{Other deformations of 2d CFTs: general $z$} \label{zneq2}

Let us next consider the more general situation in which one deforms a 2d CFT by a $(p,q)$ operator ${\cal Y}_{p,q}$ where $(p,q)$ are the
CFT scaling weights corresponding to $(v,u)$ respectively,
\be
S_{\rm CFT} \rightarrow S_{\rm CFT} + b_{p,q} \int du dv {\cal Y}_{p,q}.
\ee
Such a deformation will respect non-relativistic scale invariance with dynamical exponent $z$
under which $u \rightarrow \lambda^z u$ and $v \rightarrow \lambda^{2-z} v$
provided that
\be
(p - 1)  (z-2) = (q - 1) z. \label{p,q}
\ee
For $z \neq 2$ scale invariance is respected provided that $q=1$ with $p$ arbitrary.
For $z \ge 2$ the condition can be satisfied for discrete values of the weights $(p,q)$.
Note that the ratio $u^{2-z} v^{-z}$
is scale invariant, whilst $(u v)$ scales with dimension two (as in the relativistic theory). As a simple example
in the case of $z=3$ a classical action one can write down is:
\be
S = \int du dv (\pa_u \Phi) (\pa_v \Phi) \left (1 + b_{4,2} (\pa_u \Phi) (\pa_v \Phi)^3 \right )
\ee
where $b_{2,4}$ characterizes the deformation of the original relativistic CFT by a dimension $(2,4)$ operator.
Dual holographic geometries which respect such non-relativistic scale invariance with a generic dynamical exponent $z \neq 1$ are
given by:
\be
ds^2 = \frac{d r^2}{r^2} + \frac{1}{r^2} \left (2 du dv - b^2 \frac{du^2}{r^{2(z-1)}} \right ). \label{generic}
\ee
The symmetry group consists of:
\bea
{\cal H} &:& u \rightarrow u + a, \qquad {\cal M}: v \rightarrow v + a, \\
{\cal D} &:& r \rightarrow (1- a) r, \qquad u \rightarrow (1- z a) u, \qquad v \rightarrow (1+ (z - 2)  a) v, \nn
\eea
and thus the $v$ coordinate scales non-trivially except when $z=2$. Note that for $z > 2$ the coordinate $v$ scales as a negative power of the dilatation.

For generic $z$ such non-relativistic backgrounds can be straightforwardly realized as solutions of Einstein gravity coupled to massive vectors,
although much of the literature has concentrated on the Schr\"{o}dinger case for which $z=2$. The case of critical speeding up, namely $z<1$, for which $b^2 < 0$, will be explored in detail in other work \cite{Caldeira}; in this case the spacetime is asymptotically anti-de Sitter and many of the conceptual subtleties of the $z > 1$ case
are absent. Spacetimes with generic $z$ can also be realized as solutions of TMG: the metric (\ref{generic}) solves the TMG field equations when $\mu = (2 z - 1)$. These TMG solutions were discussed in \cite{Olmez:2005by} and fit into the classification given in \cite{Chow:2009km} as pp-waves. For $b > 0$, these solutions with $u$ compactified were recently discussed in \cite{Anninos:2010pm}.

Now consider operators ${\cal O}_{h,\bar{h}}$ of conformal dimensions $(h, \bar{h})$ in the original CFT
in Euclidean signature with $\sqrt{2} (v,u) \to (w,\bar{w})$. Under the non-relativistic scaling
symmetry with exponent $z$ such that $w \to \l^{2-z}, \bar{w} \rightarrow \lambda^z \bar{w} $,
this operator scales as:
\be
{\cal O}_{h,\bar{h}} (w,\bar{w}) \rightarrow \lambda^{-h (2-z) - \bar{h} z}
{\cal O}_{h,\bar{h}} (\l^{2-z} w, \l^z \bar{w}),
\ee
and thus the non-relativistic scaling dimension is
\be
{\Delta}_{\rm nr} = h (2 -z) + \bar{h} z. \label{por}
\ee
Non-relativistic scale invariance constrains the two point functions to be of the form
\be
\< {\cal O}_{\Delta_{\rm nr}} (w, \bar{w}) {\cal O}_{\Delta_{\rm nr}'} (0,0) \> = \frac{1}{\bar{w}^{(\Delta_{\rm nr} + \Delta_{\rm nr}')/z}} f ( {\cal X}), \label{gfr}
\ee
where $f( {\cal X})$ is an arbitrary function of the scale invariant quantity
\be \label{ratio}
{\cal X}=w^{-z} \bar{w}^{2-z}.
\ee
Note that for general $z$ operators
of different scaling dimension are not precluded from having a non-zero two point function, although when $z=2$ the additional special conformal symmetry
ensures that the correlator is only non-zero when $\Delta_{\rm nr} = \Delta'_{\rm nr}$.

Consider now the 2-point function (\ref{rel_2pt}) of an operator ${\cal O}_{h,\bar{h}}$ in a relativistic CFT.
A simple computation shows that it can be written in the form (\ref{gfr}) with the non-relativistic
dimension ${\Delta}_{\rm nr}={\Delta}_{\rm nr}'$ given in (\ref{por}) and
%\be
$f ( {\cal X}) = c_{\cal O} {\cal X}^{2h/z}$.
%\ee
Using conformal perturbation theory, the leading correction to this two point function is given by
\be
\delta \< {\cal O}_{h,\bar{h}} (w) {\cal O}_{h,\bar{h}} (0) \> \sim \frac{b_{p,q}}{w^{2h-p} \bar{w}^{2 \bar{h} - q}}
\int d^2 y \left ( \frac{1} { (w-y)^p y^p (\bar{w} - \bar{y})^q \bar{y}^q } \right ) .
\ee
Consider the case where the deforming operator ${\cal Y}_{p,q}$ has integral spin, i.e. $p = q + n$ with $n$ an integer. Then the (renormalized) correction
can be written as:
\be
b_{p,q} \frac{1}{w^{2h- q -n} \bar{w}^{2 \bar{h} - q}} (-1)^n \frac{ \Gamma(q)^2}{\Gamma(q+n)^2} \pa_{w}^{2n}
\int d^2 y {\cal R}
\left( \frac{1}{|y-w|^{2q}}\right) {\cal R} \left(\frac{1}{|y|^{2q}} \right).
\ee
When $q$ is not an integer computing the integral leads to:
\be
(-1)^{n+1} \pi \frac{ \Gamma(1-q)^2 \Gamma(2q+2n-1)}{\Gamma(q+n)^2 \Gamma(1-2q)} b_{p,q}
\frac{1}{w^{2h + n + p-1} \bar{w}^{2 \bar{h} + p - 1}}.
\ee
This implies that the two point function to first order in the deformation can be written as:
\bea
\< {\cal O}_{h,\bar{h}} (w, \bar{w}) {\cal O}_{h,\bar{h}} (0,0) \> &=& \frac{1}{w^{2h} \bar{w}^{2 \bar{h}}} \left(
c_{\cal O} + {\alpha} b_{p,q} w^{1-p} \bar{w}^{1-q} \right), \\
&=& \frac{1}{w^{2h} \bar{w}^{2 \bar{h}}} \left(
c_{\cal O} + {\alpha} b_{p,q} {\cal X}^{(p-1)/z} \right), \nn
\eea
where ${\alpha}$ is a numerical constant and (\ref{p,q}) is used in the last equality. Thus,
the two point function indeed preserves non-relativistic scale invariance at this order,
with the two point function being of the form (\ref{gfr}) with the same values of
${\Delta}_{\rm nr}={\Delta}_{\rm nr}'$ (given in (\ref{por})) and
\be
f({\cal X}) = {\cal X}^{2h/z} \left(
c_{\cal O} + {\alpha} b_{p,q} {\cal X}^{(p-1)/z} \right)
\ee
One can also understand why the corrections must behave as $b_{p,q} w^{1-p} \bar{w}^{1-q}$ as follows: the deformed action will remain invariant
under the original dilatation symmetries provided that the coupling $b_{p,q}$ is also transformed. In particular, under the scaling
$\bar{z} \rightarrow \bar{\lambda} \bar{z}$ the deformation to the action is invariant provided that $b_{p,q} \rightarrow \bar{\lambda}^{q-1} b_{p,q}$.
Since the corrections to the correlation functions respect this symmetry they must be organized in powers
of $b_{p,q} w^{1-p} \bar{w}^{1-q}$.

For integral $q$ the analysis is more complicated as
we need to use the renormalized expressions for the distributions, and we obtain instead for the integral:
\be
 (-1)^n 2 \pi \frac{ \Gamma(q)^2}{\Gamma(q+n)^2} b_{p,q} \pa_{{w}}^{2n} \Box_w^{2q-2} \left ( |w|^{-2} \ln(m^2 |w|^2) \right ),
\ee
which gives a correction to the two point function proportional to:
\be
b_{p,q} \frac{1}{w^{2h + (p-1)} \bar{w}^{2 \bar{h} + (q-1)}} \ln (\td{m}^2 |w|^2),
\ee
with $\td{m}^2$ a rescaled mass scale.
This implies that the two point function to first order in the deformation can be written as:
\bea
\< {\cal O}_{h,\bar{h}} (w, \bar{w}) {\cal O}_{h,\bar{h}} (0,0) \> &=& \frac{1}{w^{2h} \bar{w}^{2 \bar{h}}} \left (
c_{\cal O} + {\beta} b_{p,q} w^{(1-p)} \bar{w}^{(1-q)}  \ln (\td{m}^2 |w|^2) \right ), \label{integ} \\
&=& \frac{1}{w^{2h} \bar{w}^{2 \bar{h}}} \left (
c_{\cal O} + {\beta} b_{p,q} {\cal X}^{(p-1)/z}  \ln (\td{m}^2 |w|^2) \right ), \nn
\eea
where $\beta$ is a computable numerical constant and (\ref{p,q}) is used in the last equality.
Note that the correction to the two point function preserves the $z=2$ non-relativistic scale invariance when $q=1$ for any $p$. In this case not only $f({\cal X})$ but also the non-relativistic dimension of the operator
get corrections and the 2-point function is given by (\ref{gfr}) with
\bea
f( {\cal X}) &=& {\cal X}^{-2h/z} \left ( c_{\cal O} + \beta b_{p,q} {\cal X}^{(p-1)/z} \ln ( \tilde{m}^2 {\cal X}^{-1/z} ) \right ) + {\cal O}(b_{p,q}^2); \\
{\Delta}_{\rm nr}={\Delta}_{\rm nr}' &=& h (2 -z) + \bar{h} z - \beta b_{p,q} {\cal X}^{(p-1)/z} + {\cal O}(b_{p,q}^2). \nn
\eea
When $z=2$, namely the deformation is by a $(p,1)$ operator, the scale invariant
quantity ${\cal X}$ depends only on the coordinate $v$ and one can check that these
formulas reduce to the ones we presented in the previous subsection.

\subsection{Massive vector model with $z=2$}

The specific case of a deformation by a $(2,1)$ operator corresponds to the $z=2$ massive vector model. The above considerations
indicate that generically under such a deformation operators will acquire anomalous dimensions depending on the lightcone momentum as
functions of $b_{2,1} k_v$.

In the bulk calculations in later sections, we will focus on scalar operators dual to minimally coupled
scalar fields, as well as the operators dual to the bulk metric and massive vector field. We indeed find that such scalar operators and
the vector operators acquire anomalous dimensions, and that these anomalous dimensions are functions of $(b_{2,1} k_v)^2$. We can understand
the latter straightforwardly as follows. If one considers scalar operators dual to minimally coupled scalar fields in the bulk, the three point
function between two scalar operators and the vector operator necessarily vanishes. It vanishes because minimally coupled implies that
there is no three point coupling\footnote{Strictly speaking, the vanishing of the bulk three point coupling does not
always guarantee vanishing of the corresponding three point function, see \cite{Skenderis:2006uy,Skenderis:2006di}.
The boundary counterterms required by holographic renormalization can induce
non-zero three point functions even when the bulk coupling is zero; this happens for extremal correlators, but none of the correlators discussed
here is extremal.} between the bulk scalar fields and vector field: the bulk action under consideration is
\be \label{bulk_vec}
S = \int d^{d+1} x \sqrt{-g} (R - 2 \Lambda - \frac{1}{4} F^2 - \half m^2 A^2 - \half (\partial \Phi)^2 - \half m^2_{\Phi} \Phi^2 ).
\ee
Expanding the field equations to cubic order in fluctuations couples the scalar to the vectors via the metric, and thus there is generically
a non-trivial four point function between two scalars and two vectors.
The first correction to the scalar two point function therefore follows from this four point function between two scalar operators and two
insertions of the vector operator. The leading correction will hence occur at order $(b_{2,1} k_v)^2$, but to compute it via conformal perturbation theory
we would need to know the explicit form of the four point function (since it is of course not completely determined by conformal invariance).
The bulk action implies that only correlation functions involving two scalars and an even number of vectors is not zero, and therefore subsequent
corrections should be organized in powers of $(b_{2,1} k_v)^2$. Indeed this feature is seen in the holographic dual in section
\ref{sec:bulkcomputations}, along with the stronger result that the anomalous dimension resums into a closed form. It would be interesting to
derive the latter result from a null dipole realization.

A similar story holds for the deforming vector operator itself: the action implies there is no cubic coupling between
three vectors, and thus the three point function between three vector operators is zero. Since the bulk
action (\ref{bulk_vec}) is quadratic in the massive vector fields, only
correlators with an even number of vector operators are non-zero, and thus that the corrections
to the vector two point function in the deformed theory should be organized in powers of $(b_{2,1} k_v)^2$.

Let us now consider the scaling dimensions of the vector operator (in $d=2$). Consider first the theory at $b_{2,1} = 0$.
The mass of the bulk vector field is such that it
corresponds to a vector operator of dimension three in the dual conformal field theory. Splitting the vector operator into holomorphic and
anti-holomorphic components, one sees that
there is a $(2,1)$ operator $X_v$ and a $(1,2)$ operator $X_u$. Since in the conformal field theory
the two point functions are:
\be
\< X_v (w,\bar{w}) X_{v} (0,0) \> = \frac{c_{\cal {X}}}{w^{4} \bar{w}^2}; \qquad
\< X_u (w,\bar{w}) X_u (0,0) \> = \frac{c_{\cal {X}}}{w^{2} \bar{w}^4},
\ee
with $c_{\cal X}$ normalizations, the scaling dimensions with respect to the Schr\"{o}dinger symmetry are:
\be
\Delta_s(X_v) = 2; \qquad \Delta_s(X_u) = 4. \label{dime}
\ee
This is in agreement with the non-relativistic scaling dimensions of the source and vev coefficients found holographically in (\ref{bzero}).
Switching on the deformation by the operator $X_v$, we expect that the scaling dimensions of both $X_u$ and $X_v$ at non-zero $k_v$ are
corrected, with the corrections being organized in powers of $(b_{2,1} k_v)^2$. The holographic calculation of
section \ref{sec:linearizedvector} gives explicit expressions for these scaling dimensions at finite $b_{2,1}$. Just
as in the case of TMG, the dimensions are expressed as a square root, with the dimensions of both operators
being greater than two, at non-zero $k_v$.

\subsection{Higher spacetime dimensions}

There has been considerable interest in using massive vector models in dimensions higher than three to model non-relativistic theories.
As explained in subsection \ref{sec:deformation},
to leading order in $b$ the standard AdS/CFT dictionary implies that switching on $b$ corresponds
to switching on a source for a vector operator ${X}^{a}$ of dimension
\be
\Delta_v = d + z - 1
\ee
in the $d$-dimensional dual CFT, i.e.
\be
S_{\rm CFT} \rightarrow S_{\rm CFT} + \int d^{d}x b {X}_{v}. \label{high-d}
\ee
Corresponding type IIB and eleven-dimensional supergravity solutions for general $z$ are given in \cite{Donos:2009xc}.
In this section we will focus on the case of $z=2$ in dimension $d$, and discuss the leading corrections
to operator correlation functions in the deformed theory. As already mentioned, the leading corrections
to the deforming operator itself and to scalar operators dual to minimally coupled scalar fields occur at order $b^2$, and are derived from four-point functions in the conformal theory.

Corrections at order $b$ can however occur for complex scalar operators as these
can have non-zero three point functions with the deforming vector operator.
Let us consider a complex scalar operator ${\cal O}$ with dimension $\Delta$ in the original conformal field theory.
It has a two point function in the conformal field theory given by
\be
\< \bar{{\cal O}} (x) {\cal O} (0) \> = \frac{c_{\Delta}}{ |x|^{2 \Delta}},
\ee
where $c_{\Delta}$ is the operator normalization and we are working in Euclidean signature. Noticing that
\be
|x|^2 = w \bar{w} + x^i x_i,
\ee
we see that the operators are also primary from the perspective of Schr\"{o}dinger symmetry: under $\bar{w} \rightarrow \lambda^2 \bar{w}$, $x^{i} \rightarrow \lambda x^i$ with $w$ invariant, the operators scale with non-relativistic scaling dimension $\Delta_s = \Delta$.

Next consider the correction at first order in $b$ to this two point function in the deformed theory (\ref{high-d}).
The leading correction to the two point function is:
\be
\< \bar{{\cal O}} (x) {\cal O} (0) \> = b \int d^{d} y \< \bar{{\cal O}} (x) X_v (y) {\cal O} (0) \>.
\ee
This can be computed by first noting that
three point functions between a complex scalar and a vector are fixed by conformal invariance to have the form
(derived holographically in \cite{Arutyunov:1999en}):
\be \label{eq:3pt_sc-v}
\< \bar{{\cal O}} (x) X^{\mu} (y) {\cal O} (z) \> = %C_{ {\cal O}^{+} X {\cal O}^{-}}
\frac{i C}{|x-z|^{2\Delta + 1 - \Delta_v} |x - y|^{\Delta_v -1} |y-z|^{\Delta_v -1}}
\left ( \frac{ (x-y)^{\mu}}{|x-y|^2} - \frac{(z-y)^{\mu}}{|z-y|^2} \right ),
\ee
where $C$ is a (real) normalization factor.

As in the previous sections, the expression for the three point function needs to be regularized to take into account contact term contributions to the integral. In the case
of $z=2$ the deforming vector operator has dimension $\Delta_v = d + 1$. Using this fact, the three point function can be rewritten as:
\be
\< \bar{{\cal O}} (x) {X}^{\mu} (y) {\cal O} (0)  \> =  - \frac{i C}{d |x|^{2\Delta - d}} \left ( \frac{1}{|y|^d} \pa_{x_{\mu}}  \frac{1}{|x-y|^d} + \frac{1}{|x-y|^d} \pa_{y_{\mu}}  \frac{1}{|y|^d} \right ).
\ee
In differential regularization one replaces $1/|x|^d$ by the well-defined expression:
%\bea
%{\cal R} \left ( \frac{1}{(x^2)^{d/2}} \right ) &=& \lim_{\lambda \rightarrow d/2} \left ( \frac{1}{(x^2)^{\lambda}} - \frac{m^{2\lambda -d}}{(d-2 \lambda)} S_{d-1}
%\delta^{(d)}(x) \right ) \label{ren-higher} \\
%&=&
\be
{\cal R} \left ( \frac{1}{|x|^d} \right )=- \frac{1}{2 (d-2)} \Box \frac{1}{|x|^{d-2}} \left ( \ln (m^2 |x|^2) + \frac{2}{d-2} \right ), \label{ren-higher}
\ee
%\nn\eea
which is the generalization of (\ref{2d}) to arbitrary dimensions $d > 2$. (The last term in \eqref{ren-higher} is scheme dependent.)
%, as given in \cite{Osborn:1993cr}. Here $S_{d-1} = 2 \pi^{d/2}/\Gamma(d/2)$ denotes the volume of the unit $(d-1)$ sphere.
Then the leading correction to the two point function is:
\be
\delta \< \bar{{\cal O}} (x) {\cal O} (0) \> = - \frac{2 b}{d |x|^{2 \Delta -d}} i C  \pa_{w}
\int d^d y {\cal R} \left ( \frac{1}{|y|^{d}}\right) {\cal R}\left(\frac{1}{|x-y|^{d}} \right ).
\ee
The integral can be computed analogously to the two dimensional case via the convolution:
\be
\int d^d y {\cal R} \left ( \frac{1}{|y|^{d}}\right) {\cal R}\left(\frac{1}{|x-y|^{d}} \right )
= \frac{1}{ (2 \pi)^d} \int d^d k e^{i \vec{k} \cdot \vec{x}} \td{D}_{d}^2(k),
\ee
where
\be
\td{D}_{d} (k) = \int d^d x e^{-i \vec{k} \cdot \vec{x} } {\cal R} \left ( \frac{1}{|x|^d} \right )%; \\
= - \frac{2 \pi^{d/2}}{(d-2) \Gamma(d/2 -1)} \ln (|k|^2/\mu^2), \nn
\ee
with $\mu^2 = 4 m^2 \exp (\Psi(1) + \Psi(d/2 -1))$. Note that in $d=4$ this reduces to $\mu = 2 m/\gamma'$ since the Euler constant
$\gamma' = \exp(- \Psi(1)) = \exp(\gamma)$.
To derive this formula, one can use the expansion of the identity (\ref{ft-imp}) with $2 \lambda = d - 2 -2a$:
\be
\int d^d x e^{-i \vec{k} \cdot \vec{x}} \frac{1}{|x|^{d-2}} (m |x|)^{2a} = \pi^{d/2} 2^{2+2a} \frac{\Gamma(1+a)}{\Gamma(d/2-1+a)} |k|^{-2} (|k|/m)^{-2a}
\ee
in powers of $a$. Equating terms at each order in $a$ results in the identities:
\bea
\int d^d x e^{-i \vec{k} \cdot \vec{x}} \frac{1}{|x|^{d -2}} &=& \frac{4 \pi^{d/2}}{|k|^2 \Gamma (d/2-1)}; \\
\int d^d x e^{-i \vec{k} \cdot \vec{x}} \frac{\ln (m^2 x^2)}{|x|^{d -2}} &=&  -  \frac{4 \pi^{d/2}}{|k|^2 \Gamma (d/2-1)} \ln (|k|^2/\mu^2); \nn  \\
\int d^d x e^{-i \vec{k} \cdot \vec{x}}  \frac{\ln^2 (m^2 |x|^2)}{|x|^{d -2}} &=&  \frac{4 \pi^{d/2}}{|k|^2 \Gamma (d/2-1)} \left (  \ln^2 (|k|^2/\mu^2)  + \frac{\pi^2}{6}
- \Psi^{(1)} (d/2-1) \right ), \nn
\eea
where $\Psi^{(m)}(x)$ is the polygamma function, the $(m+1)$-th derivative of the gamma function. These
identities generalize similar ones for $d=4$ derived in \cite{Freedman:1991tk}.

Using the third of these identities we note
that
\bea
&&\int d^d y {\cal R} \left ( \frac{1}{|y|^{d}}\right) {\cal R}\left(\frac{1}{|x-y|^{d}} \right ) \\
&& \qquad = - \frac{\pi^{d/2}}{(d-2)^2 \Gamma(d/2-1)} \Box \frac{1}{|x|^{d-2}} \left ( \ln^2 (m^2 |x|^2) - \frac{\pi^2}{6} + \Psi^{(1)}(d/2-1) \right ). \nonumber
\eea
Note however that away from $x=0$,
\be
\Box \left ( \frac{\ln^2 (m^2 |x|^2)}{|x|^{d-2}} \right ) = \frac{1}{|x|^{d}} \left ( 4 (d-2) \ln (m^2 |x|^2) + 8 (3-d) \right ),
\ee
and therefore the leading correction to the two point function has the form:
\be
\< \bar{{\cal O}} (x) {\cal O} (0) \> = \frac{c_{\Delta}}{ (x^2)^{\Delta}} + i c_1 \frac{b}{|x|^{2 \Delta - d}} \pa_w \left [ {\cal R} \left (  \frac{\ln (M^2 |x|^2)}{|x|^{d}} \right ) \right ],
\ee
where $M$ is a rescaled mass scale and $c_1$ is a real numerical constant proportional to the three point function constant $C$:
\be
c_1 = \frac{8 \pi^{d/2}}{d (d-2) \Gamma(d/2-1)} C.
\ee
The corrected two point function can be rewritten as:
\be
\< \bar{{\cal O}} (x) {\cal O}(0) \> = \frac{c_{\Delta}}{ |x|^{2 \Delta}} + i c_1 b
\frac{d}{2 \Delta} \pa_w \left [ {\cal R} \left (  \frac{\ln (\td{M}^2 |x|^2)}{|x|^{2 \Delta}} \right ) \right ],
\ee
where $\td{M}^2 = M^2 \exp (1/\Delta - 2 /d)$. In the case where $\Delta$ is non-integral, the Fourier transform is:
\be
\< \bar{{\cal O}} (k) {\cal O} (-k) \> = c_{\Delta} \pi^{d/2} 2^{d- 2 \Delta} \frac{\Gamma(d/2 -\Delta)}{\Gamma(\Delta)} |k|^{2 \Delta - d} \left (1  + \frac{d}{2 \Delta}  k b c_1 \ln (|k|^2/\td{\mu}^2) \right ),
\ee
where $k$ is the momentum conjugate to $w$. (When $\Delta$ is
integral we need to Fourier transform the renormalized expression, which can be done analogously to previous sections.) Analytically continuing back to Lorentzian signature, the time-ordered correlator
is
\be
\< {\cal T} \bar{{\cal O}} (k) {\cal O} (-k) \> = c  (2 k_u k_v  + k_i k^i- i \ep)^{\Delta_s - d/2}, \label{s-mon}
\ee
where $c=c_{\Delta} \pi^{d/2} 2^{d- 2 \Delta} \Gamma(d/2 -\Delta)/\Gamma(\Delta)$ in the renormalization scheme where $\tilde{\mu}^2=1$. The scaling dimension $\Delta_s$ is to leading order
in $b$ given by:
\be
\Delta_s = \Delta + \gamma, \qquad \gamma = \frac{1}{\sqrt{2}} (b k_v) \frac{4 \pi^{d/2} C}{\Delta (d-2) \Gamma(d/2-1)}
\ee
and the anomalous dimension hence depends on $(b k_v)$. % and on the global $U(1)$ charge $j$ of the operator.
%Following the discussion in appendix \ref{app_FT}, one may Fourier transform the correlator (\ref{s-mon})
%to coordinate space $(u,x^i)$ arriving at the expected form of 2-point function for a Schr\"{o}dinger invariant
%theory.

%\be
%\< \bar{{\cal O}} (k_v,u,x^i) {\cal O} (-k_v,0,0) \> =  c (b k_v) \frac{1}{(2 \pi)^{d-1}} \int d^{d{-}2}\!k dk_u e^{-i k^i x_i - i k_u u}
%(2 k_u k_v  + k_i k^i - i \ep)^{\Delta_s - d/2}.
%\ee
%Integrating first over $k_u$ results in
%\bea
%\< \bar{{\cal O}} (k_v,u,x^i) {\cal O} (-k_v,0,0) \> &=&  c (b k_v) \frac{1}{(2 \pi)^{d-1}}
%\int d^{d{-}2}\!k e^{-i k^i x_i + i \frac{k^i k_i u}{2 k_v}} {\cal D} (k_v,u); \\
%{\cal D}(k_v,u) &=&
%- \theta(k_v) \theta(-u) \frac {\pi}{\Gamma(d/2 - \Delta_s)}
%\left ( \frac{2k_v}{i} \right )^{\Delta_s - d/2} \frac{1}{u^{\Delta_s + 1 - d /2}} \nn \\
%&& - \theta(- k_v) \theta(u) \frac {\pi }{\Gamma(d/2 - \Delta_s)}
%\left ( \frac{-2 k_v }{i} \right )^{\Delta_s - d/2} \frac{1}{u^{\Delta_s + 1 - d/2}}. \nn
%\eea
%The remaining integral over $k_i$ is Gaussian and gives:
%\bea
%\< \bar{{\cal O}} (k_v,u,x^i) {\cal O} (-k_v,0,0) \> &=& C (b k_v) \left ( \theta(k_v) \theta(-u) + \theta(- k_v) %\theta(u) \right ) |k_v|^{\Delta_s -1}
%|u|^{-\Delta_s} e^{ - i \frac{k_v x^i x_i}{2 u}}. \nn \\
%C( b k_v) &=& c (b k_v) \frac{\pi^{1-d/2} 2^{\Delta_s -d }} {\Gamma (d/2 - \Delta_s) i^{\Delta_s + d - 1}},
%\eea
%This is of the form expected for operators in a Schr\"{o}dinger invariant theory with lightcone momentum (charge) $k_v$ and scaling dimension $\Delta_s$.

\bigskip

To describe the theory at finite $b$ the null dipole picture should be of use: in the context of the $S^5$ reduction of type IIB, the dual theory should be
the null dipole deformation of ${\cal N} = 4$ SYM. In this case the global $R$ symmetry charges are used to determine the dipole vectors
of the various fields as follows. Choose a constant element $B \in su(4)$, where $su(4)$ is the Lie algebra of $SU(4)$.
Then denote the elements of $B$ in the ${\bf 4}$ representation as $U_{j \bar{k}}$, where $j,\bar{k} = 1,\cdots, 4$, and
denote the elements of $B$ in the ${\bf 6}$ representation as the antisymmetric matrices
$M_{IJ}$, where $I,J = 1,\cdots,6$. Let $u_{I}^l$ be an eigenvector
of $M_{IJ}$ with eigenvalue $L_l$. The ${\cal N} = 4$ SYM complex valued scalar fields $\Phi^l = \sum_{I} u_{I}^l \Phi^I$ are then assigned
(null) dipole vectors $L_l$. Similarly the fermionic fields are assigned dipole vectors determined by the eigenvalues of the matrix
$U_{j \bar{k}}$.

Suppose for example one chooses $M_{IJ}$ such that only $M_{12} = - M_{21}$ is non-zero. Next complexify the six scalars into three complex
scalars $\Phi_c^a$, with $a=1,2,3$ where $\Phi_c^1 = \Phi^1 + i \Phi_2$ and so on. Then $\Phi_c^1$ has non-zero dipole charge,
given by $L_{\mu} = - i b \d_{\mu u}$, while the other two complex scalars have zero charge.
The dipole product (\ref{dipole}) can be expanded
to leading order in $b$ as:
\be \label{dipole-prod}
\Phi_c^1 \ast \Phi_c^{1 \dagger} = \Phi_1^2 + \Phi_2^2 + b (\pa_v \Phi_1 \Phi_2 - \pa_v \Phi_2 \Phi_1) + \cdots
\ee
The analogue of the Seiberg-Witten map in this case relates fields $\Phi$ in the dipole theory
to fields $\Phi_o$ in the ordinary theory as:
\be
\Phi = \Phi_o {\cal P} \left ( \exp (i L^{\mu} \int^{1}_{0} A_{\mu} (x + t L) dt) \right ).
\ee
Substituting into the ${\cal N} = 4$ SYM action and expanding to leading order in $b_u$ results in:
\be
S = S_{ {\cal N} = 4} + \int d^4 x b {\cal V}^{12}_v + \cdots
\ee
where the vector operator $V^{IJ}_{\mu}$ transforms as the ${\bf 15}$ of the $SO(6)$ R symmetry group
and has the form:
\be
{\cal V}^{[IJ]}_{\mu}
= {\rm Tr} \left ( F_{\mu \nu} \phi^{[I} D^{\nu} \phi^{J]} + \sum_{K} \phi^K D_{\mu} (\phi^{[K} \phi^{I} \phi^{J]}) + \cdots \right ).
\ee
Here $I,J=1,\cdots,6$ are R symmetry indices, $D_{\mu}$ is the covariant derivative with respect
to gauge fields $A_{\mu}$ of the field strength $F_{\mu \nu}$ and square parentheses denote antisymmetrisation. The ellipses denote
fermionic terms. At linear order in $b$ the null dipole theory is indeed equivalent to a deformation
of ${\cal N} = 4$ SYM by a constant null source $b$ for an operator of relativistic dimension $(d+1) = 5$. This operator
however has Schr\"{o}dinger dimension four, using the fact that $\Delta_s (A_v) = 0$ whilst $\Delta_s (\phi^I) = \Delta_s (A_u) = \Delta_s (A_i) = 1$.
Expanding to higher order in $b$ will lead to a series of deformations respecting Schr\"{o}dinger symmetry. Note that the amount of supersymmetry preserved by
the dipole deformation depends on the explicit choice of $M_{IJ}$.

It is important to note that all these deformations are marginal with respect to the Schr\"{o}dinger symmetry. This immediately
follows from the fact that every term in the null dipole product of any two fields has the same Schr\"{o}dinger dimension, even though
the relativistic conformal dimensions of the terms are different. One can see this property in (\ref{dipole-prod}), since a term at order $b^n$ involves
$n$ derivatives in the $v$ direction, which leaves the Schr\"{o}dinger dimension unchanged but increases the usual dimension. More generally, the
product of any two fields $(\Phi_{q_1}, \Phi_{q_2})$ with dipole charges $L_1 = - i n_1 b$  and $L_2 = - i n_2 b$ respectively, in $k_v$ momentum space is:
\be
\Phi_{q_1}(k^1_v, u, x^i) \ast \Phi_{q_2} (k^2_v, u, x^i) = e^{i b (n_2 k^1_v - n_1 k^2_v)} \Phi_{q_1} (k^1_v,u,x^i)
\Phi_{q_2} (k^2_v, u, x^i),
\ee
with the dipole dependence contained in the phase factor, which is invariant under Schr\"{o}dinger transformations.

Given that the series of deformations considered here is expected to resum
into a null dipole theory by the analogue of the Seiberg-Witten map \cite{Seiberg:1999vs}, one might ask what this implies
for the renormalizability and unitarity of the theory at finite $b$. This question has not been explored in previous literature, although
it has been argued that analogous theories with light-like noncommutativity are well-defined, despite the non-locality in one null direction \cite{Aharony:2000gz}.
The reason is that lightcone quantization in which the other (local) null coordinate is treated as the time coordinate results in a lightcone Hamiltonian
which is Hermitian. Naively at least the same argument would apply also to null dipole theories, since they are local in the $(u,x^i)$ coordinates.

\subsection{Stress energy tensor}
\label{sec:nrstresstensor}

Next we turn to the stress energy of the deformed theory. It is useful to start the discussion by
considering again a simple explicit model which is Schr\"{o}dinger invariant:
\be
S = \int d^2 x \left ( \pa_u \Phi \pa_v \Phi + b \pa_u \Phi (\pa_v \Phi)^2 \right ).
\ee
This model exhibits the Schr\"{o}dinger group of symmetries \eqref{group} presented in the introduction.
Let us now consider the Noether current ${\cal J}^i$ for each symmetry $Q$. For $H$ the corresponding symmetry current is:
\be
{\cal H}^{u} = 0; \qquad
{\cal H}^{v} = (\pa_u \Phi)^2 (1 + 2 b \pa_v \Phi).
\ee
For $M$ the symmetry current is:
\be
{\cal M}^{u} = (\pa_v \Phi)^2 (1 + b (\pa_{v} \Phi)); \qquad
{\cal M}^{v} = b \pa_u \Phi (\pa_v \Phi)^2.
\ee
The dilatation current is:
\be
{\cal D}^{u} = 0; \qquad {\cal D}^{v} = u  (\pa_u \Phi)^2 (1 + 2 b \pa_v \Phi),
\ee
whilst the special conformal current is similar:
\be
{\cal C}^{u} = 0; \qquad {\cal C}^{v} = u^2  (\pa_u \Phi)^2 (1 + 2 b \pa_v \Phi).
\ee
The currents are conserved onshell because of the field equation:
\be
\pa_{u} \pa_v \Phi + b \pa_u \Phi \pa_v^2 \Phi + 2 b \pa_{u} \pa_v \Phi \pa_v \Phi = 0.
\ee
Usually the currents associated with (conformal) isometries can
be expressed in terms of a symmetric conserved (traceless) stress energy tensor $T_{ij}$ as
\be
{\cal J}^{i} = T^{ij} \zeta^Q_j
\ee
where $\zeta^Q_j$ is the vector field generating each symmetry, i.e. $\delta_{Q} x^i = - \zeta^{Qi}$.

In the case at hand the symmetry currents can all be written as ${\cal J}^{i} = t^{ij} \zeta^{Q}_j$ where the tensor $t_{ij}$
is:
\begin{align}
t_{uu} &= (\pa_u \Phi)^2 (1 + 2 b \pa_v \Phi); & t_{vu} &= 0; \label{tt} \\
t_{uv} &=  b \pa_u \Phi (\pa_v \Phi)^2; & t_{vv} &= (\pa_v \Phi)^2 (1 + b(\pa_{v} \Phi)). \nn
\end{align}
Under the dilatation symmetry $u \rightarrow \lambda^2 u$ the tensor $t_{ij}$ scales as:
\be
t_{uu} \rightarrow \lambda^{-4} t_{uu}; \qquad
t_{uv} \rightarrow \lambda^{-2} t_{uv}; \qquad
t_{vv} \rightarrow \lambda^{0} t_{vv}. \label{t-scaling}
\ee
This tensor is conserved but it is not symmetric and it is not traceless. The latter implies that the
theory is invariant only under chiral dilatation invariance, rather than the corresponding anti-chiral dilatation
invariance. In a relativistic theory one can always find an improvement term $\theta_{ij} = \partial^l Y_{lij}$,
where $Y_{lij}$ is antisymmetric in its first two indices, such that $T_{ij} = t_{ij} + \theta_{ij}$ is
symmetric. In the example at hand one can easily prove that such improvement tensor does not exist\footnote{
Assuming $Y$ to exist one obtains from $T^{uv}= T^{vu}$ and (\ref{tt}) that
$b \partial_u \Phi (\partial_v \Phi)^2 = \partial_v Y^{vuv} - \partial_u Y^{uvu}$
should hold. This equation, however, does not hold as the deformation is not a total derivative.}.
This illustrates a general result:

{\it Any theory in Minkowski spacetime that possesses a conserved, symmetric stress
energy tensor $T_{ij}$ is Lorentz invariant.}

This can be shown as follows. Minkowski space has an isometry corresponding to Lorentz transformations
with Killing vector $\xi^i = \omega_{ij} x^j$ and $\omega_{ij}$ antisymmetric.
Using $T_{ij}$ one can construct the corresponding conserved Lorentz current,
\be
J_L^i = T^{ij} \omega_{jk} x^k
\ee
This is conserved because
\be
\partial_i J_L^i = (\partial_i T^{ij}) \omega_{jk} x^k + T^{ij} \omega_{ij} =0.
\ee
For the Schr\"{o}dinger theory, invariance under translations implies the existence
of a conserved stress energy tensor. However, this must be non-symmetric
because otherwise the theory would be Lorentz invariant.

When one couples a theory which is Lorentz invariant to background gravity the
symmetric and conserved stress energy tensor arises from the variation of the action
with respect to the metric. In a non-Lorentz invariant theory one
can couple the model to background gravity using vielbeins. The coupling to gravity amounts to
replacing curved indices by flat target indices using vielbeins, i.e.  $A_i$ becomes $e_i^m A_m$,
where $m$ is a curved space index. The fact that the underlying theory is not Lorentz invariant
means that some of the tangent space indices are open, i.e. not contracted.
In this case the non-symmetric energy momentum tensor will naturally emerge from the
variation of the action with respect to the vielbein.

For the toy model given above coupling to the vielbein gives
\be
S = \int d^2 x \; e \left ( e^{mi} e_{i}^n \pa_m \Phi \pa_n \Phi + b e^{m}_{\hat{u}} e^{n}_{\hat{v}} e^{p}_{\hat{v}}
(\pa_m \Phi)(\pa_n \Phi) (\pa_p \Phi) \right ) \equiv \int d^2 x \; e {\cal L},
\ee
where $e$ is the determinant of the vielbein and the tangent space metric is $g_{ij} = \eta_{ij}$ such
that $\eta_{\hat{u} \hat{v}} =1$. For clarity the tangent space indices are denoted $(\hat{u}, \hat{v})$.
Defining:
\be
t_{mn} = \frac{e_n^j}{e} \frac{\delta {\cal L}}{\delta e^{mj}}
\ee
gives:
\be
t_{mn} = \pa_{m} \Phi \pa_n \Phi - g_{mn} {\cal L} + b \left ( e_{\hat{u} n} e^{p}_{\bar{v}} e^{q}_{\hat{v}} +
2 e_{\hat{v}n} e^p_{\hat{u}} e^q_{\hat{v}}  \right ) \pa_m \Phi \pa_p \Phi \pa_q \Phi,
\ee
which in a flat background in which $g_{mn} = \eta_{mn}$ reproduces the tensor $t_{ij}$ above. The simplest
way to show that $t_{mn}$ is conserved is the following. Pulling
$t_{mn}$ back into tangent space reproduces $t_{ij}$; the latter is manifestly conserved using the field
equation:
\be
\Box \Phi + D_{\hat{u}} (b (\pa_{\hat{v}} \Phi)^2) + 2 D_{\hat{v}} (b \pa_{\hat{u}} \Phi \pa_{\hat{v}} \Phi )  = 0.
\ee
Since
\be
D^{m} t_{mn} = D^{m} (e_{m}^{i} e_{n}^j t_{ij}) = e_{n}^j D^{i} t_{ij},
\ee
conservation of the pulled back tensor $t_{ij}$ implies conservation of $t_{mn}$.

To summarize, there are two related stress energy operators in the deformed field theory. The first is the symmetric tensor $T_{ij}$ which couples
to the metric, but is not conserved in the deformed theory at finite $b$. The second is the operator $t_{mn}$ that couples to the vielbein, which
is not symmetric but is conserved. It is the latter operator which is natural in the deformed theory, and thus in the holographic analysis one
should trade metric fluctuations for vielbein fluctuations. In other words, one should specify boundary conditions for the vielbein, rather than the metric;
this point was noted in \cite{Ross:2009ar}.

It is interesting to connect this discussion with the conventional description of non-relativistic theories. Suppose one formulates a Schr\"{o}dinger
invariant theory in $(d-1)$ dimensions, with background coordinates $(u,x^i)$. Then, the natural operators are the energy current ${\cal E}$ with components
$({\cal E}_{u}, {\cal E}_i)$; the mass current $\rho$ with components $(\rho_{u}, \rho_i)$ and the (symmetric) stress tensor $\pi_{ij}$. In the $d$-dimensional
realization of the Schr\"{o}dinger symmetry discussed here, the operator $t_{mn}$ should include these operators. It will however also include additional
operators associated with the extra lightcone dimension, which are not usually discussed in the lower dimensional realizations. These extra
operators are necessarily present because the Schr\"{o}dinger theory is obtained via a deformation of a relativistic theory in $d$ dimensions.

\bigskip

Before leaving the field theory discussion, it is also useful to summarize which operators are expected to be marginal, irrelevant and
relevant in the deformed Schr\"{o}dinger invariant theory.
In the cases of both TMG and the massive vector, the operator which deformed the theory to Schr\"{o}dinger is
irrelevant at finite $k_v$; its scaling dimension in both cases is given holographically (for $d=2$) as:
\be
\Delta_s = 1 + \sqrt{1 + b^2 k_v^2}.
\ee
In the massive vector case the dimension of the other vector operator $X_u$ is such that it is also irrelevant for non-zero $k_v$. Therefore
these operators are all irrelevant in the Schr\"{o}dinger theory, except for the deforming operator itself with $k_v=0$.

Now consider the stress energy tensor $T_{ij}$, at $b=0$. The non-relativistic scaling dimensions of its components are:
\be
\Delta_s (T_{uu}) = 4; \qquad
\Delta_s (T_{uv}) = 2; \qquad
\Delta_s (T_{vv}) = 0, \label{se-dimensions}
\ee
and thus the components are irrelevant, marginal and relevant, respectively. Away from $b=0$ this is not the conserved
operator, but the components of $t_{ij}$ have the same scaling dimensions, see (\ref{t-scaling}), and these are protected. Thus from the
perspective of the Schr\"{o}dinger symmetry group, the operator $T_{uu}$ is irrelevant.

Since the operators $X_{vv}(k_v) $, $X_{v}(k_v)$, $X_{u}(k_v)$ and $T_{uu}$ are irrelevant, switching on finite sources for these operators
in the deformed theory would be expected to change the UV structure of the theory. Correspondingly, in the holographic dual, finite sources for these
operators would be expected to change the asymptotic structure of the dual spacetime. In the subsequent sections we will consider
solutions of the bulk linearized equations, and we will indeed see the bulk fields dual to these operators blow up faster
at the boundary than the background, demonstrating the fact that these modes change the asymptotic structure of the spacetime.

\section{TMG and the null warped background} \label{sec:hol-prel}
In this section we consider TMG and the null warped $AdS_3$ background in more detail.
The results presented below will provide a basis for the holographic computation of correlation functions
in the next sections. In subsection \ref{sec:varprincipletmg} we investigate the variational principle for TMG for
finite values of the cutoff. In subsection \ref{sec:dilatationop} we consider the dilatation operator for linearized
fluctuations around the null warped background.

\subsection{The variational principle for TMG}
\label{sec:varprincipletmg}
In this subsection we analyze the variational principle for TMG for finite values of the cutoff. Recall
that for ordinary Einstein gravity such an analysis results in the addition of the Gibbons-Hawking term to the
Einstein-Hilbert action. For TMG the corresponding variational principle is worked out in detail in this section. Note that
the results in this section are completely general and independent of the choice of background metric.
They can therefore also be used to obtain, for example, the correct on-shell action for black hole solutions in TMG.

We begin by decomposing the background metric in a radial ADM form:
\be
ds^2 = N^2 dr^2 + \g_{ij} (dx^i + N^i dr) (dx^j + N^j dr).
\ee
The connection coefficients are then given by:
\begin{align}
\G^r_{rr} &= \frac{\dot N}{N} + N^i B_i &  \G^i_{rr} &= \g^{ik} \dot N_i - N^i \G^r_{rr} - N^2 B^i - F^i_{\ph{i}j} N^j \nn \\
\G^i_{jr} &= F_j^{\ph{j}i} - N^i B_j & \G^r_{ij} &= \frac{1}{N}K_{ij}\\
\G^r_{ri} &= B_i & \G^i_{jk} &= \G^i_{jk}[\g] - \frac{1}{N} N^i K_{jk} \nn
\end{align}
where we defined:
\be
F_{ij} = \cdel_i N_j - N K_{ij}, \qquad \qquad B_i = \frac{1}{N} (\del_i N + K_{ij}N^j).
\ee
A dot denotes a radial derivative. Indices are raised with $\g^{ij}$ and $\epsilon_{ij}$ and covariant
derivatives $\cdel_k$ are defined using $\g_{ij}$ as well. Henceforth all $\G^i_{jk}$ are those associated to $\g_{ij}$. In our
conventions the extrinsic curvature is given by:
\be
\label{eq:Kij}
K_{ij} = \frac{1}{2N} (\cdel_i N_j + \cdel_j N_i - \del_r \g_{ij}).
\ee
We henceforth suppose that $N > 0$. The Einstein-Hilbert action plus Gibbons-Hawking term then becomes:
\be
S_{\text{EH}}= \frac{1}{16 \pi G_N} \int d^3 x \sqrt{-\g} N(R[\g] - 2 \Lambda + K^2 - K_{ij}K^{ij}).
\ee
The Chern-Simons action becomes:
\be
\label{eq:SCSADM}
\begin{split}
S_{\text{CS}} &= \frac{1}{16 \m \pi G_N} \int d^3 x \sqrt{-\g} \e^{ij} \Big( K_i^k \dot K_{kj} + N [2 \cdel_k \cdel_i K_j^k] \\
&\qquad + N^l [\cdel_k (K_i^k K_{jl}) - 2 K_{kl}\cdel_i K^k_j + \hf \e_l^{\ph{l}k} \del_k R] - \hf \G_{il}^k \del_r \G_{jk}^l\Big)
\end{split}
\ee
plus a radial boundary term of the form $\int d^2 x \sqrt{-\g} \e^{ij} K_{ik} (\del_j N^k)/(2 N)$ which vanishes once we gauge-fix $N^k = 0$ below.
Here spatial total derivatives have been omitted. This ADM form of the Chern-Simons action was also found in
\cite{Deser:1991qk,Buchbinder:1992pe,Hotta:2008yq}.

The Chern-Simons action can be manipulated as follows. First of all, we define $A^{kl}$ via:
\be
\label{eq:Aij}
\int d^3 x \sqrt{-\g} \e^{ij} \G_{il}^k \del_r \G_{jk}^l = \int d^3 x \sqrt{-\g} A^{kl} \dot \g_{kl}
\ee
up to total spatial derivatives. Furthermore, the first term in \eqref{eq:SCSADM} can be rewritten. To this end we decompose the extrinsic curvature as:
\be
\label{eq:Kleftright}
K_{ij} = k_{ij} + \bar k_{ij} + \hf k \g_{ij}
\ee
with:
\be
\label{eq:componentsK}
\begin{split}
k &= \g^{ij} K_{ij}\\
k_{ij} &= P_i^{\ph{i}k} (K_{kj} - \hf k \g_{kj}) = \hf (\delta_i^k + \e_i^{\ph{i}k})(K_{kj} - \hf k \g_{kj}); \\
\bar k_{ij} &= \bar P_i^{\ph{i}k} (K_{kj} - \hf k \g_{kj}) = \hf (\delta_i^k - \e_i^{\ph{i}k})(K_{kj} - \hf k \g_{kj}).
\end{split}
\ee
Notice that $k_{ij}$ and $\bar k_{ij}$ are symmetric, traceless and chiral (or antichiral), so:
\be
\label{eq:propertiescomponentsK}
\begin{split}
\g^{ij} k_{ij} &= 0, \qquad \qquad k_{ij} = k_{ji}, \qquad \qquad \bar P_i^{\ph{i}k} k_{kj} = 0; \\
\g^{ij} \bar k_{ij} &= 0, \qquad \qquad \bar k_{ij} = \bar k_{ji}, \qquad \qquad P_i^{\ph{i}k} \bar k_{kj} = 0.
\end{split}
\ee
These equations imply that $k_{ij}$ and $\bar k_{ij}$ each have a single independent component. Using \eqref{eq:Kleftright} we may rewrite:
\be
\label{eq:KdotK}
\begin{split}
\int d^3 x \sqrt{-\g} \e^{ij} K_i^k \dot K_{kj} &= \int d^3 x \sqrt{-\g} \Big(
-2 k^j_k \del_r \bar k^k_j - N^l \cdel_l (k^j_k \bar k^k_j) \\&\qquad + N K k^j_k \bar k^k_j \Big) + \int d^2 x \sqrt{-\g} k^j_k \bar k^k_j.
\end{split}
\ee
The last boundary term should be canceled by a boundary term in the action.

We now regard $k_i^j$, $\bar k_i^j$ and $k$ as independent variables, together with $\g_{ij}$, $N^i$ and $N$. On the other hand, $K_{ij}$
should henceforth be understood as in \eqref{eq:Kleftright}. We need to introduce a single Lagrange multiplier $\Pi^{ij}$ enforcing \eqref{eq:Kij}:
\be
\label{eq:constraint}
\int d^3 x \sqrt{-\g} \Pi^{ij} (\dot \g_{ij} + 2N K_{ij} - \cdel_i N_j - \cdel_j N_i).
\ee
The combined action $S$ consists of the Einstein-Hilbert term, the Chern-Simons term, the
constraint \eqref{eq:constraint} and two boundary terms: the Gibbons-Hawking term and a new term that cancels the last term in \eqref{eq:KdotK}.
The total action then becomes:
\be
\label{eq:tmgadm}
S = \frac{1}{16\pi G_N} \int d^3 x \sqrt{-\g} \Big(- \frac{2}{\m}  k^j_k \del_r \bar k^k_j  + (\Pi^{jk}
- \frac{1}{2\m} A^{jk}) \dot \g_{jk} + N {\cal H} + N^l {\cal P}_l \Big)
\ee
with:
\begin{align}
{\cal H} &= R[\g] - 2 \Lambda + \hf k^2 - 2 k^j_k \bar k^k_j + 2 \Pi^i_j (k^j_i + \bar k^j_i)
+ \Pi_k^k k + \frac{2}{\m} \cdel^k \cdel_i (k^i_k - \bar k^i_k)  + \frac{1}{\m} k k^j_k \bar k^k_j \nonumber \\
{\cal P}_l &=  2 \cdel^i \Pi_{il} + \frac{1}{\m} \Big[ \cdel_l (k^j_k \bar k^k_j) + \e^{ij} \cdel_k (K_i^k K_{jl})
- 2 \e^{ij}K_{kl} \cdel_i K^k_j + \hf \e_l^{\ph{l}k} \del_k R \Big].
\end{align}
The variation of the action is given by:
\be
\label{eq:Stmgfirstvar}
\delta S = \frac{1}{16 \pi G_N} \Big[ \int d^3 x (\text{eom}) + \int d^2 x \sqrt{-\g} \Big(- \frac{2}{\m}  k^j_k \delta \bar k^k_j
+ (\Pi^{jk} - \frac{1}{2\m} A^{jk}) \delta \g_{jk} \Big) \Big]
\ee
so we find a well-defined variational principle if we define $\g_{ij}$ and $\bar k_i^j$ as the boundary data. Notice that we may
also lower the index on $\bar k_i^j$ since using $\bar k_{ij} = \bar k_i^k \g_{kj}$ rather than $\bar k_i^j$ is a simple change
of variables that does not affect the variational principle.

Below we will need to evaluate the first variation of the on-shell action and we therefore need the on-shell expression for $\Pi_{ij}$.
This can be obtained by varying the action \eqref{eq:tmgadm} with respect to $k_{ij}$, $\bar k_{ij}$ and $k$. In the gauge where $N^i=0$
and $N = N(r)$ (which we use below) we obtain:
\be
\label{eq:piij}
\Pi_{ij} =  \frac{1}{\m N} \del_r (\bar k_{ij}- k_{ij}) + k_{ij} + \bar k_{ij} - \g_{ij}(\hf k  + \frac{1}{2\m} k_l^m \bar k_m^l)
+ \frac{1}{2\m} k (\bar k_{ij} - k_{ij}).
\ee
The other equations of motion combine to \eqref{eq:eomtmg} and \eqref{eq:componentsK} with \eqref{eq:Kij}, as expected.

\subsection{The dilatation operator in the null warped background}
\label{sec:dilatationop}
A convenient way to obtain the correct counterterms in the usual AlAdS
backgrounds is the radial Hamiltonian method
\cite{Papadimitriou:2004ap,Papadimitriou:2004rz}. In this method
one expands the conjugate momenta to the bulk fields in terms of
eigenfunctions of a covariant \emph{dilatation operator}, denoted
$\d_{\cal D}$, which is asymptotically equal to the radial derivative.

As an example, consider a $(d+1)$-dimensional asymptotically locally AdS
background of the form:
\be
\label{eq:aladsm}
ds^2 = d\bar{r}^2 + \g_{ij} dx^i dx^j  \qquad \qquad \g_{ij} =
e^{2 \bar{r}} g_{(0)ij} + \ldots
\ee
where $\bar{r}$ is a new radial coordinate relative to
\eqref{eq:alads}: $r = e^{-\bar{r}}$ so the boundary is now at $\bar{r}
\to \infty$. To keep the notation uncluttered in this and the following section
the coordinate $\bar{r}$ will henceforth be denoted as $r$.
In anticipation of what follows we also consider
a massive symmetric transverse traceless
second rank tensor, $\Phi^\mu_\nu$, satisfying
\be \label{symm_tr}
(\nabla_\mu \nabla^\mu + (2-m^2)) \Phi^\mu_\nu=0, \qquad
\Phi^\mu_\mu=0 \qquad \nabla_\mu \Phi^\mu_\nu=0,
\ee
where $m^2=\Delta (\Delta-d)$. This field is dual to an operator of
dimension $\Delta$.  In a spacetime of the form
\eqref{eq:aladsm} the symmetric tensor has a radial expansion which to leading
order has the form:
\be
\label{eq:aladsphi}
\Phi^i_j(r,x^i) = e^{(\D-d)r} \phi_{(0)}{}^i_j%(x^k)
+ \ldots
%\qquad \qquad \D = \frac{d}{2} + \hf \sqrt{d^2 + 4 m^2}
\ee
Consider now an object like the regulated on-shell action $S$ for this
configuration. It is a functional of the boundary data for the fields
at the cutoff surface, so we may write:
\be
S[\g_{ij},\Phi^i_j]
\ee
where it is understood that $\g_{ij}$ and $\Phi^i_j$ are the induced fields on the
cutoff surface. Since the action does not depend explicitly on the
radial coordinate $r$, its radial dependence is inherited completely
from the radial dependence of $\g_{ij}$ and $\Phi^i_j$. For its radial
derivative we may therefore write:
\be
\dot S = \int d^d x \, \dot \g_{ij} \fdel[S]{\g_{ij}} + \int d^d x \,\dot
\Phi^i_j \fdel[S]{\Phi^i_j}
\ee
where a dot denotes a radial derivative.
Asymptotically we may use \eqref{eq:aladsm} and \eqref{eq:aladsphi}
to find that:
\be \label{eq:dil_sym}
\del_r \sim \int d^d x \,2 \g_{ij} \fdel{\g_{ij}}
+ \int d^d x\, (\D -d) \Phi^i_j \fdel{\Phi^i_j} = \delta_{\cal D}
\ee
at least when acting on a functional like the on-shell action. The
tilde symbolizes equality up to terms that vanish as $r \to
\infty$. The functional operator $\delta_{\cal D}$ is precisely the field
theory dilatation operator and this relation between the
radial derivative and dilatation operator reflects the fact that the
RG scale becomes geometric in gauge/gravity duality. This relation
is useful for
the renormalization of the action in the following way. Normally the
divergences in the bare on-shell action are organized in terms of
degree of divergence, \ie in terms of a power series in $e^r$. However
we may also organize these terms in terms of eigenfunctions of the
dilatation operator $\delta_{\cal D}$. Just as in the ordinary method the
equations of motion are then used to determine the exact form of this
expansion. The advantage of expanding in eigenfunctions of $\delta_{\cal D}$
is that $\delta_{\cal D}$ is fully covariant and as a consequence the
divergences will be organized directly in a covariant expansion as
well. This greatly simplifies the analysis of the counterterms which
one can essentially pick to be (minus) the divergent components in
this expansion. We will see an example worked out below. (In the
presence of a conformal anomaly this story is altered somewhat as we
will also demonstrate below.)

Let us now formulate a similar operator for the null warped
background. For concreteness we are working in
TMG; the analysis for the massive vector model is completely
analogous.

For TMG the on-shell action $S$ depends not only on the induced metric
$\g_{ij}$ but also on a component of the extrinsic curvature $\bar
k^i_j$.
%and we again include a massive scalar field $\Phi$ as the only
%other nonzero field in the theory.
For the on-shell action we
therefore write this time:
\be
S[\g_{ij},\bar k_i^j].
\ee
For the
radial derivative on such a functional we find:
\be
\label{eq:delrtmg}
\del_r = \int d^2 x\, \dot \g_{ij} \fdel{\g_{ij}} + \int d^2 x \, \dot{\bar{k}}_i^j \fdel{\bar k_i^j}% + \int d^2 x \dot \Phi \fdel{\Phi}.
\ee
Now, for the AlAdS spacetimes we substituted at this point the general
leading-order radial behavior of the fields (given in
\eqref{eq:aladsm} and \eqref{eq:aladsphi}) to obtain a covariant
dilatation operator which was valid for all AlAdS spacetimes. On the
other hand, for the null warped case we do not have a precise
definition of an `asymptotically null warped spacetime' and we do not
know what the leading-order behavior of the fields for the general
solution should be. Therefore we will from now on narrow down our
analysis to a specific class of solutions. Specifically, the analysis
below is valid for solutions that asymptote to the null warped AdS solution
(\ref{null_warped}). The null warped background metric can be written as:
\be
\label{eq:nrmetricdomainwall}
ds^2 = dr^2 + \g_{ij} dx^i dx^j \qquad \qquad \g_{ij} = \z_{ij} + b_{ij}
\ee
where we introduced
\be
\label{eq:zijbij}
\z_{ij} dx^i dx^j = e^{2 r} 2 du dv \qquad \qquad b_{ij} dx^i dx^j = - e^{4r} b^2 du^2
\ee
Notice that $\z_{ij}$ is simply $e^{2r} \eta_{ij}$ and in particular
is not equal to the induced metric $\g_{ij}$ on slices of constant
$r$. On this background we find that
\be
K_{ij} = \hf \dot \g_{ij} = \g_{ij} + b_{ij}
\ee
from which it is easy to find that $\bar k_{ij} = b_{ij}$
(in conventions where $\sqrt{-\g} \e^{uv} = +1$) and therefore:
\be
\bar k_i^j = \g^{jk} b_{ki} = - e^{2r} b^2 \delta_i^u \delta^j_v.
\ee
Note that $\bar k_i^j$ is a second rank symmetric transverse traceless
tensor that satisfies (\ref{symm_tr}) for $d=2, \Delta=4$ in an
$AdS_3$ background.

Let us now go back to \eqref{eq:delrtmg}. Using the above
asymptotics we obtain
\be \label{eq:delrorignal}
\del_r \sim \int d^2 x \,(2 \g_{ij} + 2 \bar k_{ij}) \fdel{\g_{ij}} + \int d^2 x \,2 \bar k_i^j \fdel{\bar k_i^j}
%+ \int d^2 x\, (\D_s - 2) \Phi \fdel{\Phi} = \delta_{\mathcal D}
\ee
Let us further change variables from $(\g_{ij}, \bar k_i^j)$ to
 $(\z_{ij}, \bar k_i^j)$. This yields
\be
\label{eq:deltaDposspace0}
\del_r \sim \int d^2 x\,\left( 2 \z_{ij}\fdel{\z_{ij}} +
2 \bar k_i^j \fdel{\bar k_i^j}\right) =\delta_{\mathcal D}
\ee
which is precisely the dilatation operator that corresponds to a deformation
of the CFT (on 2d Minkowski spacetime) by a dimension 4 operator,
cf (\ref{eq:dil_sym}).

The most useful representation of the
radial derivative however is obtained by changing variables from
$(\z_{ij}, \bar k_i^j)$ to $(\z_{ij}, b^{ij} = \zeta^{ik} \bar k_k^j)$.
This results in
\be
\label{eq:deltaDposspace}
\del_r \sim \int d^2 x\, 2 \z_{ij}\fdel{\z_{ij}} =\delta_{\mathcal D}.
\ee
Note that $b^{vv} = - b^2$ and $b^{uu}=b^{uv}=0$, so after all
these change of variables
\be
S[\gamma_{ij}, \bar k^i_j] \to S[\zeta_{ij}, b^2]
\ee
and in this functional the radial derivative is represented by
(\ref{eq:deltaDposspace}). We will use this form of the operator
$\delta_{\mathcal D}$ to organize the divergences in the next
section.

\bigskip

It is also interesting to use this discussion to understand how to couple
a metric to the dual deformed theory. Recalling that the deformation is by a $(3,1)$ operator,
a simple toy model that captures this behavior is the scalar field theory used in section \ref{sec:fieldthy}:
\be
S = \int d^2 x \left ( \pa_u \Phi \pa_v \Phi + b \pa_u \Phi (\pa_v \Phi)^3 \right ).
\ee
For this model the coupling to the metric is the following
\be
S[\z, \bar{k}] = \int d^2 x \sqrt{-\z} [\z^{ij} \partial_i \Phi \partial_j \Phi
+ \bar{k}^i_j (\partial_i \Phi \partial_k \Phi \partial_l \Phi \partial_m \Phi \z^{jk} \z^{lm})]
\ee
Under Weyl transformations $(\pa_i \Phi \pa_k \Phi \pa_l \Phi \pa_m \Phi \z^{jk} \z^{lm})$
transforms as a dimension 4 operator. Then
\be
T_{ij} = \frac{2}{\sqrt{-\z}} \frac{\d S}{\d \z^{ij}}.
\ee
Note that this operator is not conserved except when $b=0$.

\section{Scalar field in the Schr\"odinger background}
\label{sec:bulkcomputations}

In section \ref{sec:fieldthy} we discussed the effect of irrelevant
deformations preserving Schr\"{o}dinger symmetry from the field theory perspective.
In this section we holographically compute the two-point
function of a scalar operator to illustrate the consequences of the
finite irrelevant deformation; this computation displays the same structure as found in
the field theory. For computational simplicity, we work in $d=2$, so three bulk
dimensions, but the generalization to higher dimensions would be straightforward.

\subsection{Action and equations of motion}
For a massive scalar field the action is:
\be
S = - \hf \int d^3 x \sqrt{-G} \Big(\del_\m \Phi \del^\m \Phi + m^2 \Phi^2 \Big).
\ee
The equation of motion in the metric \eqref{eq:nrmetricdomainwall} becomes:
\be
\label{eq:eomscalar}
\ddot \Phi + 2 \dot \Phi + \square_\z \Phi  - (m^2 - b^2 \del_v^2) \Phi = 0
\ee
where a dot denotes a radial derivative and $\square_\z = 2 e^{-2r}\del_u \del_v$. This
equation of motion is precisely the equation of motion for a massive scalar in $AdS$:
\be
\ddot \Phi + 2 \dot \Phi + \square_\z \Phi  - M^2 \Phi = 0
\ee
with the effective mass squared $M^2 = m^2 - b^2 \del_v^2$ which depends on the lightcone momentum.\footnote{
In this paper we suppress real-time issues. In \cite{Blau:2010fh} it was shown that the
Scr\"odinger spacetimes do not admit a global time function and it 
was argued that the initial value problem for (\ref{eq:eomscalar}) is not well-defined, unless one removes the $k_v=0$ modes. In the context of AdS/CFT
the proper set up to investigate the initial value problem is the framework of \cite{Skenderis:2008dh}.
This framework was adapted to the
Schr\"odinger case in \cite{Leigh:2009eb}, but the analysis for $k_v=0$ is not given there.
This is an interesting issue that we leave for future work.} If we Fourier transform
$\tilde{\Phi}(k_u,k_v) = \int du dv  \exp(ik_u u + i k_v v) \Phi(u,v)$ then the asymptotic solution to this equation is of the form:
\be
\label{eq:radexpansionphi}
\Phi(r,k_u,k_v) = e^{(\Delta_s - 2)r} \Big(\phi_{(0)}(k) + \ldots + e^{- (2\D_s - 2)r} \phi_{(2\D_s - 2)} (k) + \ldots \Big)
\ee
where here and below we drop the tilde from $\tilde{\F}$ and
$\D_s = 1 + \sqrt{1 + m^2 + b^2 k_v^2}$. For $b^2 = 0$ one would find that $\D_s = \Delta$ with $\Delta = 1 + \sqrt{1 + m^2}$
the scaling dimension of the dual operator. For nonzero $b^2$ we shall see how $\D_s$ will be the scaling dimension under the
dilatation operator $\mathcal D$ that appears in the Schr\"odinger group \eqref{group}; the fact that the scaling dimension depends
on the lightcone momentum $k_v$ was noted previously in \cite{Son:2008ye,Balasubramanian:2008dm,Goldberger:2008vg}.
The generalization to arbitrary dimensions is:
\be
\Delta_s = \frac{d}{2} + \sqrt{\frac{d^2}{4} + m^2 + b^2 k_v^2}.
\ee
Notice that $\D_s$ explicitly depends on the
lightcone momentum $k_v$. The expression \eqref{eq:radexpansionphi} is valid for generic values of $\D_s$; exceptions occur
when $\D_s \in \{1,2,3,\ldots\}$ and we discuss these separately below.

At this point it is important to make a distinction between $b^2$ positive and negative. For $b^2$ negative (which is
allowed in TMG but not in the massive vector model), there is a critical value of the lightcone momentum $k_v^2 = m^2/|b^2|$ above
which the effective mass squared drops below the BF bound and the corresponding operator dimension becomes complex. For $b^2$
positive the effective mass squared is always positive, and the operator dimension grows linearly with large momentum $k_v$.

We now proceed to analyze the massive scalar using Hamiltonian methods of holographic renormalization. At a technical level
the similarity of \eqref{eq:eomscalar} with the equation of motion in pure AdS means that the analysis will proceed along the
same lines as for pure AdS, but with an important conceptual difference: {\it the counterterms depend explicitly on $\D_s$ and therefore
on the lightcone momentum $k_v$ in such a way that they are non-local in the $v$ direction}.

\subsection{The dilatation operator and asymptotic expansions}

The on-shell regularized
action $S$ for $\Phi$ is a functional of $\g_{ij}$, $\bar
k_i^j$ and $\Phi$. In this section we treat the scalar in a
fixed null warped AdS background, so we may replace
$\g_{ij}$ and $\bar k_i^j$ with their background values and
furthermore perform a change of variables to $\z_{ij}$ and
$b^{ij}$. We then write:
\be
S[\z_{ij}, b^{ij}, \Phi]
\ee
where $b^{vv} = -b^2$. Again, all radial dependence of such a functional resides in the radial dependence of its
arguments $\z_{ij}$, $b^{ij}$ and $\Phi$. The radial derivative of $S$ can therefore be written as:
\be
\label{eq:delr}
\begin{split}
\del_r &= \int d^2 k \, \dot \z_{ij}(k) \fdel{\z_{ij}(k)} + \int d^2 k \, \dot b^{ij}(k) \fdel{b^{ij}(k)} + \int d^2 k \, \dot \Phi(k) \fdel{\Phi(k)}\\
&= \int d^2 k \, 2 \z_{ij} (k) \fdel{\z_{ij}(k)} + \int d^2 k \, \Pi (k)\fdel{\Phi(k)}\\
\end{split}
\ee
where we substituted the explicit expressions for the radial derivatives and defined $\Pi = \dot \Phi$.
Notice that we are again working in Fourier space. By Hamilton-Jacobi or by explicit computation we obtain that:
\be
\Pi = \frac{1}{\sqrt{-\z}}\fdel[I]{\Phi}
\ee
where $I$ is the onshell action and we used that $\det(G) = \det(\z)$. From this expression it follows that $\Pi$, just like $I$, can also be
regarded as a functional of $\z_{ij}$, $b^{ij}$ and $\Phi$ and therefore the radial derivative acting on $\Pi$ can also be written in the form \eqref{eq:delr}:
\be
\label{eq:dotpi}
\dot \Pi = \int d^2 k \, 2 \z_{ij} (k) \fdel[\Pi]{\z_{ij}(k)} + \int d^2 k \, \Pi (k)\fdel[\Pi]{\Phi(k)}.
\ee
Having found the asymptotic solution, the next step in the Hamiltonian holographic renormalization is to expand the
divergences in eigenfunctions of the dilatation-like operator $\delta_{\mathcal D}$ defined in \eqref{eq:deltaDposspace}. We write it in Fourier space as:
\be
\label{eq:deltaD}
\delta_{\mathcal D} = \int d^2 k \, 2 \z_{ij} (k) \fdel{\z_{ij}(k)} + \int d^2 k \, (\D_s -2) \Phi (k) \fdel{\Phi(k)}
\ee
We now want to organize $\Pi$ in terms of eigenfunctions of the dilatation operator $\delta_{\mathcal D}$. We therefore write:
\be
\label{eq:dilexpansionpi}
\Pi = \Pi_{(2 - \D_s)} + \Pi_{(4 - \D_s)} + \Pi_{(6 - \D_s)} + \ldots + \Pi_{(\D_s)}  + \ldots
\ee
where by definition
\be
\label{eq:dilweight}
\delta_{\mathcal D} \Pi_{(s)} = -s \Pi_{(s)}\,.
\ee
From \eqref{eq:radexpansionphi} we obtain to leading order in the radial expansion that:
\be
\label{eq:pilo}
\Pi = (\D_s -2) \Phi + \ldots
\ee
and therefore we immediately find that $\Pi_{(2-\D_s)} = (\D_s - 2)\Phi$ which is in fact how we obtained the leading-order
weight $(2 - \D_s)$ in \eqref{eq:dilexpansionpi}. Furthermore, substituting this in \eqref{eq:delr} shows immediately that:
\be
\del_r \sim \delta_{\mathcal D}.
\ee
To find an expression for the subleading terms we rewrite the equation of motion \eqref{eq:eomscalar} as:
\be
\dot \Pi + 2 \Pi - k^2_\z \Phi  - (m^2 - b^2 k_v^2) \Phi = 0
\ee
where $k^2_\z \equiv \z^{ij} k_i k_j = 2 e^{-2r} k_u k_v$. We then replace $\dot \Pi$ with \eqref{eq:dotpi},
substitute \eqref{eq:dilexpansionpi} and use \eqref{eq:dilweight} to collect terms of equal dilatation weight. This results
in the following expressions for the subleading terms:
\be
\label{eq:dilexpansioncoeff}
\begin{split}
\Pi_{(4 - \D_s)} &= \frac{1}{2 \D_s -4} k^2_\z \Phi\\
\Pi_{(6 - \D_s)} &= \frac{-1}{(2\D_s -4)^2 (2\D_s - 6)} k_\z^4 \Phi\\
\Pi_{(2m - \D_s)} &= c(m,\D_s) k_\z^{2m-2} \Phi
\end{split}
\ee
where the coefficients $c(m,\D_s)$ can be determined recursively. The above expansion continues up to the terms with weight $-\D_s$. At this order
we find that the expression for $\Pi_{\D_s}$ is undetermined for generic values of $\D_s$, whereas if $\D_s \in \{1,2,3,\ldots\}$ we can
only satisfy the equation of motion if we include in the expansion an inhomogeneous term of the form:
\be
\Pi = \ldots + r \tilde \Pi_{(\D_s)} + \ldots
\ee
with
\be
\delta_{\mathcal D} \tilde \Pi_{(\D_s)} = - \tilde \Pi_{(\D_s)}
\ee
after which we obtain that:
\be
\delta_{\mathcal D} \Pi_{(\D_s)} = - \D_s \Pi_{(\D_s)} + \tilde \Pi_{(\D_s)}.
\ee
The equation of motion then fixes:
\be
\tilde \Pi_{(\D_s)} = \tilde c(\D_s) \, k_\z^{2\D_s - 2}\Phi
\ee
where the prefactor $\tilde c(\D_s)$ is also determined by the equation of motion. In those cases $\Pi_{\D_s}$ is still undetermined.

\subsection{Renormalization of the action}
The bare on-shell action has the form:
\be
\label{eq:Sbarescalar}
S_{\text{bare}} = \hf \int d^2 k \, \sqrt{-\z} \Phi(-k) \Pi(k) \,
\ee
where $\sqrt{-\z} = e^{2r}$ so in particular it is independent of $k$. Substituting the expansion \eqref{eq:dilexpansionpi}, we find
divergences that for generic $\D_s$ can be removed by adding the counterterm action:
\be
\label{eq:Sctscalar}
S_{\text{ct}} = - \hf \int d^2 k \, \sqrt{-\z} \Phi(-k) \sum_{(2 - \D_s) \leq s < \D_s} \Pi_{(s)} (k)
\ee
Notice that for $b^2=0$ these counterterms would be local functions of the boundary data $\Phi$, which
can be seen directly from the explicit expressions \eqref{eq:dilexpansioncoeff}. This is of course required
by the locality and the renormalizability of the dual theory. Transforming back to position space we would then
find the standard covariant counterterms. To emphasize this fact we use the `covariant' notation $\sqrt{-\z}$ rather
than simply $e^{2r}$, even when we write the counterterms in Fourier space like in \eqref{eq:Sctscalar}. 
For nonzero $b^2$ the explicit coefficients involving $\D_s = 1 + \sqrt{1 + m^2 + b^2 k_v^2}$ make the counterterms nonlocal
in the $v$-direction. On the other hand, the counterterms do remain local in the $u$-direction.

We mentioned before that we need an inhomogeneous term in the expansion \eqref{eq:dilexpansionpi} at the integer values of $\D_s$. To see
how this affects the holographic renormalization we consider the particular example where $\D_s \approx 3$. If $\D_s$ is slightly
less than 3, we need the counterterms:
\be
S_{\text{ct}, \D_s \lesssim 3} = - \hf \int d^2 k \, \sqrt{-\z} \Big( (\D_s - 2) \Phi^2 + \frac{k_\z^2 \Phi^2}{2\D_s - 4}\Big)
\ee
where $\Phi^2 = \Phi(-k) \Phi(k)$. At $\D_s = 3$ we need:
\be
\label{eq:sctDelta3}
S_{\text{ct}, \D_s = 3} = - \hf \int d^2 k \, \sqrt{-\z} \Big( \Phi^2 + \frac{1}{2} k^2_\z \Phi^2 - \frac{1}{4} k_\g^4 \Phi^2 r \Big)
%I_{\text{ct}, \D_s = 3} = - \hf \int d^2 k \, \sqrt{-\z} \Big( (\D_s - 2) \Phi^2 + \frac{k^2_\z \Phi^2}{2\D_s - 4} - \frac{k_\g^4 \Phi^2}{(2\D_s -4)^2} r \Big)
\ee
and when $\D_s$ is slightly bigger than three we find:
\be
S_{\text{ct}, \D_s \gtrsim 3} = - \hf \int d^2 k \, \sqrt{-\z} \Big( (\D_s - 2) \Phi^2 - \frac{k_\z^2 \Phi^2}{2\D_s - 4}
- \frac{k_\z^4\Phi^2}{(2 \D_s - 6)(2\D_s -4)^2}  \Big)
\ee
Of course, we may include both of the above counterterms involving $k^4_\z$ in the first case as well,
as it does not play a role when $\D_s \lesssim 3$. However, the last counterterm in the case $\D_s \gtrsim 3$ becomes
singular when $\D_s = 3$ and conversely the inhomogeneous counterterm (so the last term in \eqref{eq:sctDelta3} is divergent for $\D_s \gtrsim 3$.

% To be very precise, we may Fourier transform:
% \be
% \Phi(u,v,r) = \frac{1}{(2\pi)^2} \int d\omega e^{i \omega u + i k v}  \tilde \Phi(\omega,k,r)
% \ee
% after which the first two counterterms take the form:
% \be
% \begin{split}
% - \hf \int d^2 x e^{2r} (\D_s - 2) \Phi^2 &= \frac{1}{8\pi^2}\int d\omega dk e^{2 r} \Big(1 - \sqrt{1 + m^2 - b k^2}\Big)
% \tilde \Phi(-\omega,-k,r) \tilde \Phi(\omega,k,r)\\
% - \hf \int d^2 x e^{2r} (\D_s - 2) \Phi^2 &= \frac{1}{8\pi^2}\int d\omega dk \frac{\omega k }{1 -  \sqrt{1 + m^2 - b k^2}}
% \tilde \Phi(-\omega,-k,r) \tilde \Phi(\omega,k,r)\\
% \end{split}
% \ee
% Notice the polynomial behavior in $\omega$ but the nonlocal behavior in $k$.

Adding \eqref{eq:Sbarescalar} and \eqref{eq:Sctscalar} we find the renormalized action to be:
\be
S_{\text{ren}} = S_{\text{bare}} + S_{\text{ct}} = \hf \int d^2 k \, \sqrt{-\z} \Phi (-k) \Pi_{(\D_s)}(k)
\ee
and the one-point function is then given by:
\be
\vev{\op(k)} = \lim_{r \to \infty}  \frac{e^{\D_s r}}{\sqrt{-\z}} \fdel[S_{ren}]{\Phi(k)} = \lim_{r \to \infty} e^{\D_s r} \Pi_{(\D_s)}(k).
\ee
If $\D_s(k)$ is non-integral the one point function for the operator ${\cal O}(u,k)$ can be explicitly written as:
\be
\label{eq:renvev}
\vev{\op(k)} = - (2\D_s -2) \phi_{(2\D_s -2)}(k).
\ee
The one point function at integral $\D_s(k)$ receives additional contributions.

To evaluate the two point function, we note that the complete regular solution of the field equations
can be expressed in terms of a Bessel function:
\be
\begin{split}
\Phi(k,\r) &= \phi_{(0)}(k) \frac{2^{2 - \D_s} |k|^{\D_s - 1}}{\G(\D_s -1)} e^{-r} K_{\D_s - 1} (|k| e^{-r})\\
&= \phi_{(0)}(k) e^{(\D_s - 2)r} \Big(1 + \ldots + (|k|/2)^{2\D_s - 2}\frac{\G(1 - \D_s)}{\G(\D_s - 1)} e^{-(2\D_s - 2)r}+ \ldots\Big)
\end{split}
\ee
where $|k|^2 = 2 k_u k_v$ and the expressions are again valid for the non-integral values of $\D_s$.
Substituting this regular solution into \eqref{eq:renvev} we find that:
\be
\< {\cal O} (k) \>  = -2 (\Delta_s -1) %(k_\e/2)^{2\D_s - 2}
(|k|/2)^{2\D_s - 2}\frac{\G(1 - \D_s)}{\G(\D_s - 1)} \phi_{(0)}(k),
\ee
and thus the two point function is given by:
\be
\label{eq:2ptscalar}
\< {\cal O} (k) {\cal O} (- k) \>  = -2 (\Delta_s - 1) (|k|/2)^{2\D_s - 2}\frac{\G(1 - \D_s)}{\G(\D_s - 1)}
\ee
(Note that this expression is not valid at $\D_s \in \{1,2,3,\ldots\}$ where the radial behavior changes and logarithmic counterterms are required.)

\subsection{Interpretation of the result}
The result \eqref{eq:2ptscalar} is of the expected form for a Schr\"{o}dinger invariant theory:
as reviewed in section \ref{sec:fieldthy}, the two point function for an operator of dimension $\D_s$ should be of the form
\be
\< {\cal O}_{\D_s} (u, k_v) {\cal O}_{\D_s} (0,-k_v) \> = c_{\D_s,k_v} \delta_{\Delta,\D_s}
u^{-\D_s}, \label{t2pf}
\ee
where $c_{\D_s,k_v}$ is the normalization obtained from \eqref{eq:2ptscalar}. Recalling that the Fourier transform
of $u^{-\D_s}$ is proportional to $k_u^{\D_s -1}$ we see the expected behavior.
The normalization of the correlator does not however
agree with earlier computations such as \cite{Fuertes:2009ex,Volovich:2009yh,Leigh:2009eb} in which holographic renormalization was not carried
out. (The form of the correlators found in these works otherwise agrees with that found here, see also
\cite{Chen:2009hg,Chen:2009cg} for further computations of correlation functions in warped $AdS_3$.)
Already in an AdS background, i.e. $b=0$,
the two point functions have incorrect normalizations if holographic renormalization is not carried out; this incorrect
normalization was pointed out in \cite{Freedman:1998tz} and the correct normalization derived by holographic
renormalization is given in \cite{Skenderis:2002wp}.

Our computations also agree with the picture given in section \ref{sec:fieldthy}. In particular, the
counterterms are nonlocal but the nonlocality is restricted to the $v$ lightcone direction. Therefore
it is meaningful to compute the correlation function as a function of $u$
and the result \eqref{t2pf} is scheme-independent away from $u = 0$.

Note that an alternative way to derive the counterterms in the case of asymptotically AdS spacetimes is to
impose the appropriate variational problem for spacetimes with a (non-degenerate) conformal boundary
\cite{Papadimitriou:2005ii}. As noted earlier, the Schr\"odinger spacetimes are outside this framework
and these results do not carry over automatically. In a recent paper \cite{Papadimitriou:2010as}
the variational problem was analyzed for spacetimes with general asymptotics and it was found that in general
non-local boundary terms may be required. In the context of holography,
one must understand the physics behind possible non-local boundary terms, i.e. whether or not
such non-local counterterms can be associated with a dual (non-local) quantum field theory.

Let us note in addition that the counterterms are actually of precisely the same form as in AdS, except for the implicit lightcone momentum
dependence in the quantity $\Delta_s$. Schr\"{o}dinger symmetry does not enforce this; more general counterterms with arbitrary $k_v$
dependence would also respect the required symmetry. It seems possible that this specific form of the counterterms, together
with the simple expression for the scaling dimension relative to that in the original CFT, follows from the null dipole picture.

\section{Linearized analysis for TMG}
\label{sec:linearizedtmg}
In this section we consider the linearized TMG bulk equations of motion around the Schr\"{o}dinger background \eqref{eq:nrmetric} and present the most
general solution of the linearized equations. In particular, this solution contains sources for both the dual energy momentum tensor and for the deforming operator
$X_{vv}$. The general linearized solution allows us to compute two-point functions in the
deformed field theory of both the energy-momentum tensor and the deforming operator $X_{vv}$.
To obtain these two-point functions from an on-shell action the usual procedure of holographic renormalization needs to be carried out.
%and in this section we will carry out the holographic renormalization for a subsector of the full set of solutions.

We should emphasize that to carry out holographic renormalization one needs to obtain the {\it most general} asymptotic solution of the field equations. In earlier
work the most general solutions were not considered, but this corresponds to switching off or constraining sources in the dual field theory. Systematic
renormalization requires the complete set of divergences to be isolated, and then these divergences should be
canceled by appropriate counterterms. In the
case at hand, the counterterm action would be expected to be non-covariant, respecting only the Schr\"{o}dinger symmetry, and non-local in the lightcone direction.

The section is split into the following parts. We first consider the linearized equations of motion around the null warped background and
solve them explicitly. Then we substitute these solutions into the on-shell action, isolate the divergences and use holographic renormalization
to render it finite. We may then functionally differentiate this finite on-shell action with respect to the sources to obtain correlation functions.

\subsection{Linearized bulk equations and solutions}
\label{sec:linearizedsolntmg}
In this section we consider the linearized TMG bulk equations of motion around the background \eqref{eq:nrmetric} and present the general solution.

\paragraph{Background and equations of motion\\}
We work in coordinates in which the background metric takes the form:
\be
\label{eq:nullwarpedbg}
G_{\m \n} dx^\m dx^\n = \frac{d\r^2}{4\r^2} + \g_{ij} dx^i dx^j \qquad \qquad
\g_{ij} dx^i dx^j = - \frac{b^2 du^2}{\r^2} + \frac{2 du dv}{\r}
\ee
where $\rho = r^2$ with $r^2$ the coordinate appearing in
equation \eqref{eq:nrmetric}. We use conventions where $N = 1/(2\rho)$ and $\sqrt{-\g} \e^{uv} = + 1$. We then find that on the background:
\be
\label{eq:bgK}
k = 2 \qquad \qquad k_{ij} = 0 \qquad \qquad \bar k_{ij} = b_{ij}
\ee
with $b_{uu} = -b^2/\r^2$ the only nonzero component of $b_{ij}$. From \eqref{eq:piij} we obtain:
\be
\label{eq:bgPi}
\pi_{ij} = - \g_{ij}\,.
\ee
Finally, from \eqref{eq:Aij} it is trivial to see that:
\be
\label{eq:bgAij}
A_{ij} = 0
\ee
on the null warped background.

We now consider a small deformation $\delta G_{\m \n} = H_{\m \n}$ and we will work in a
radial-axial gauge so that $H_{\r \r} = 0$ and $H_{\r i} = 0$. We then define $h_{ij} = \rho \delta \g_{ij} = \rho H_{ij}$.
We introduce the following notation for the first-order variation of the other fields:
\be
\begin{split}
\Pi_{ij} &= - \g_{ij} + \pi_{ij}[h]; \\
k &= 2 + \k[h]; \\
k_{ij} &= \k_{ij}[h]; \\
\bar k_{ij} &= b_{ij} + \bar \k_{ij}[h].
\end{split}
\ee
Since we work at the linearized level, $\pi_{ij}, \k, \bar \k_{ij}$ and $\k_{ij}$ are all linear in $h_{ij}$. We
mentioned below equation \eqref{eq:propertiescomponentsK} that $k_{ij}$ and $\bar k_{ij}$ have a single independent component. To see
what this implies for their fluctuations $\k_{ij}$ and $\bar \k_{ij}$ we linearize the equations \eqref{eq:propertiescomponentsK}
around the background \eqref{eq:nullwarpedbg}. This results in:
\begin{align}
\label{eq:linearizedconstraintsk}
\k_{uu} &= \frac{b^4}{4\r^2} \k_{vv}, & \k_{uv} &= -\frac{b^2}{2\r} \k_{vv}, \\
\label{eq:linearizedconstraintsbark}
\bar \k_{uv} &= - \frac{b^2}{2\r^2} h_{vv}, & \bar \k_{vv} &= 0,
\end{align}
so the independent components of $\k_{ij}$ and $\bar \k_{ij}$ are $\k_{vv}$ and $\bar \k_{uu}$.

It is now convenient to express the fluctuations in momentum space as:
\be
h_{ij}(x) =  \int\frac{d^2 k}{(2\pi)^2} e^{i k_u u + i k_v v} h_{ij}(k)
\ee
Using the equations of motion \eqref{eq:eomtmg} we then find at the linearized level:
\begin{eqnarray}
\label{eq:rr}
0 &=& 6 k_v k_u \r^2 h_{uv}-3 k_v^2 \r^2 h_{uu}- 6 b^2 h_{vv}+(-b^2 k_v -3 k_u \r) (-b^2 k_v + k_u \r) h_{vv} \\ &&
\nonumber +2 \r \left(\left(6 -b^2 k_v^2\right) \r h_{uv}'-k_v^2 \r^2 h_{uu}'+\left(k_u^2 \r^2 + b^2 (2+k_v k_u \r )\right) h_{vv}' - 2 b^2 \r h_{vv}''\right)
\\
\label{eq:ru}
0 &=& 2 k_v k_u \r^2 ( b^2 k_v+k_u \r) h_{uv}+ k_v^2 \r^2 (-b^2 k_v-k_u \r) h_{uu} \\ && \nonumber +(-b^2 k_v-k_u \r)
\left(- 6 b^2+k_u^2 \r^2\right) h_{vv} +4 \r^2 \left((b^2 k_v- 3 k_u \r) h_{uv}' -b^2 k_u h_{vv}' \right. \\ && \nonumber
\left. +2 k_v \r \left(2 h_{uu}' + b^2 h_{uv}''+\r h_{uu}''\right) + b^2 (b^2 k_v + k_u \r ) h_{vv}''\right)
\\
\label{eq:rv}
0 &=& -2 k_u k_v^2 \r^2 h_{uv}+ k_v^3 \r^2 h_{uu} + k_v \left(6 b^2+k_u^2 \r^2 \right) h_{vv}\\&& \nonumber+4 \r
\left(-3 k_v \r h_{uv}'+(-b^2 k_v+2 k_u \r ) h_{vv}' - \r (b^2 k_v + 2 k_u \r) h_{vv}''\right)
\\
% uu
\label{eq:uu}
0 &=& 2 b^2 k_v k_u \r^2 h_{uv} - b^2 k_v^2 \r^2 h_{uu} - b^2 \left(-b^2 \left(6 -b^2 k_v^2\right)
+ 2 b^2 k_v k_u \r +2 k_u^2 \r^2 \right) h_{vv}\\ &&\nonumber-2 \r \left(\r \left(-b^2 \left(2 + b^2 k_v^2\right)+k_u^2 \r^2\right)
h_{uv}'+ k_v \r^2 (-b^2 k_v-k_u \r ) h_{uu}' \right. \\ &&\nonumber \left. -b^2 \left(-k_u^2 \r^2  -b^2(-2+k_v k_u \r)\right) h_{vv}'
+4 \r^2 \left(2 b^2 h_{uv}''+\r \left(3 h_{uu}'' + b^2 h_{uv}^{(3)}+\r h_{uu}^{(3)}\right)\right)\right)
\\
% uv
\label{eq:uv}
0 &=& -b^2 \left(-12- b^2 k_v^2-k_v k_u \r \right) h_{vv}-  2 b^2 k_v^2 \r^2 h_{uv}' - 10 b^2 \r h_{vv}' \\ &&\nonumber+ \r^2
\left(-k_v^2 \r h_{uu}'+k_u (2 b^2 k_v+k_u \r ) h_{vv}' +4 \left(3 \r h_{uv}'' + b^2 h_{vv}'' -b^2 \r h_{vv}^{(3)}\right)\right)
\\
% vv
\label{eq:vv}
0 &=& - \frac{b^2 k_v^2 h_{vv}}{\r^2}+2 k_v^2 h_{uv}'-2 k_v k_u h_{vv}'+8 \r h_{vv}^{(3)}
\end{eqnarray}
From the trace of the equation of motion (which is $R = -6$) we find:
\be
-2 k_v k_u \r^2 h_{uv} +k_v^2 \r^2 h_{uu}  + 10 b^2 h_{vv} + k_u^2 \r^2 h_{vv} -4 \r^2 h_{uv}'  - 8 b^2 \r h_{vv}'+ 8 \r^3 h_{uv}'' + 4 b^2 \r^2 h_{vv}'' = 0.
\ee
Let us consider the case where $k_v$ is nonzero. We can then use the trace equation to write $h_{uu}$ in terms of
$h_{vv}$ and $h_{uv}$ and their derivatives. We then substitute this into \eqref{eq:rv} to find an expression for
$h_{uv}''$ in terms of $h_{uv}$, $h_{uv}'$ and $h_{vv}$ and its derivatives. Similarly, we can use \eqref{eq:vv} to solve for
$h_{uv}'$ in terms of $h_{vv}$ and its derivatives. Substituting now all these expressions into the first radial derivative
of \eqref{eq:rr} results in the ordinary differential equation:
\be \label{eq:TMG_eq}
(-b^2 k_v^2 -2 k_u k_v \r) h_{vv}'' + 8 \r h_{vv}^{(3)}+ 4\r^2 h_{vv}^{(4)} = 0.
\ee
Having solved this equation, one can obtain the other components from equation \eqref{eq:vv} and the trace equation.
Note that
\be \label{eq:TMG_T}
h_{vv}''=0
\ee
is a trivial solution of (\ref{eq:TMG_eq}). Actually (\ref{eq:TMG_T}) is one of the equations obtained
by linearizing three
dimensional Einstein gravity. We will call the solution obtained by solving (\ref{eq:TMG_T}) the
`T' solution and the solution obtained from the regular non-trivial solution of (\ref{eq:TMG_eq}) the
`X' solution. We will distinguish the two sets of solutions with superscripts $T$ and $X$.

\paragraph{The `T' solution\\}
The first solution takes the form:
\be
\begin{split}
\label{eq:hmetriccoeff}
h^T_{uu} &= \frac{1}{\r} h_{(-2)uu} + \tilde h_{(0)uu} \log(\r) + h_{(0)uu} + \r h_{(2)uu}\\
h^T_{uv} &= \frac{1}{\r} h_{(-2)uv}  + \tilde h_{(0)uv} \log(\r)  +  h_{(0)uv}+ \r h_{(2)uv}\\
h^T_{vv} &= h_{(0)vv} + \r h_{(2)vv}
\end{split}
\ee
with:
\begin{align}
k_v^2 h_{(-2)uu}  &= -b^2\Big(k_u k_v h_{(0)vv}  - 4 h_{(2)vv} \Big) & h_{(-2)uv} &= - \hf b^2 h_{(0)vv}\\
k_v \tilde h_{(0) uu} &= b^2  k_u h_{(2)vv} & \tilde h_{(0)uv} &= \frac{b^2}{2} h_{(2)vv}\\
h_{(2)uv} &= \qt \tilde R_{(0)} & k_v h_{(2)uu} &= \qt k_u \tilde R_{(0)} \\
k_u h_{(2)vv} &= \qt k_v \tilde R_{(0)} &  &
\end{align}
and with
\be
\tilde R_{(0)} = k_u^2 h_{(0)vv} - 2 k_v k_u h_{(0)uv} + k_v^2 h_{(0)uu} \label{curv}
\ee
the linearized scalar curvature associated to a metric perturbation $\eta_{ij} + h_{(0)ij}$.

These modes are expected to correspond to
switching on a source for the energy-momentum tensor in the boundary theory. As already noted in the introduction,
these solutions blow up faster at the boundary than the background metric. In earlier work, the boundary conditions used
only allowed a subset of these solutions. For example, in the original paper of Son \cite{Son:2008ye}, an asymptotically
Schr\"{o}dinger metric of the type
\be
ds^2 = \frac{d \rho^2}{4 \rho^2} + \frac{1}{\rho} \left ( - \frac{b^2 e^{-2 \Phi}}{\rho} du^2 + 2 e^{- \Phi} du (dv - a_0 dv) \right )
\ee
was proposed. (For convenience of comparison the notation here follows that of \cite{Son:2008ye}.)
Here the functions $\Phi(\rho,u,v)$ and $a_0 (\rho,u,v)$ were proposed to correspond to the energy and mass currents of the dual field theory, respectively,
and were assumed to have a finite limit as $\rho \rightarrow 0$. Clearly the linearization of this ansatz does not match the solutions
above: the $h_{vv}$ fluctuations have been switched off entirely.
Constraining the asymptotics by not allowing for these fluctuations necessarily constrains or sets to zero certain of the operator sources in the
field theory; related constraints were noticed in earlier discussions of renormalization for Schr\"{o}dinger in \cite{Herzog:2008wg}.
Before analyzing the complete set of solutions of the linearized equations
and setting up the holographic dictionary one cannot determine whether switching off $h_{vv}$ corresponds to switching off or constraining sources.
While it would be consistent to set certain operator sources to zero, it is not consistent to constrain sources.
It is also not consistent to set the vector fluctuation to zero, as proposed in \cite{Son:2008ye}; in the analysis of the massive vector model in the next
section we will see that the linearized field equations are coupled, and solutions involve both metric and vector fluctuations.

Let us also note that if the $h_{vv}$ fluctuations are switched off, then the remaining solution is asymptotically
Schr\"{o}dinger, in the sense that the fluctuations fall off at least as fast as the background as $\rho \rightarrow 0$.
This connects with the discussion around (\ref{se-dimensions}): the operator $T_{uu}$ is irrelevant, whilst the other
components of $T_{ij}$ are marginal or relevant, and (at $b=0$) $h_{(0)vv}$ acts as a source for $T_{uu}$.
Switching on $h_{vv}$ therefore seems related to switching on the linearized source for $T_{uu}$, which would be expected
to change the asymptotic structure of the background spacetime.

The precise holographic interpretation of these metric modes is however rather subtle and will be discussed elsewhere. For now we will set all of these modes to zero,
and focus on the second independent type of fluctuations:

\paragraph{The `X' solution\\}
The second solution takes the form:
\be
h^X_{vv} = h_{(4 - 2s)vv} \r^{2-s} {}_1 F_2(1 - s; 2 - 2s, 3 - s; \hf k_v k_u \r) + h_{(2s + 2)vv} \r^{s+1} {}_1 F_2(s; 2s, 2 + s; \hf k_v k_u \r)
\ee
with
\be
s = \hf + \hf \sqrt{1 + b^2 k_v^2}
\ee
and with ${}_1 F_2$ the hypergeometric function. When we Taylor expand the hypergeometric functions around $\r \to 0$ we find
subleading terms whose form depends on whether $s \in \{1,2,3,\ldots\}$ or not. Namely in the first case we find, just as for
the scalar field, logarithmic terms in the expansion. We will not treat these cases here but note that one may deal with them
in the same way as we did in section \ref{sec:bulkcomputations}. For non-integral $s$ the subleading terms become an ordinary power series:
\bea
%\begin{split}
h_{vv} &=& h_{(4-2s)vv} \r^{2-s} (1 + \a_1(s) k_u k_v\r + \a_2(s) (k_u k_v \r)^2 + \ldots + \a_n(s) (k_u k_v \r)^n + \ldots)\label{eq:subleadinghvv}  \\
&&  \qquad + h_{(2s + 2)vv} \r^{s+1} (1 + \b_1(s) k_u k_v \r + \b_2(s) (k_u k_v \r)^2 + \ldots + \b_n(s) (k_u k_v \r)^n + \ldots) \nn
%\end{split}
\eea
where $\a_n(s)$ and $\b_n(s)$ are rational functions of $s$. Their explicit form follows directly from the
expansion of the hypergeometric function but it will not be needed here.

To compute a two point function we also need to identify which solution is regular throughout the bulk spacetime.
To check regularity as $\r \to \infty$ we use the asymptotic behavior:
\be
\begin{split}
&\lim_{\r \to \infty} \r^{2-s} {}_1 F_2(1 - s; 2 - 2s, 3 - s; \hf k_v k_u \r) \\
&\qquad = \lim_{\r \to \infty}  \r^{1/4} e^{2 \sqrt{k_v k_u \r/2}} \frac{1}{4\sqrt{\pi}} (2 - s)  \Gamma(3 - 2s) \Big(\frac{k_v k_u}{2}\Big)^{s - 7/4} + \ldots
\end{split}
\ee
and we find that the divergent pieces in $h^X_{vv}$ cancel only if we relate the two independent solutions as:
\be
\label{eq:vevhvv}
h_{(2s + 2)vv} = \frac{(s-2) \Gamma(3-2s)}{(s+1) \Gamma(1+2s)} \Big(\frac{k_v k_u}{2}\Big)^{2s-1} h_{(4 - 2s)vv}\,.
\ee

The solution for the other components $h_{uv}$ and $h_{uu}$ can be obtained by completely
solving the linearized equations of motions. Their asymptotic behavior as $\r \to 0$ takes the form:
\be
\label{eq:expansionhuuhuv}
\begin{split}
% h^X_{uu} &= \r^{-s} h_{(-2s)uu} + \r^{-s + 1} \frac{k q}{4-4 s} h_{(-2s)uu}  + \ldots + \r^{1 - s} h_{(2k)uu} + \ldots\\
% h^X_{uv} &= \r^{-s + 1}  \frac{(2s -3) k_v^2}{(2s -2) (8 + (-b^2) k_v^2)} h_{(-2s)uu} + \ldots + \r^{s} h_{(2k)uv}\\
% h^X_{vv} &=  \r^{-s + 2} \frac{k_v^2}{8 (-b^2) + (-b^2)^2 k_v^2} h_{(-2s)uu} + \ldots + \r^{s + 1} h_{(2k) vv} + \ldots
h^X_{uu} &= \r^{-s} h_{(-2s)uu} (1+ \ldots) + \r^{s - 1} h_{(2s - 2)uu} (1+ \ldots)\\
h^X_{uv} &= \r^{-s + 1} h_{(2 - 2s) uv} (1+ \ldots) + \r^{s} h_{(2s)uv} (1 + \ldots)\\
% h^X_{vv} &= \r^{-s + 2} h_{(4 - 2s) vv} + \ldots + \r^{s + 1} h_{(2s + 2) vv} + \ldots
\end{split}
\ee
with the leading coefficients given by:
\be
\label{eq:sourcehuuhuv}
\begin{split}
h_{(-2s)uu} &= \frac{16}{k_v^4} s(s-2) (s^2 -1) h_{(4 - 2s)vv}\\ % \Big(b^4 - 8 \frac{b^2}{k_v^2} \Big)
h_{(-2s)uv} &= -b^2 \frac{(2s -3)}{(2s -2)} h_{(4 - 2s)vv}
\end{split}
\ee
and analogous formulas hold for the expression of $h_{(2s-2)uu}$ and $h_{(2s)uv}$ in terms of $h_{(2s+2)vv}$. The
ellipses in \eqref{eq:expansionhuuhuv} represent subleading terms which have a similar form to those in $h_{vv}$: in each case the
ellipses represent an infinite power series of the form
\be
\label{eq:powerseriessubleading}
\sum_{n=1}^{\infty}\a_n(s) (k_u k_v \r)^n\,.
\ee
The coefficients $\a_n(s)$ are again rational functions of $s$ whose precise form is different for each of the occurrences of
such a series in \eqref{eq:expansionhuuhuv}, but whose values are easily calculable starting from the expansion of $h_{vv}$.
% When $k_v = 0$ we find that equation \eqref{eq:rr} (-b^2)ecomes simply:
% \be
% V''' = 0
% \ee
% We find (-b^2)ack the solution la(-b^2)elled $M$ a(-b^2)ove (with $k_v = 0$) as well as a new solution which has the form:
% \be
% \begin{split}
% h_{uu} &= - 2 (-b^2)^2 h_{(4)vv} \log(\r) + \frac{5}{72} (-b^2) k_u^2 h_{(4)vv} \r^2 + \frac{1}{2880} (k_u)^4 h_{(4)vv} \r^4 \\
% h_{uv} &= \frac{(-b^2)}{2} h_{(4)vv} \r - \frac{1}{36}k_u^2  h_{(4)vv}  \r^3 \\
% h_{vv} &= h_{(4)vv}\r^2
% \end{split}
% \ee

Using the linearized version of \eqref{eq:Kij} (with $N^i = 0, N = 1/(2\r)$) and \eqref{eq:componentsK} we obtain that for this solution:
\be
\begin{split}
\k_{vv} &= 2 (s-2) h_{(4-2s)vv} \r^{-s+1} \Big( 1 + \ldots\Big) - (1+s) h_{(2s+2)vv} \r^{s} \Big( 1 + \ldots\Big)
\end{split}
\ee
as well as:
\be
\label{eq:barkappauu}
\begin{split}
\bar \k_{uu} &= \frac{8}{k_v^4} s^2 (s-1) \left(s^2 + 5s -14\right) h_{(4-2s)vv} \r^{-s-1} \Big( 1 + \ldots\Big) \\
& \qquad - \frac{4}{k_v^4} (s-1)^2 s \left(s^2 - 7s - 8\right) h_{(2s+2)vv} \r^{s-2} \Big( 1 + \ldots\Big)\\
\end{split}
\ee
and we recall that the other components of $\k_{ij}$ and $\bar \k_{ij}$ are given in \eqref{eq:linearizedconstraintsk}
and \eqref{eq:linearizedconstraintsbark}. For $\pi_{ij}$ we linearize \eqref{eq:piij} and obtain:
\be
\begin{split}
\pi_{uu} &= - \frac{32}{3 k_v^4}s (6 - 11 s + 14 s^2 - 13 s^3 + 4 s^4) h_{(4-2s)vv} \r^{-s-1} \Big( 1 + \ldots\Big) \\
& \qquad + \frac{16}{3 k_v^4} (s-1)s (4s^3 - 3s^2 -s +6) h_{(2s+2)vv} \r^{s-2} \Big( 1 + \ldots\Big) \\
\pi_{uv} &= -\frac{4}{3 k_v^2} s (1 + 10 s - 10 s^2 + 2 s^3) h_{(4-2s)vv}\r^{-s} \Big( 1 + \ldots\Big)
\\ & \qquad - \frac{2}{3 k_v^2} (s-1)(-3 - 4 s + 4 s^2 + 2 s^3)h_{(2s+2)vv}  \r^{s-1} \Big( 1 + \ldots\Big) \\
\pi_{vv} &= \frac{2}{3}(2s^2 - 4s -3) h_{(4-2s)vv}\r^{-s+1} \Big( 1 + \ldots\Big)
\\ & \qquad + \frac{1}{3}(2s^2 - 5) h_{(2s+2)vv}  \r^{s} \Big( 1 + \ldots\Big)
\end{split}
\ee
Again the ellipses in the above five equations represent subleading terms of the form \eqref{eq:powerseriessubleading}
with $\a_n(s)$ calculable rational functions of $s$ whose precise form depends on the quantity under consideration.

\subsection{On-shell action}
Let us begin with the on-shell action evaluated on the null warped background. We substitute the
background values \eqref{eq:bgK} and \eqref{eq:bgPi} in \eqref{eq:tmgadm} to find that the on-shell action when evaluated on the background takes the form:
\be
S_{[0],\text{bare}} = \frac{1}{16 \pi G_N} \int d^2 x \int_{\rho_0} d\rho \frac{2}{\rho^2}
\ee
where the subscript $[0]$ denotes the fact that this is the on-shell action to zeroth order in the perturbation and $\rho_0$ is a cutoff to
regulate the action. As $\rho_0 \to 0$ we find a divergence which is canceled by the counterterm:
\be
\label{eq:S0ct}
S_{[0],\text{ct}} = - \frac{1}{8 \pi G_N} \int d^2 x \sqrt{-\g}.
\ee
With this counterterm included, the combined action $S_{[0],\text{bare}} + S_{[0],\text{ct}}$ is finite and actually vanishes as $\rho_0 \to 0$.

We next consider the on-shell action evaluated to first order in the perturbations. We need to add the first-order
variation of the on-shell action, given in \eqref{eq:Stmgfirstvar}, to the first-order variation of \eqref{eq:S0ct}. This results in the following expression:
\be
\label{eq:s1bare}
S_{[1],\text{bare}} = \frac{1}{16 \pi G_N}  \int d^2 x \sqrt{-\g} \Big(\frac{2}{3}  k^j_k \delta \bar k^k_j
+ (- \Pi^{jk} - \g^{jk} + \frac{1}{6} A^{jk}) \delta \g_{jk} \Big)
\ee
where we added an extra sign with respect to \eqref{eq:Stmgfirstvar} because the radial boundary is at the lower
end of the $\rho$ integration and we have set $\m = 3$. Substituting the background values given in \eqref{eq:bgK},
\eqref{eq:bgPi} and \eqref{eq:bgAij} we find that all conjugate momenta vanish and
and therefore $S_{[1]}$ vanishes as well.

With $S_{[0]}$ and $S_{[1]}$ vanishing, we conclude that the lowest-order term in the on-shell action is second order in the variations. From
the variation of \eqref{eq:s1bare} we find that:
\be
\label{eq:S2bare}
S_{[2],\text{bare}} = \frac{1}{32\pi G_N} \int d^2 x \frac{1}{\rho} \Big(\frac{2}{3} \k^j_k \bar \k^k_j - (\pi^{jk} + h^{jk})h_{jk}).
\ee
Notice in particular that the second variation of the term $\int \sqrt{-\g} A^{ij} \delta \g_{ij}$ vanishes for our background, as
can be directly seen from its definition \eqref{eq:Aij}.

Let us for now set the `T' modes to zero. We then substitute the expansions for the `X' modes
given in subsection \ref{sec:linearizedsolntmg} in \eqref{eq:S2bare} to find its radial expansion:
\begin{align}
S_{[2],\text{bare}} &= \frac{1}{32 \pi G_N} \int d^2 k \Big[ \r^{-2s + 1} \frac{32}{3 k_v^4} (-2+s) (-1+s) s^2 \left(-12+7 s+s^2\right)
h_{(4-2s)vv}^2 \Big( 1 + \ldots\Big)  \nn \\& - \frac{32}{3 k_v^4} (-2 - s + s^2) (-3 s + 4 s^2 - 2 s^3 + s^4) h_{(4-2s)vv}(-k) h_{(2+2s)vv}(k)  \nn \\& + O(\r) \Big]
\label{eq:s2bareexp}
\end{align}
where $h_{(4-2s)vv}^2 = h_{(4-2s)vv}(-k)h_{(4-2s)vv}(k)$. The dots again represent subleading terms which take the usual
form \eqref{eq:powerseriessubleading} with certain rational coefficients $\a_n(s)$. These terms are divergent and need to be canceled with a suitable counterterm action. The $O(\r)$ symbol represents the terms that vanish as the cutoff $\r \to 0$.

To find the counterterms one may follow the usual procedure of holographic renormalization. However we discussed above that the current framework is not well adapted to discuss the `T' modes and we would therefore like to set these to zero. Unfortunately this introduces an ambiguity in the definition of the counterterm action because then all components of $h_{ij}$ and $\bar \k_{ij}$ essentially have the same component at leading order in their radial expansion. One may therefore replace for example $h_{uu}$ with $\bar \k_{uu}$ in a counterterm (and adjust the overall coefficient) without affecting the fact that it appropriately cancels a certain divergence. When we include the `T' modes however this ambiguity is lifted, since a counterterm which seemed suitable once the `T' modes are set to zero may actually induce extra divergences once they are nonzero. For this reason one cannot consistently perform the holographic renormalization without switching on the `T' modes as well and a more careful analysis of the counterterms will therefore be given elsewhere.

Nevertheless the structure of the divergences still allows us to obtain certain nontrivial results without performing the complete holographic renormalization. From a short analysis of the divergences and the possible counterterms one obtains that the renormalized on-shell action necessarily takes the same functional form as the term of order one in \eqref{eq:s2bareexp}, albeit with a different normalization which can be an arbitrary function of $s$. We therefore write:
\be
\label{eq:S2ren}
S_{[2],\text{ren}} = S_{[2],\text{bare}} + S_{[2],\text{ct}} = \frac{1}{32 \pi G_N}\int d^2 k \frac{c(s)}{k_v^4}h_{(4-2s)vv}(-k) h_{(2+2s)vv}(k)
\ee
with an arbitrary function $c(s)$. The fact that $c(s)$ is undetermined as long as the holographic renormalization is not appropriately performed is similar to the well-known fact that for AlAdS spacetimes the correct normalization of the two-point function is obtained only once one properly holographically renormalizes the action. We repeat that we have set the `T' modes to zero by hand which is why they do not appear in \eqref{eq:S2ren}.

Just as for the empty AdS background we will take the leading component of $\bar \k_{uu}$ to be the source of a dual operator $X_{vv}$. So if we rewrite \eqref{eq:barkappauu} as:
\be
\bar \k_{uu} = \r^{-s-1} \left ( \bar \k_{(0)uu} \Big( 1 + \ldots\Big) + \bar \k_{(4s -2)uu} \r^{2s-1} \Big( 1 + \ldots\Big) \right )
\ee
with
\be
\begin{split}
\bar \k_{(0)uu} &= \frac{8}{k_v^4} s^2 (s-1) \left(s^2 + 5s -14\right) h_{(4-2s)vv}\\
\bar \k_{(4s-2)uu} &= - \frac{4}{k_v^4} (s-1)^2 s \left(s^2 - 7s - 8\right) h_{(2s+2)vv}
\end{split}
\ee
then we should take $\bar \k_{(0)uu}$ to be the source for $X_{vv}$.

Using the above equation as well as \eqref{eq:vevhvv} we can rewrite the renormalized action \eqref{eq:S2ren} in terms of $\bar \k_{(0)uu}$:
\be
S_{[2],\text{ren}} = \frac{1}{32 \pi G_N}\int d^2 k \, \tilde c(s)\, \bar \k_{(0)uu} (-k)  k_v^4 (k_v k_u)^{2s-1} \bar \k_{(0)uu}(k)
\ee
with undetermined normalization $\tilde{c}(s)$. The two-point function of $X_{vv}$ then takes the form:
\be
\vev{X_{vv}(k) X_{vv}(-k)} = - i \frac{\pi}{4 G_N} \tilde c(s) k_v^4 (k_v k_u)^{2s-1}.
\ee
%We can again do a partial Fourier transform in the $u$ direction to obtain:
%\be
%\vev{X_{vv}(u,k_v) X_{vv}(u,-k_v)} = u^{-2s}
%\ee
This is  the behavior expected for the two-point function of an operator of
weight $\D_s = 2s$ under the Schr\"{o}dinger dilatation symmetry $\mathcal D$.
As $b^2 \to 0$, the correlation function also reduces to that of a $(3,1)$ operator in a CFT, as expected.

\section{Linearized analysis for the massive vector model}
\label{sec:linearizedvector}

In this section we present the general linearized solution to the vector model equations of motion in three dimensions. We should note
that in five dimensions gravity coupled to a massive vector is not by itself a consistent truncation of ten-dimensional supergravity;
additional scalar fields need to be included \cite{Maldacena:2008wh}. The linearized equations of this consistent truncation do not coincide with
those of the gravity plus vector model, but the generic features of the linearized solutions of this system are expected to be similar to those in the model
analyzed here.

We choose a radial gauge such that:
\be
ds^2 = \frac{d\rho^2}{4\rho^2} + \left(\g_{ij} + \frac{h_{ij}}{\rho}\right) \, dx^i dx^j \;, \;\;\;\;\; A_\mu = A_{\mu}^{bg} + \A_{\mu},
\ee
where $\g_{ij}$ is the background metric \eqref{eq:nullwarpedbg}, $A^{bg} = \frac{b}{\rho} du$ is the background gauge field, and
$h_{ij}$ and $\A_{\mu}$ parameterize small metric and gauge field perturbations.
The linearized Einstein equations can then be written as:
\begin{eqnarray}
&& R_{ij}[h] + \tr(\gamma^{-1} h') \gamma_{\, ij} - \rho \left( 2\, h_{ij}'' - 4\, b^{2} \rho^{-2} \delta_{(i}^{u}h_{j)v}'\, +
b^{2} \rho^{-2} \tr( \gamma^{-1} h' ) \delta_{i}^{u} \delta_{j}^{u}\, \right) \nonumber \\
& &
\qquad = 3 b^{2} \rho^{-2} h_{vv} \gamma_{ij} + b^4  \rho^{-3} h_{vv} \delta_{i}^{u}
\delta_{j}^{u}\, + 4 \, b\, \rho^{-1} {\cal A}_{(i} \delta_{j)}^{u}\, + 4 b\, \left( \delta_{(i}^{u} {\cal F}_{j)\rho}
- {\cal F}_{ v \rho}\, \gamma_{ij} \right); \nonumber \\
& & \partial_{i} \left( \tr(\gamma^{-1} h') \right) - \gamma^{jk }\partial_{k} h_{ij}' -
\frac{1}{2} \, b^{2} \rho^{-2} \partial_{i} h_{uvv}\, + b^{2} \rho^{-2} \delta_{i}^{u}\, \partial_{k} \left( h_{v}^{k}
- \frac{1}{2} \tr(h) \delta_{v}^{k} \right) \\
& & \qquad = b \rho^{-1} {\cal F}_{iv} - 4 b\, \rho^{-1} {\cal A}_{\rho}\, \delta_{i}^{u}; \nonumber \\
& & \frac{1}{2} b^{2}\, \partial_{\rho} \left( \rho^{- 2}\, h_{vv} \right) = \frac{1}{2} \tr( \gamma^{-1} h'' ). \nonumber
\end{eqnarray}
Note also that $R_{ij}[h] - \frac{1}{2} \gamma_{ij} R[h] = 0$, and
${\cal F}_{ij} = \partial_{i} {\cal A}_{j} - \partial_{j} {\cal A}_{i}$.
The linearized vector field equations are
\begin{eqnarray}
\partial_{i} \left( \gamma^{ij} {\cal F}_{j \rho} \right) -  b\, \rho^{-2}  \partial_{i} \left( h_{v}^{i} - \frac{1}{2}, \tr( h )
\delta_{v}^{i} \right) &=& {4 \over \rho}\, {\cal A}_{\rho}; \\
\partial_{i} \left( \gamma^{ij} \gamma^{kl} {\cal F}_{jl} \right) + 4 \partial_{\rho} \left( \rho \gamma^{ik}
{\cal F}_{\rho i} \right) + 4 b\, \rho^{- 1} \partial_{\rho} \left( h_{v}^{k} - \frac{1}{2} \tr( h ) \delta_{v}^{k} \right) &=& {4 \over \rho}\,
\gamma^{ki} {\cal A}_{i}, \nonumber
\end{eqnarray}
whilst the linearized divergence equation is:
\be
\partial_{i} \left( \gamma^{ik} {\cal A}_{k} \right) + 4 \rho\, {\cal A}_{\rho}' - b\, \rho^{- 1} \partial_{i} \left( h_{v}^{i} - \frac{1}{2} \tr( h )
\delta_{v}^{i} \right) = 0.
\ee
In these equations $\tr(h) \equiv \tr (\gamma^{-1} h) = b^2 h_{vv} \rho^{-1} + 2 h_{uv}$.

It is useful in solving these equations
to carry out a Fourier transform with respect to the boundary coordinates $x^i$, by defining
\be
h_{ij}(x,\rho) = \int \frac{d^2k}{(2\pi)^2} \, e^{i k_u u + i k_v v}
\td{h}_{ij} (k,\rho) \;, \;\;\;\;\;\A_{\mu}(x,\rho) = \int \frac{d^2k}{(2\pi)^2} \, e^{i k_u u + i k_v v} \td{\A}_{\mu}(k,\rho).
\ee
In order to keep notation uncluttered, we will drop the tilde in what follows. In solving the
linearized equations of motion around the null warped background\footnote{We have used Mathematica to do this computation.} we find that the
solutions again split into two independent sets, the `X' modes and `T' modes, as follows:
\be
h_{ij} (k,\rho) = h_{ij}^T (k,\rho)+h_{ij}^X (k,\rho) \;, \;\;\;\;\;\A_\mu (k,\rho) = \A_\mu^T (k,\rho)+\A_\mu^X (k,\rho).
\ee
As in the previous section, the `T' modes are associated with the dual stress energy tensor whilst the `X' modes are associated with the deforming vector operator.

\subsubsection*{The `T' modes}

This part of the solution can be written as:
\bea
h_{uu}^T(\rho,k) & = & \frac{1}{\rho} \, h_{(-2)uu} - \tilde{h}_{(0)uu} \ln \rho  + h_{(0)uu}+ \rho \, h_{(2)uu} \non \\
h_{uv}^T(\rho,k) & = & \frac{1}{\rho} \, h_{(-2)uv} - \tilde{h}_{(0)uv} \ln \rho  + h_{(0)uv} + \rho \, h_{(2)uv} \non \\
h_{vv}^T(\rho,k) & = & h_{(0)vv} + \rho \, h_{(2)vv} \\
\A_\mu (\rho,k ) &= & \frac{1}{\rho} \A_{(0)\mu}+ \A_{(2)\mu} \non
\eea
The various fields carrying $(n)$ subscripts are purely functions of $k^i$. The equations of motion relate the various functions as follows:
\be
\A_{(0)u} = - \frac{1}{2b} \, h_{(-2)uu} \;, \;\;\;\;\; \A_{(0)v} = - \frac{1}{b} \, h_{(-2)uv} \;, \;\;\;\;\; \A_{(0)\rho} = - \frac{F}{4}
\ee
\be
 A_{(2)i} = - \frac{\p_i F}{4} \;, \;\;\;\;\; A_{(2)\rho} =0\;, \;\;\;\;\; F \equiv k_u A_{(0)v} - k_v A_{(0)u}
\ee
Thus, the `T' modes of the gauge fields are completely determined in terms of $h_{(-2)ij}$ or vice versa. Now let us turn to
the metric solution. The equations of motion give\footnote{Here the equations of motion have been solved order by order in $\rho$.}
\be
h_{(-2)uv} = - \frac{b^2}{2}\, h_{(0)vv}\;, \;\;\;\;\; 2 k_u k_v \tilde{h}_{(0)uv}
= k_v^2 \tilde{h}_{(0)uu} \;, \;\;\;\;\; b^2 h_{(2)vv} = 2 \tilde{h}_{(0)uv}.
%= -2\, b \,k_v\, F.
\ee
%
%\be
% h_{uv}^{(2)} = \frac{1}{4} \, \tilde{R}_{(0)} =\frac{2}{b} \, k_u F =  k_u k_v h_{(-2)uu} - 2 k_u^2 h_{(-2)uv} \ee
The quantity $\tilde{R}_{(0)}$ has been defined in \eqref{curv}. Notice that the last
relation implies a nonlocal relationship between the coefficients $h_{(-2)ij}$ and $h_{(0)ij}$. Finally, the remaining equations of motion give
\be k_v h_{(2)uu} = k_u h_{(2)uv}\;,\;\;\;\;\;
 k_v h_{(2)uv} = k_u h_{(2)vv}
\ee
Combining the last equations we find that $\tilde{R}_{(0)}$ determines $h_{(2)ij}$ and $\tilde{h}_{(0)ij}$ as
\be
h_{(2)uu} = \frac{k_u}{4 k_v} \tilde{R}_{(0)} \;, \;\;\;\;\;
h_{(2)uv} = \frac{1}{4} \tilde{R}_{(0)} \;, \;\;\;\;\;h_{(2)vv} = \frac{k_v}{4 k_u} \tilde{R}_{(0)} \label{first-rel} \ee
\be
\tilde{h}_{(0)uu} = \frac{b^2}{4 } \tilde{R}_{(0)} \;, \;\;\;\;\;\tilde{h}_{(0)uv} = \frac{b^2 k_v}{8 k_u} \tilde{R}_{(0)} \ee
and a particular linear combination of the $h_{(-2)ij}$ as
\be
k_u k_v h_{(-2)uu} - 2 k_u^2 h_{(-2)uv} = b^2 \tilde{R}_{(0)}.
\ee
Using the first relation in \eqref{first-rel}, we obtain
\be
k_u h_{(-2)uu} = b^2 (k_v h_{(0)uu}-2 k_u h_{(0)uv}).
\ee
The `T' modes should correspond to the energy momentum tensor but are subject to the same subtleties as the `T' mode solutions of TMG.

\subsubsection*{The `X' modes}

The second set of independent solutions are the `X' modes; one can view these as physical modes, since unlike the `T' modes they propagate in the bulk.
We define the following function
\be
g(\rho) = \rho\A_\rho^X,
\ee
and substitute this into the equations of motion. We find
\bea
\A_u^X &=& \frac{2 i  k_u \rho}{k_v^2b^2} \, (k_v k_u g + 2 g' - 2 \rho g'')
+ \frac{i }{k_v \rho} \, (4 g + 3 k_v k_u \rho g - 4 \rho^2 g'') + \frac{i b^2 k_v g}{ \rho} \non \\
\A_v^X &=& - \frac{2 i   \rho}{k_v b^2} \, (k_v k_u g + 2 g' - 2 \rho g'')  - i k_v  g \;, \;\;\;\;\;\;\;\;\;\;\;\A_{\rho}^X =  \frac{ g}{\rho}.
\eea
Only one metric component is nonzero for this part of the solution, namely
\be
h_{uu}^X = \frac{8 i  }{b k_v^3  } (2 k_v k_u g + 4 g' +k_v k_u  \rho g' - 4 \rho g''-2 \rho^2 g''') + \frac{4 i b g'}{k_v }.
\ee
The equations of motion further require that $g(\rho)$ satisfy the fourth order equation:
\bea
&& \rho^4 g^{(4)} + 4 \rho^3 g^{(3)} - \left( \half b^2 k_v^2 \rho^2  + k_v k_u \rho^3 \right) g'' \nn \\
&& \qquad
- k_v k_u \rho^2 g' + \left( \half b^2 k_v^2 + \frac{b^4 k_v^4}{16}  + \frac{b^2 k_v^3 k_u \rho}{4}+ \frac{k_v^2 k_u^2 \rho^2}{4} \right)\, g =0 \label{eqg}
\eea
A feature of this equation is that it depends on $\rho$ only though the combination:
\be
x = \rho \, k_v\, k_u.
\ee
Note in particular that this variable $x$ is invariant under the dilatation symmetry.
In terms of the new variable, the equation \eqref{eqg} becomes:
\be \label{eq:g_x}
x^4 g^{(4)}(x) + 4 x^3 g^{(3)}(x) - (2 \a x^2 + x^3) g''(x) - x^2 g'(x) + (2 \a + \a^2 + \a x + \frac{x^2}{4}) g(x) =0
\ee
where
\be
\a \equiv \frac{b^2 k_v^2}{4}
\ee
and all primes now denote derivatives with respect to $x$.
Unfortunately, the exact set of solutions of this equation has not been found. The equation becomes exactly
solvable in terms of hypergeometric functions in the two limiting cases, (i) $x \ll a$ and (ii) $x \gg a$.
In particular, the approximation in case (ii) leads to the same equation as in the undeformed theory, $b=0$,
and we will discuss the exact solution for this case below.

However, one can obtain the general asymptotic solution of (\ref{eq:g_x}).
There are four independent
set of such solutions which to leading order
in the radial variable $x$ are:
\bea
g(x) &=& x_{(2-\Delta_1)} (\alpha) x^{\half - \half \sqrt{1+b^2 k_v^2}}
+ x_{(\Delta_1)} (\alpha) x^{\half + \half \sqrt{1+b^2 k_v^2}} \label{asyg2} \\
&& + x_{(2-\Delta_2)} (\alpha) x^{\half - \half \sqrt{9+b^2 k_v^2}} + x_{(\Delta_2)} (\alpha)
x^{\half + \half \sqrt{9+b^2 k_v^2}}.  \nn
\eea
where
\be
\Delta_1 = 1 + \sqrt{1+b^2 k_v^2} \;, \;\;\;\;\; \Delta_2 = 1 + \sqrt{9+b^2 k_v^2}.
\ee
Furthermore, the subleading terms in each independent solution are necessarily expanded as a power series in $x$, e.g. the first of the solutions
behaves as:
\be
x^{\half - \half \sqrt{1+b^2 k_v^2}} (x_{(2 - \Delta_1)} (\alpha ) + x_{(4 - \Delta_1)} (\alpha) x + x_{(6  - \Delta_1)} (\alpha) x^2 + \cdots ).
\ee
For generic values of the product $b^2 k_v^2$ the dimensions $(\Delta_1, \Delta_2)$ are irrational, and the asymptotic expansion of each independent solution
involves only radial powers.  The notation
used for $x_{(n)}$ indicates that in this case the coefficients scale as $x_{(n)} \rightarrow \lambda^{- n} x_{(n)}$ under the dilatation symmetry which
acts as $\rho \rightarrow \lambda^2 \rho$, $u \rightarrow \lambda^2 u$.
When the dimensions take rational values, however, logarithmic terms arise in the asymptotic expansion and
in this case the coefficients do not all scale homogeneously under the dilatation symmetry.

From the form of the asymptotic solution, one would expect that the coefficients $x_{(2-\Delta_i)}$
are related to the sources for dual operators with
dimensions $\Delta_i$, while $x_{(\Delta_i)}$ are associated with the corresponding vevs; we will explain this further below. Requiring smoothness\footnote{Strictly
speaking, the regular requirement should be imposed on the vector field itself, rather than on the function $g(\rho)$. However, the only feature we use
in what follows is that regularity imposes two conditions on the four independent solutions.} of the
solution for $g(\rho)$ in the interior of the spacetime should determine $x_{(\Delta_i)}$ in terms of $x_{(2-\Delta_i)}$ completely. Since these coefficients depends only on $\a$ the
regularity conditions can depend only on the quantity $\alpha$,
and thus will result in:
\be
x_{(\Delta_i)} = f_{i}(\alpha) x_{(2 - \Delta_i)}
\ee
where $f_{i}(\alpha)$ are functions of $\alpha$. This implies that the regular solution is
of the form
\be \label{vev-rel}
g(\r) = \sum_{i=1}^2 x_{(2-\D_i)} \r^{1 - \half \D_i} \left(1 + \cdots +
\r^{\D_i - 1} f_i(\alpha) k_v^{\Delta_i - 1} k_u^{\Delta_i - 1} + \cdots\right)
\ee
where the dots indicate subleading terms.
In order to determine the explicit form of $f_i(\alpha)$ we would need to solve the full equation for $g(x)$, but fortunately the
$k_u$-dependence is completely determined without knowing the full solution. Note that for rational values of the $\Delta_i$ this argument needs to
be refined: logarithmic terms can arise in the asymptotic expansion and the coefficients $x_{(\Delta_i)}$ do not then scale homogeneously under
dilatations. This happens for example at $b=0$ where the equation for $g(x)$ can be solved exactly: the two regular solutions are
\be
g(x) = g_1 x K_2 (\sqrt{2x}) + g_1 x^{-1/2} K_1 (\sqrt{2x}),
\ee
where $K_n(z)$ is the modified Bessel function of the second kind. Using the asymptotic expansion of these functions as $x \rightarrow 0$,
we indeed find logarithmic terms; these relate to the terms logarithmic in momentum in the renormalized correlation functions, as expected
for operators of integral dimension in a CFT.

It is also interesting to see how the $b^2 \rightarrow 0$ limit is reached. Let us rewrite the leading terms in the expansion of $g(\rho)$ as:
\bea
g(\rho) &=& i k_v g_{(2-\Delta_1)} \rho^{\half - \half \sqrt{1+b^2 k_v^2}} +
i b^2 k_v g_{(2-\Delta_2)} \rho^{\half - \half \sqrt{9+b^2 k_v^2}} \\
&& + i b^2 k_v g_{(\Delta_1)} \rho^{\half + \half \sqrt{1+b^2 k_v^2}} +
i k_v g_{(\Delta_2)} \rho^{\half + \half \sqrt{9+b^2 k_v^2}} +  \cdots \nn
\eea
where the ellipses denote the subleading terms and the $(x_{(1-\Delta)}, x_{(\Delta)})$ have been rescaled relative
to (\ref{asyg2}). The notation for $g_{(n)}$ again
indicates that the coefficient scales as $g_{(n)} \rightarrow \lambda^{- n} g_{(n)}$ under the dilatation symmetry which
acts as $\rho \rightarrow \lambda^2 \rho$, $u \rightarrow \lambda^2 u$. Substituting these terms back into the vector field components we find:
\bea
\A_{v} &=& \frac{4 g_{(2-\Delta_1)}}{b^2} \rho^{\half - \half \sqrt{1+ b^2 k_v^2}} (1 - \sqrt{1+ b^2 k_v^2})
- 4 g_{(2-\Delta_2)} \rho^{\half - \half \sqrt{9 + b^2 k_v^2}} (1+ \sqrt{9 + b^2 k_v^2})  \nn \\
&+& 2 g_{(\Delta_1)} \rho^{\half + \half \sqrt{1+b^2 k_v^2}} (1 + \sqrt{1+ b^2 k_v^2})
- \frac{g_{(\Delta_2)}}{b^{2}}  \rho^{\half + \half \sqrt{9 + b^2 k_v^2}} (6 - 2 \sqrt{9 + b^2 k_v^2}) + \cdots; \nn \\
\A_{u} &=& 4 g_{(2 - \Delta_1)} \rho^{-\half - \half \sqrt{1 + b^2 k_v^2}} - 4 b^2
g_{(2 - \Delta_2)} \rho^{-\half - \half \sqrt{9 + b^2 k_v^2}} \label{bnon} \\
&-&  6 b^2 g_{(\Delta_1)} \rho^{- \half + \half \sqrt{1+b^2 k_v^2}}
+ 4 g_{(\Delta_2)} \rho^{- \half + \half \sqrt{9 + b^2 k_v^2}}
+ \cdots. \nn
\eea
Taking the limit of $b^2 \rightarrow 0$ and retaining only the leading terms results in:
\bea
\A_{v} &=& - 16 g_{(-2)} \rho^{-1} - 4 k_v^2 g_{(0)} + 4 g_{(2)} \rho + 3 k_v^2 g_{(4)} \rho^{2} + \cdots;
\label{bzero} \\
\A_{u} &=& 4 g_{(0)} \rho^{-1} + 4 g_{(4)} \rho + \cdots \nn
\eea
where we have used $\Delta_1 \rightarrow 2$ and $\Delta_2 \rightarrow 4$.

Let us next review the holographic correspondence at $b=0$, namely for linearized perturbations around the AdS
background. The mass of the vector field is such that it corresponds to a vector operator of dimension three in the
dual conformal field theory. Solving the linearized equations of motion for the vector
around AdS results in an asymptotic expansion of the vector fluctuations of the form:
\be
\A_i = \frac{a_i (u,v)}{\rho} (1 + \cdots) + \td{a}_i (u,v) \rho (1 + \cdots),
\ee
where $i = (u,v)$ and the ellipses denote subleading terms in $\rho$. Then $a_i$ is a source for the dual vector
operator $X_i$ whilst its expectation value is related to the normalizable mode $\td{a}_i$. Comparing with (\ref{bzero}) we note that:
\be
a_{v} = - 16 g_{(2 - \Delta_2)}; \qquad
a_{u} = 4 g_{(2 - \Delta_1)}; \qquad
\td{a}_v = 4 g_{(\Delta_1)}; \qquad \td{a}_u = 4 g_{(\Delta_2)}, \label{zero-limit}
\ee
and thus the data $(g_{(2-\Delta_1)}, g_{(\Delta_1)})$ is associated with the operator $X_v$ whilst
the data $(g_{(2-\Delta_2)}, g_{(\Delta_2)})$ is associated with the operator $X_u$.

\subsection{Interpretation}

From the general linearized solution of the massive vector equations one can compute the two point functions of the dual stress energy tensor and of
the deforming vector operator. This calculation however requires systematic holographic renormalization, which is rather complex and will be discussed
elsewhere. Focusing on the `X' modes one can however make a number of interesting
preliminary observations. Let us first recall what happens at $b=0$ in the AdS background;
recall that the scaling dimensions with respect to the Schr\"{o}dinger symmetry are such that
$\Delta_s(X_v) = 2$ and $\Delta_s(X_u) = 4$, according to \eqref{dime}.
From the asymptotics of the bulk vector field one sees that
$a_u$ acts as a source for the $(2,1)$ operator $X_v$ whilst $a_v$ acts as a source for the $(1,2)$ operator $X_u$.

Working now to leading order in $b$, the field theory is deformed by a source for the operator $X_v$. The general arguments made
in section \ref{sec:fieldthy} indicate that both operators, $X_u$ and $X_v$, would be expected to acquire lightcone
momentum dependent anomalous dimensions, when $k_v \neq 0$.
 Looking at (\ref{bnon}), if one continues to interpret
$g_{(2 - \Delta_2)}$ as a source for the deformed operator $X^b_u$ away from $b =0$, and similarly interprets
$g_{(2- \Delta_1)}$ as a source for the deformed operator $X^b_v$, then
\be
\Delta_s(X^b_v) = \Delta_1 = 1 + \sqrt{1 + b^2 k_v^2}; \qquad
\Delta_s(X^b_u) = \Delta_2 = 1 + \sqrt{9 + b^2 k_v^2}.
\ee
These expressions are in agreement with (\ref{dime}) at $b=0$, and indicate that the anomalous dimensions at
finite $b^2$ take this closed form. Given that the coefficients $g_{(\Delta_1)}$ and $g_{(\Delta_2)}$ have dilatation
weights $(\Delta_1,\Delta_2)$ respectively, the vevs of the operators are expected to be of the form:
\be
\< X_{u} \> \sim g_{(\Delta_2)}; \qquad \< X_v \> \sim g_{(\Delta_1)},
\ee
since for generic non-rational values of $\Delta_i$ no other terms in the asymptotic expansion of $g(\rho)$ have
this dilatation weight. Using (\ref{vev-rel}), we can immediately infer that
\be
\< X_u (k) X_u(-k) \> \sim k_u^{\Delta_2 -1}; \qquad
\< X_v (k) X_v (-k) \> \sim k_u^{\Delta_1 - 1},
\ee
which is indeed of the form expected for operators of these scaling weights in a Schr\"{o}dinger invariant theory. The $k_v$ dependent
normalization can however only be determined using exact regular solutions of the linearized field equations together with holographic renormalization.

\section{Conclusions}
\label{sec:conclusions}

In this paper we have considered holography for $d+1$ dimensional Schr\"{o}dinger backgrounds and argued that they are dual to $d$ dimensional theories on Minkowski spacetime
obtained by deforming conformal field theories by irrelevant operators, which are however
exactly marginal from the perspective of the non-relativistic conformal group. In other words, they
describe a continuous deformation from a relativistic fixed point to a non-relativistic fixed point.
On the field theory side, we used conformal perturbation theory to study this system (so
the undeformed CFT can be either weakly or strongly coupled) and showed that the deforming
operator is indeed exactly marginal. We also studied how the dimensions of operators change, to leading order
in the deformation parameter $b^2$, when we deform the theory.
These results are in agreement with the results obtained using the
gravity dual, when linearized in $b^2$. The gravitational result, however, is valid for any $b^2$ and
resums the corrections into a closed, squared root form.
An important result is that the boundary counterterms obtained by holographic renormalization are
are non-local in the lightcone $v$ direction, implying that the boundary theory is not a local
QFT.

Working at the linearized level in the metric sector, we saw that
the general solution of the equations of motion blows up faster at the boundary than the background Schr\"{o}dinger solution. This was not unexpected, since the
linearized solutions correspond to operators that
are irrelevant with respect to the non-relativistic scaling symmetry, and therefore a finite source for this operator would be expected to change the asymptotic
structure of the spacetime. Similarly components of the stress energy tensor are also irrelevant with respect the Schr\"{o}dinger dilatation symmetry.
This feature is responsible for many subtleties in setting up holographic renormalization for Schr\"{o}dinger, and needs to be understood better before generalizing
the analysis to the non-linear level. The conserved
stress energy tensor of the dual theory couples to the vielbein, rather than the metric, and hence the correct holographic setup is to fix boundary conditions
for the vielbein, rather than the metric. The holographic dictionary for the gravitational sector is thus rather technically involved, even at the linearized level,
and it will be discussed elsewhere.

Throughout this paper we have noted a number of interesting open
questions. The Schr\"{o}dinger invariant theory should admit a
realization as a null dipole theory; for each ``ordinary" CFT one
would obtain the null dipole theory by replacing ordinary products
with null dipole products.  Such a realization should allow one to
compute the scaling dimensions $\Delta_s$ in the deformed theory in
terms of the conformal dimensions $\Delta$ in the ordinary
CFT. Moreover the analog of the Seiberg-Witten map
\cite{Seiberg:1999vs} between the null dipole and ordinary theories
should allow one to understand the detailed structure of the
counterterms. In particular, we noticed that in the holographic
realization the boundary counterterms for a scalar field in
Schr\"{o}dinger were simply related to those of a scalar field in AdS, and
this may well be a consequence of an underlying null dipole structure.
It would also be interesting to understand the counterterms better from
the field theory perspective; working perturbatively in $b^2$ we
expect a series of counterterms involving increasing numbers of
lightcone derivatives to be induced. This series of terms would then
be expected to resum into the structure obtained in the null dipole
theory.

\section*{Acknowledgments}

We would like to thank A. Adams, A. Dabholkar, A. Sen and A. Strominger for
discussions.
KS and MMT would especially like to thank Allan Adams for stimulating their
interest in Schr\"{o}dinger holography. KS and
MMT would like to thank the Aspen Center of
Physics for hospitality during the initial stages of this work
and the Simons Workshop in Mathematics and Physics 2010
for hospitality during the final stages of this work.
This work is part of the research program of the `Stichting voor
Fundamenteel Onderzoek der Materie (FOM)', which is financially
supported by the `Nederlandse Organisatie voor Wetenschappelijk
Onderzoek (NWO)'. The authors acknowledge support from NWO, KS via a
Vici grant, MMT via the Vidi grant ``Holography, duality and time
dependence in string theory" and BvR via an NWO Spinoza grant.

%\bibliographystyle{utphys}
%\bibliography{biblio_balt}

\end{document}